\documentclass[11pt,letterpaper]{article}
\usepackage{jheppub}

\usepackage{graphicx}
\usepackage{bbm}
\usepackage{amsmath}
\usepackage{amssymb}
\usepackage{mathtools}
\usepackage{amsthm}
\usepackage{mathrsfs}
\usepackage{marvosym}
\usepackage{dsfont}
\usepackage[labelformat=simple]{subcaption}
\usepackage[table,dvipsnames]{xcolor}
\usepackage{braket}
\usepackage{cleveref}
\usepackage{comment}
\usepackage{float}
\usepackage{fancybox}
\usepackage{soul} 
\usepackage[skins,theorems]{tcolorbox}
\usepackage{rotating}
\usepackage{longtable}
\usepackage{lscape}
\usepackage{array}
\newcolumntype{P}[1]{>{\centering\arraybackslash}p{#1}}
\tcbset{highlight math style={enhanced,
		colframe=black,colback=white,arc=0pt,boxrule=1pt}}

\allowdisplaybreaks

\definecolor{dark-gray}{gray}{0.20}
\definecolor{gray}{gray}{0.30}
\definecolor{light-gray}{gray}{0.80}
\definecolor{dark-red}{rgb}{0.7,0,0}
\definecolor{dark-green}{rgb}{0.1,0.4,0}
\definecolor{dark-blue}{rgb}{0.3,0.3,0.7}
\definecolor{light-blue}{rgb}{0.8,0.8,1}
\definecolor{swamp}{RGB}{240, 199, 197}

\def\be{\begin{equation}}
	\def\ee{\end{equation}}
\def\bea{\begin{eqnarray}}
	\def\eea{\end{eqnarray}}

\hypersetup{
	colorlinks=true,
	linkcolor=dark-blue,
	citecolor=dark-red,
	urlcolor=dark-green,
	linktoc=page
}

\theoremstyle{remark}

\newtheoremstyle{named}{}{}{\itshape}{}{\bfseries}{.}{.5em}{#3}
\theoremstyle{named}

\def \pa{\partial}
\def\a{\alpha}
\def\d{\delta}
\def\D{\Delta}
\def\s{\sigma}
\def\cK{{\cal K}}
\def\cV{{\cal V}}

\title{\centering Integral Scaling for EFT Strings from the Bottom-Up}

\author{Muldrow Etheredge$^a$, Alvaro Herr\'aez$^a$, Dieter L\"ust$^{a,b}$, and Luca Melotti$^a$}

\affiliation{$^a$Max-Planck-Institut f\"ur Physik, Boltzmannstrasse 8, 85748 Garching bei M\"unchen, Germany}
\affiliation{$^b$ Arnold Sommerfeld Center for Theoretical Physics,\\ Ludwig-Maximilians-Universit\"at M\"unchen, 80333 M\"unchen, Germany}

\emailAdd{muldrow@mpp.mpg.de, aherraez@mpp.mpg.de, luest@mpp.mpg.de, melotti@mpp.mpg.de}

\abstract{
Near the core of an EFT string in a 4d $\mathcal{N}=1$ theory, the scalars are dynamically driven to infinite field distance and a tower of states becomes light, with mass scaling with the string tension in Planck units as $m^2\sim \mathcal{T}^{\,w}$. According to the Integral Scaling Conjecture, $w$ takes only values 1, 2 and 3. In this paper, we examine how this conjecture can follow from the brane-taxonomy rules associated with the Emergent String Conjecture. In this context, we classify the relevant types of duality frames into 74 classes, identify which lattice sites can be relevant EFT candidates, and exhaustively test integral scaling for all of these candidates. We find that it holds with $w\leq 3$ for all the leading towers and also for the subleading towers below the species scale (up to half-integral subtleties that also appear in top-down examples). We further find evidence that the oscillator modes of EFT string candidates generate the lattices of particles and strings. Moreover, $w=1$ implies a perturbative string limit, but that the converse is not true. We compare our classification with concrete type IIA, F-theory and M-theory compactifications.
}

\begin{document}
	
\makeatletter
    \let\old@fpheader\@fpheader
	\renewcommand{\@fpheader}{\vspace*{-0.1cm} \hfill MPP-2026-128}
	\makeatother
	
	\maketitle
	\setcounter{page}{1}
	\pagenumbering{roman} 
 
\pagenumbering{arabic}

\newpage
\section{Introduction and Summary}
\label{sec:intro}

The Swampland program \cite{Vafa:2005ui} aims to identify the criteria that distinguish effective field theories (EFTs) admitting an ultraviolet completion in quantum gravity from those that do not (see \cite{Brennan:2017rbf,Palti:2019pca,vanBeest:2021lhn,Grana:2021zvf,Agmon:2022thq} for reviews). Apart from trying to delineate the boundaries of the landscape, this approach has produced a web of quantitative statements about gravitational EFTs, providing a bridge between quantum gravity and the low-energy physics that may describe our universe.

A central role in the Swampland program is played by the Distance Conjecture \cite{Ooguri:2006in}, which states that infinite-distance limits in the moduli space of a consistent theory of quantum gravity are accompanied by infinite towers of states whose masses decay exponentially with the traversed distance, at a rate expected to be an order-one constant in Planck units \cite{Klaewer:2016kiy,Etheredge:2022opl}. This behavior has been verified in a plethora of string constructions (see e.g.  \cite{Grimm:2018ohb,Grimm:2018cpv,Corvilain:2018lgw,Font:2019cxq}). The nature of the light towers is addressed by the \emph{Emergent String Conjecture} (ESC) \cite{Lee:2019wij}, according to which every infinite-distance limit is, in some suitable duality frame, either a decompactification limit (with the leading tower given by Kaluza-Klein modes) or an emergent string limit (in which the leading tower is formed by the oscillation modes of a unique critical string becoming tensionless). The ESC has also passed numerous top-down tests (see e.g. \cite{Lee:2018urn,Lee:2019apr,Lee:2019tst,Lee:2020gvu,Lee:2021usk,Alvarez-Garcia:2021pxo}).

The evidence for these conjectures mostly relies on the direct study of the asymptotic regions of string theory moduli spaces. A complementary strategy is to realize the corresponding large field excursions \emph{dynamically},  as flows sourced by the backreaction of extended objects, and to ask whether the physics of these objects can inform the structure of the towers \cite{Lanza:2020qmt,Lanza:2021udy}.\footnote{Other interesting dynamical flows are those related to black holes, in particula via the attractor mechanism \cite{Ferrara:1995ih,Ferrara:1997tw,Sen:2005wa}, revisited in the Swampland literature in \cite{Bonnefoy:2019nzv,Cribiori:2022nke,Calderon-Infante:2025pls}}. Objects of low codimension are the natural candidates to source such flows and, moreover, they are known to become light along infinite-distance limits, as exemplified by the towers of tensionless strings and branes of \cite{Font:2019cxq}. In codimension one, this strategy is also realized by the running solutions of dynamical cobordism \cite{Buratti:2021yia,Buratti:2021fiv,Angius:2022aeq}, motivated by the Cobordism Conjecture \cite{McNamara:2019rup} and further explored in \cite{Blumenhagen:2022mqw,Angius:2022mgh,Blumenhagen:2023abk,Calderon-Infante:2023ler,Angius:2023xtu,Huertas:2023syg,Angius:2023uqk,Angius:2024zjv,Huertas:2024mvy,Angius:2024pqk,Calderon-Infante:2026ymy,Makridou:2026jzy}. In this work we focus on the codimension-two case which, in 4d $\mathcal{N}=1$ theories, is captured by \emph{EFT strings} \cite{Lanza:2020qmt,Lanza:2021udy,Lanza:2022zyg}. These are $\frac{1}{2}$-BPS fundamental axionic strings, magnetically charged under the axions of the theory, whose backreaction drives the saxionic partners of these axions towards infinite field distance as the string core is approached. Along these flows the string tension becomes light in Planck units and, in fact, it has been conjectured that all infinite-distance limits in the saxionic field space can be realized as EFT string flows \cite{Lanza:2021udy}. By now, EFT strings have become a standard tool to probe 4d EFTs consistent with quantum gravity, with applications that include the Weak Gravity Conjecture for strings and axionic towers \cite{Heidenreich:2021yda,Cota:2022yjw}, quantum corrections at infinite distance \cite{Klaewer:2020lfg,Kaufmann:2026mha, Kaufmann:2026tsy}, strong coupling dynamics of 4d strings \cite{Marchesano:2022avb,Wiesner:2022qys}, quantum gravity bounds from worldsheet consistency \cite{Martucci:2022krl,Martucci:2024trp}, emergence \cite{Marchesano:2022axe}, the asymptotics of the moduli space curvature \cite{Marchesano:2023thx,Marchesano:2024tod} and of Yukawa couplings \cite{Casas:2024ttx}, tameness \cite{Grimm:2022sbl}, and emergent string limits in complex structure moduli space \cite{Hassfeld:2025uoy,Monnee:2025ynn}.

A particularly intriguing outcome of this program is the observation \cite{Lanza:2021udy} that, along the flow sourced by an EFT string, the characteristic mass scale $m_{\rm tow}$ of the leading tower and the EFT string tension $\mathcal{T}$ are related by
\begin{equation}
\label{eq:intro-integral-scaling}
\left(\dfrac{m_{\rm tow}}{M_{\rm Pl}}\right)^2 \sim \left(\dfrac{\mathcal{T}}{M_{\rm Pl}^2}\right)^{w}\,, \qquad w=1,2,3\, .
\end{equation}
That is, the scaling weight $w$ is not only quantized, but it takes values in a remarkably short list. The statement that \eqref{eq:intro-integral-scaling} holds along every EFT string flow is the \emph{Integral Scaling Conjecture} (ISC) \cite{Lanza:2021udy,Lanza:2022zyg}. Its validity was recently investigated in large families of top-down compactifications in \cite{Grieco:2025bjy}, where it was verified in all the examples considered and shown to extend, in refined form, also to subleading towers (see also \cite{GriecoRuizValenzuelaToAppear} for complementary upcoming work on how integral scaling constrains the possible towers of light states from the bottom-up). Despite this evidence, the ISC has so far remained an empirical observation, as there is no bottom-up proof or precise understanding of why the scaling weights should be integers, nor of why the list should stop at $w=3$. Filling this gap is the motivation of this paper.

Such a strategy is a recurrent one in the Swampland program. Swampland statements are often motivated from patterns observed across string compactifications, and complementary bottom-up rationales are essential both to test their generality beyond the explored regions of the string landscape and to uncover the principles behind them. Recent examples include bottom-up arguments for the Distance Conjecture and its species scale counterpart \cite{Stout:2022phm, Calderon-Infante:2023ler}, as well as for the ESC \cite{Basile:2023blg,Bedroya:2024ubj, Herraez:2024kux, Kaufmann:2024gqo}. A convenient language for this approach is that of $\alpha$-vectors. To a tower with moduli-dependent characteristic mass $m_{\rm tow}$ (or to a brane with tension $\mathcal{T}_p$) one associates the vector $\vec{\alpha}=-\vec{\nabla} \log (m_{\rm tow}/M_{\rm Pl})$ (or $\vec{\alpha}_p=-\vec{\nabla} \log (\mathcal{T}_p/M_{\rm Pl}^p)$), with the gradient taken with respect to the canonically normalized scalars. In these terms, the Distance Conjecture becomes a convex hull condition on these $\alpha$-vectors of the available towers \cite{Calderon-Infante:2020dhm,Etheredge:2022opl}, in analogy with the convex hull formulation of the Weak Gravity Conjecture \cite{Cheung:2014vva}. A similar condition captures the asymptotic decay of the species scale \cite{Dvali:2007hz,Dvali:2007wp,Dvali:2009ks,Dvali:2010vm} (i.e., the quantum gravity cutoff lowered with respect to the Planck scale by the towers, and systematically studied in the context of gravitational EFTs and string compactifications in \cite{vandeHeisteeg:2022btw,Cribiori:2022nke, Castellano:2022bvr,vandeHeisteeg:2023ubh, vandeHeisteeg:2023dlw,Castellano:2023aum,Calderon-Infante:2025ldq}) in terms of its own species scale vector \cite{Calderon-Infante:2023ler}, whose products with the $\alpha$-vectors of the leading towers follow a universal pattern \cite{Castellano:2023stg,Castellano:2023jjt}. These relations were recently systematized in the \emph{taxonomy} of infinite-distance limits \cite{Etheredge:2024tok}, whose approach can be roughly summarized as follows. Assuming the ESC, the $\alpha$-vectors of the towers that dominate the consecutive asymptotic regimes of a given limit (the so-called \emph{principal towers}) obey universal dot-product rules, which organize each duality frame into a \emph{frame simplex}. This framework was extended in \cite{Etheredge:2025ahf} to a \emph{brane taxonomy}, in which the $\alpha$-vectors of towers and branes are further constrained to populate lattices determined by the frame data. It is this framework that we take as our starting point.

In this paper we show that the ISC follows from the brane-taxonomy rules which, even if they incorporate extra assumptions, are ultimately rooted in the ESC. Our assumptions are the following. First, we assume the taxonomy rules of \cite{Etheredge:2024tok,Etheredge:2025ahf} to hold for all states parametrically below the species scale in each duality frame. Second, we identify \emph{EFT string candidates} with string lattice sites whose $\alpha$-vectors have norm $\sqrt{2/n_{\bf e}}$ for a positive integer $n_{\bf e}\leq 7$. As we review in section \ref{sec:review}, this behavior follows from the homogeneity of the K\"ahler potential in the asymptotic regimes of 4d $\mathcal{N}=1$ theories \cite{Lanza:2021udy,Grieco:2025bjy}, and the restriction $n_{\bf e}\leq 7$ covers all known top-down realizations. Under these assumptions, proving the ISC reduces to a tedious but finite check. We classify all the frame simplices that can arise in up to seven-dimensional slices of the moduli space of a 4d EFT, finding exactly 74 of them: 44 \emph{geometric} (all principal towers of KK type) and 30 \emph{stringy} (one principal tower of string oscillators). Of these, only 48 (18 geometric and all 30 stringy) contain EFT string candidates within the simplex or on its boundary. For each candidate in each of these frames we verify the integral scaling relation, conveniently recast as the exact statement \cite{Grieco:2025bjy}
\begin{equation}
\label{eq:intro-alpha-scaling}
\vec{\alpha}_{\rm str}\cdot \left(2\, \vec{\alpha}_{\rm tow}\right) = w\, |\vec{\alpha}_{\rm str}|^2\, ,
\end{equation}
for every tower that remains below the species scale along the flow, treating recursively the special loci (facets and pericenters) of each simplex.

We find that integral scaling holds in every single case, with integer $w\leq 3$ for the leading towers, thus completing the argument. Certain subleading towers can instead display half-integral weights, in a way that is both allowed by the taxonomy rules and realized in top-down examples \cite{Grieco:2025bjy}. Our scan also points to a stronger, lattice-level statement: the $\alpha$-vectors associated to the oscillation modes of EFT string candidates act as generators of the lattices of both particle and string $\alpha$-vectors. We prove this for all 1d and 2d frame simplices and verify it in many higher-dimensional ones, suggesting a promotion of the convex hull statements of \cite{Grieco:2025bjy} to full lattice statements, which are compatible with the half-integral scaling mentioned above. Moreover, the scaling weight acquires a sharp physical interpretation.  $w=1$ occurs precisely when the EFT string itself is the emergent string of the limit, whereas $w>1$ corresponds either to a decompactification or to an emergent string limit in which the lightest string is not the EFT string sourcing the flow. We comment on the extension of the analysis to bound states of EFT strings, showing that whenever a single monomial of the K\"ahler potential dominates within a frame simplex, integral scaling for bound states is inherited from that of their elementary constituents, with additive scaling weights. Finally, we find evidence that EFT string candidates with $n_{\bf e}>7$ always yield $w>3$ for the leading tower. This suggests that the ISC and the taxonomy rules combined bound the degree of homogeneity of the asymptotic K\"ahler potential as $n\leq 7$, in accordance with eleven being the maximal decompactification dimension, which would in turn justify a key assumption for the sharpened supersymmetric axion Weak Gravity Conjecture of \cite{Etheredge:2026rio}. This also fits nicely with \cite{Grieco:2025bjy}, where they argue that $n\leq 7$ implies $w\leq 3$.

We complement this bottom-up analysis with explicit top-down examples: type IIA compactifications on Calabi--Yau threefolds, F-theory compactifications on elliptically fibered fourfolds, and M-theory on $G_2$ orbifolds. These examples show how the frame simplices and lattice sites from the taxonomy classification are populated (or not) by actual EFT strings and towers, how certain growth sectors and duality frames are glued together along a given slice of moduli space, and how bound states are realized. In doing so, they also provide first steps towards a precise dictionary between the assumptions of the taxonomy framework, formulated in lattice language, and physical data such as the amount of supersymmetry, the compactification geometry and the available brane content. Our approach is complementary to that of \cite{Grieco:2025bjy}. Rather than starting from a given asymptotic form of the K\"ahler potential, we begin with a specific compactification and identify the frame simplices realized in its different growth sectors, recovering the corresponding scaling relations.

The rest of the paper is organized as follows. In section \ref{sec:review} we review the taxonomy rules and the main properties of EFT strings, recasting the ISC in the language of $\alpha$-vectors. We also summarize our assumptions in section \ref{ss.assumptions}. Section \ref{s:IntegralScalingProof} contains our main results, namely the classification of frame simplices, the scan over EFT string candidates and the verification of the integral scaling relations, together with comments on the extension to bound states and the generalization to lattice statements. In section \ref{sec:topdown} we analyze the top-down examples. We present our conclusions and outlook in section \ref{sec:conclusions}. Appendix \ref{ap:Homogeneousfunctions} derives the norm of EFT string $\alpha$-vectors from the homogeneity of the K\"ahler potential, and appendix \ref{ap:tables} collects the complete tables of our scan.

\section{EFT Strings, Taxonomy Rules, and Assumptions}
\label{sec:review}

\subsection{Brief Summary of Taxonomy Rules}
\label{ss.TaxonomyRules}

We now review the taxonomy rules of \cite{Etheredge:2024tok, Etheredge:2025ahf}.

Consider a geodesic $\gamma$ that travels to an infinite-distance-limit in a moduli space. By the Emergent String Conjecture \cite{Lee:2019wij}, one will find a set of light towers that are either KK-modes or string oscillators, with mass $m_1$. If the towers are KK-modes, the theory decompactifies to a higher-dimensional theory, and the geodesic either continues in the higher-dimensional theory's moduli space or terminates in the higher-dimensional theory (for example, if the higher-dimensional theory has no moduli, such as in M-theory). If the geodesic continues in the higher-dimensional moduli space, then the geodesic will result in the higher-dimensional theory having either a tower of KK-modes or string oscillators, with mass $m_2$. If the towers are KK-modes, then the geodesic decompactifies the theory to an even higher-dimensional theory, and this process continues until either the last tower is a set of string oscillators, or the geodesic terminates.

The set of towers one obtains this way are called the \textbf{\emph{principal towers}}. In lower-dimensional Planck units, they furnish a set of mass scales, $m_1<m_2<\dots<m_N$. The logarithms of these masses in Planck units furnish a set of radion and dilaton coordinates for an $N$-dimensional flat slice of moduli space called the \emph{\textbf{principal plane}}. The logarithmic moduli-derivatives, or $\alpha$-vectors defined as
\begin{align}
    \vec\alpha=-\nabla \log \frac{m}{M_{\rm Pl}} \, ,
\end{align}
 of these towers form a basis for the principal plane, and satisfy dot product rules \cite{Etheredge:2024tok}.\footnote{Here, and throughout this paper, these vectors are defined with respect to the moduli space metric, which appears in the low-energy Lagrangian density using the convention $\mathcal L=M_{\rm Pl}^2 \left(\frac 12R-\frac 12{{\cal G}_{ij}}(\partial \phi^i)(\partial \phi^j)\right).$} Namely, for a $d$-dimensional theory, given two principal towers with $\alpha$-vectors $\vec \alpha_{i,j}$, they satisfy\footnote{As explained in detail in \cite{Etheredge:2024tok}, this requires some additional technical assumptions, such as assuming that the decompactification limits are to theories satisfying Einstein's vacuum equations, and thus does not have warped decompactifications and sliding as described in \cite{Etheredge:2023odp}. See \cite{Etheredge:2024tok} for a complete discussion. In this paper, we take these rules as our starting point and assume they hold.}
\begin{align}        \label{tax_rules}
    \vec \alpha_i\cdot \vec \alpha_j=\frac 1{d-2}+\frac 1{D_i-d}\delta_{ij},
\end{align}
where, if the $i$th tower is a KK-mode, then $D_i$ is the spacetime dimension it decompactifies to, or $D_i=\infty$ if the $i$th-tower is a string oscillator mode. The convex hull of these principal-towers forms what is called the \emph{\textbf{frame simplex}}, and the rays within the frame simplex are said to be in the same \emph{\textbf{duality frame}}. In each spacetime dimension, requiring 11d to be the maximum decompactification dimension and 10d as the maximum dimension with perturbative string limits, there are only a finite number of frame simplices one can construct. In particular, in 4d theories, there are only 74 frame simplices one can construct. This will be key to our analysis.

In \cite{Etheredge:2025ahf}, the taxonomy rules for the principal towers were generalized into taxonomy rules governing also the heavier towers, and also branes. In particular, given a principal KK tower $\vec \alpha_\text{KK,$D$}$ decompactifying to $D$-dimensions, or a principal string oscillator tower $\vec \alpha_\text{ osc}$, the radion and dilaton lattice rules for the $\vec \alpha_p$-vector of a $p$-dimensional brane, defined as
\begin{align}       \label{alpha_brane}
    \vec \alpha_p=-\nabla \log \frac{{\cal T}_p}{M_{\rm Pl}^p},
\end{align}
are, in the case of a non-oscillator brane,
\begin{align}
    \vec \alpha_p\cdot \hat \alpha_\text{KK,$D$}&=\frac{p(D-2)-P(d-2)}{\sqrt{(D-d)(D-2)(d-2)}},\label{e.radionrule}\\
    \vec \alpha_p\cdot \hat \alpha_\text{osc}&=\frac p{\sqrt{d-2}}-\frac{\sqrt{d-2}}2(P-p), \label{e.dilatonrule}
\end{align}
where $P$ is an integer. For radion lattices, $P$ can often be thought of as the dimension of the brane in the $D$-dimensional theory, though KK-modes and monopoles have exotic values of $P$. For dilaton lattices, $P$ is related to the number of string-couplings appearing in the tension of the brane in string-units. These rules result in $\alpha$-vectors of branes and strings resulting on a lattice \cite{Etheredge:2025ahf}.

\subsection{EFT Strings}
\begin{figure}
    \centering
\includegraphics[width=0.65\textwidth]{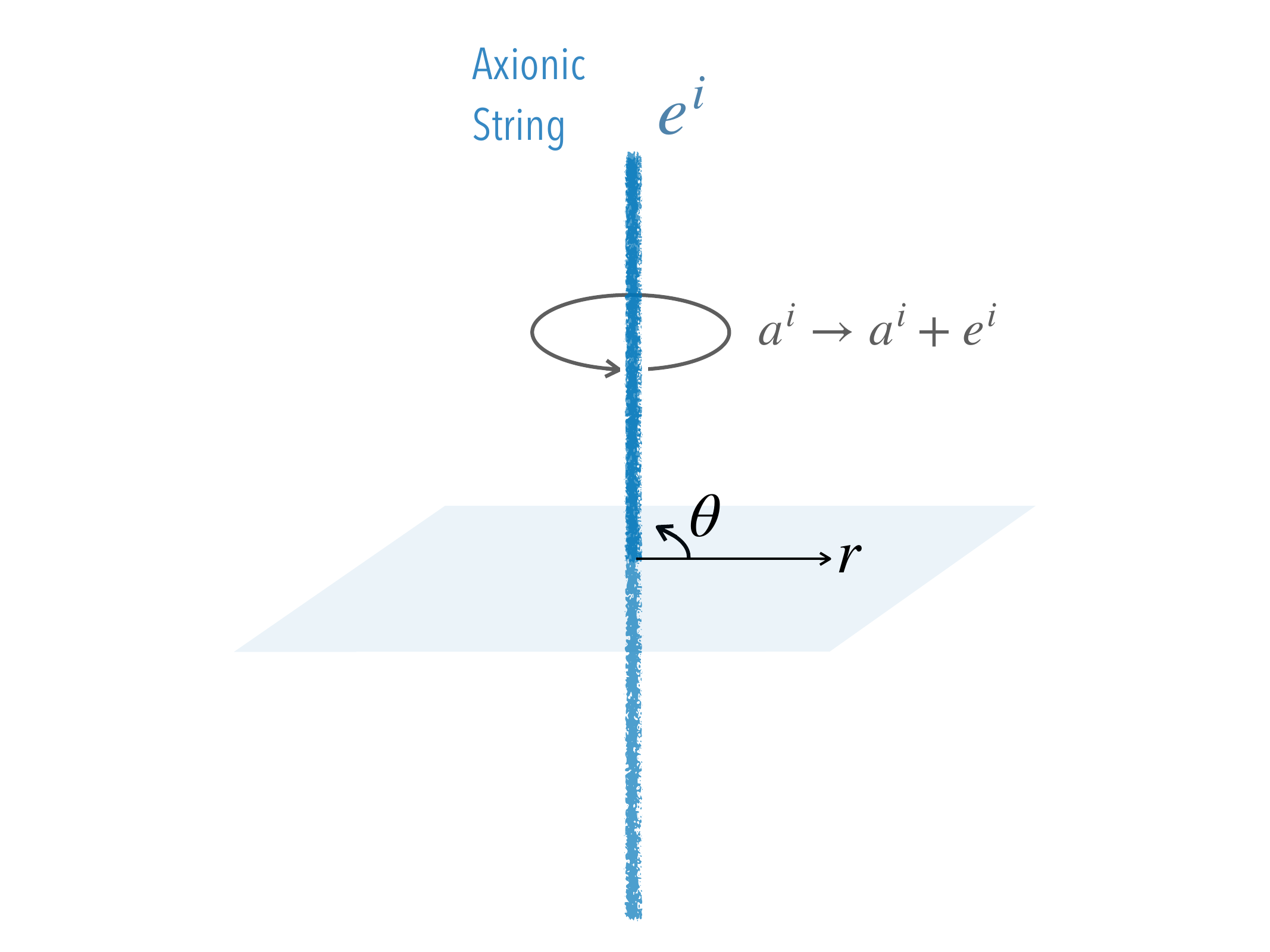}
    \caption{}
    \label{fig.EFTstring}
\end{figure}
In this section we summarize some key properties of EFT strings, as originally introduced in \cite{Lanza:2020qmt,Lanza:2021udy,Lanza:2022zyg}, and connect them to the taxonomy rules and notation introduced in section \ref{ss.TaxonomyRules}. For concreteness, we focus on 4d ${\cal N}=1$ theories. We label the scalar component of chiral fields by $t^i=a^i+is^i$, where $a^i$ represents an axion and $s^i$ its saxionic partner.\footnote{Cases with extended supersymmetry may also be included in this analysis by re-expressing them in $\mathcal{N}=1$ language.} The theory is then defined by a K\"ahler potential, $K(t^i)$, which enjoys axionic shift-symmetries (generically broken by non-perturbative effects) that become exact as we approach asymptotic regions of moduli space that lie near infinite distance. Following \cite{Lanza:2020qmt,Lanza:2021udy,Lanza:2022zyg}, we focus on fundamental, codimension-2 solutions, which are $\frac{1}{2}$-BPS. These EFT strings are \emph{fundamental axionic strings}, since they are magnetically charged under axions whose shifts are implemented as the string is encircled. For an EFT string with charges $e^i \in \mathbb{Z}_{\geq 0}$ under the axions $a^i$, solving the BPS flow equations results in the following field profiles
\begin{equation}
    \label{eq:EFTstringscalarflows}
    s^i\, =\, s^i_0 +e^i \sigma\, , \qquad a^i \, = \, a^i_0+e^i\dfrac{\theta}{2\pi}\,,  \qquad \sigma \equiv -\log \dfrac{r}{r_0}\, .
\end{equation}
As shown in Figure \ref{fig.EFTstring},  $r$ and $\theta$ are the radial and angular coordinates in the plane transverse to the EFT string, $r=0$ is the location of the string, and we have also defined the coordinate $\sigma$ for future use using some arbitrary reference radius, $r_0$. Following the scalar flow of this solution one sees that as the core of the string is approached, i.e. $\sigma\to \infty$, the saxionic partners of the axions under which the EFT string is charged are driven to infinite distance. Similarly, it can be seen how a shift of the axion $a^i$ by $a^i\to a^i+e^i$ is implemented when the string is encircled. It is thus consistent to use properties of the theory near the asymptotic regions of field space to study EFT string solutions near their core, since the scalars are dynamically driven precisely to those regions.

The tension of these EFT string solutions depends on the distance from the core at which it is measured, as explained in detail in \cite{Lanza:2021udy}, but it can be written in terms of the running scalars as
\be
\frac{{\cal T}_{\bf e}}{M_{\rm Pl}^2} = e^i \ell_i \, , \qquad \ell_i = -\frac{1}{2} \frac{\pa K}{\pa s^i} \, ,
\ee
where we introduced the so-called dual saxions, $\ell_i$. Crucially, the tension of EFT strings becomes light in Planck units as the infinite distance points in moduli space are approached along the flow. 

In fact, near an asymptotic region, the non-perturbative corrections that break the axionic shift symmetry are of order $\mathcal{O}(e^{2\pi i m_i t^i})$ for some integers $m_i \in \mathbb{Z}$ that define the so-called BPS instanton cone, $\mathcal{C}_I$, which contains the corresponding non-vanishing instanton contributions. As explained in \cite{Lanza:2021udy}, the asymptotic region of field space where these corrections can be neglected, and thus we recover the corresponding axionic shift symmetry, is the deep interior of the so-called \textit{saxionic cone}, which is defined as
\be
\D = \{ s^i \in \mathbb{R} \ / \ m_i s^i > 0 \, , \, \forall m_i \in {\cal C}_I \} \, .
\ee
We can then use \eqref{eq:EFTstringscalarflows} to restate such condition in terms of the charges of the EFT strings as $m_i e^i\geq 0$ for all non-vanishing $m_i$. This defines the set of charges whose flows stay inside such asymptotic region, the so-called EFT string cone \cite{Lanza:2022zyg}
\begin{equation}        \label{C_S^EFT}
    \mathcal{C}^{\rm EFT}_S=\left\{ e^i \in \mathbb{Z}_{\geq 0 } \ / \  m_i e^i\geq 0 \quad  \forall m_i\ \in \mathcal{C}_I \right\} \, .
\end{equation}
Importantly, this means that the set of EFT strings enjoys a conical structure, with the cone being generated by \emph{elementary} EFT strings, which can be thought of as EFT strings magnetically charged only under one axion, and with unit charge. All other EFT strings, including possible bound states with multiple charges, can be generated as linear combinations of these elementary ones.

Since EFT strings in the aforementioned cone drive the saxions towards infinite distance points where the axionic shift symmetry of the K\"ahler potential becomes exact, one can write the moduli space metric as
\begin{equation} 
\label{Gij}
{\cal G}_{ij} = \frac{1}{2} \frac{\pa^2 K(s^i)}{\pa s^i \pa s^j} \, .
\end{equation} 
The physical charge of the string (in 4d Planck units as) is then  ${\cal Q}^2_{\bf e} = {\cal G}_{ij} e^i e^j \, $. Furthermore, using \eqref{alpha_brane} one computes the $\alpha$-vector for the EFT string
\begin{equation}
\label{eq:EFTstringalphavector}
    (\a_{\rm str})^i=-{\cal G}^{ij}\dfrac{\partial}{\partial s^j}\log \left( \dfrac{{\cal T}_{\bf e}}{M_{\rm Pl}^2}\right)=\dfrac{e^i}{{\cal T}_{\bf e}/M_{\rm Pl}^2}\, ,
\end{equation}
which points in the direction of the charge vector, $e^i$, and has norm $|\vec{\a}_{\rm str}|={\cal Q}_{\bf e} M_{\rm Pl}^2/{\cal T}_{\bf e}$.

Moreover, being at the asymptotic regions of moduli space we assume the following form for the K\"ahler potential
\begin{equation}
\label{Kahler_pot}
    K = -\log P(s^i) 
\end{equation}
where $P(s)$ is a homogeneous function of positive integer degree $n$ of the saxions. This includes all known examples from supersymmetric string compactifications as long as we restrict the analysis to asymptotic limits near infinite distance, where non-perturbative corrections can be neglected and the K\"ahler potential can be approximated by the homogeneous polynomial.\footnote{To be precise, in general this may hold only monomial-by-monomial in different growth sectors, each of them having different degree $n_{\bf e}$ for the leading monomial, see e.g., \cite{Grimm:2018cpv}.} Following the flow of the saxions along the EFT solution with charges $e^i$ given in \eqref{eq:EFTstringscalarflows} we get the scaling
\begin{equation}
    P(s^i)\sim \sigma^{n_{\bf e}}
\end{equation}
with $n_{\bf e}$ some integer such that $0 < n_{\bf e} \leq n$. This form of the K\"ahler potential allows us to re-write the norm of the corresponding EFT string as (see Appendix \ref{ap:Homogeneousfunctions} for a more general derivation of this result based on the homogeneity of $P(s^i)$).
\begin{equation}
\label{eq:EFTstringnorm}
    |\vec{\a}_{\rm str}|=\sqrt{\dfrac{2}{n_{\bf e}}}\, .
\end{equation}
This property of the $\a$-vectors of EFT strings, already observed in \cite{Grieco:2025bjy}, provides the defining equation that we will use for \emph{EFT string candidates} throughout the paper.

According to the Distance Conjecture \cite{Ooguri:2006in}, infinite towers of states become light as infinite distance points are approached. One of the original motivations to study dynamical realizations of such infinite distance field excursions was trying to understand whether the properties of such flows could inform the nature of the towers becoming light \cite{Lanza:2020qmt}. A particularly intriguing connection between the mass of the lightest tower of states along the field trajectory dictated by an EFT string flow and the tension of such EFT string  was observed in \cite{Lanza:2021udy}
\begin{equation}
\label{eq:IntegralScaling}
   \left(\dfrac{m_{\rm tow}}{M_{\rm Pl}} \right)^2 \sim\left(\dfrac{{\cal T}_{\bf e}}{M_{\rm Pl}^2} \right)^w\, \qquad w=1,\, 2, \, 3,\, 
\end{equation}
and promoted to a general property of EFT strings and leading towers. This is the so-called \emph{Integral Scaling Conjecture}. The generality and validity of this conjecture was recently explored in large families of top-down string compactifications in \cite{Grieco:2025bjy}, where it was noticed that for \emph{elementary} EFT string flows this integral scaling is not only fulfilled by the leading towers (as the original conjecture proposed), but also for the subleading light towers that are \emph{generators} of the tower convex hull \cite{Calderon-Infante:2020dhm, Etheredge:2024tok}.\footnote{We will come back to the observation that only the \emph{generators} of the convex hull of particles fulfill integral scaling, as opposed to all light towers, in section \ref{s:IntegralScalingProof}, where we discuss particular cases of half-integral scaling for these subleading towers which are allowed by taxonomy and also realized in the top-down examples of \cite{Grieco:2025bjy}.} The goal of this work is to characterize whether these integral scaling relations, or related versions thereof, can be proven from a bottom-up perspective using the framework of (brane) taxonomy \cite{Etheredge:2024tok, Etheredge:2025ahf} (see also \cite{GriecoRuizValenzuelaToAppear} for future complementary work on how integral scaling relations constrain towers of light states from a bottom-up perspective). 

Using \eqref{eq:EFTstringalphavector} and \eqref{eq:EFTstringnorm}, one can rewrite the Integral Scaling Relation \eqref{eq:IntegralScaling} between an EFT string (with charges $e^i$) and a light tower (with $\a$-vector $\vec\a_i$), along the flow generated by such EFT string, as \cite{Grieco:2025bjy}
\begin{equation}
\label{e.wi}
    \vec \alpha_{\rm str}\cdot (2\vec \alpha_i)=w_i |\vec \alpha_{\rm str}|^2\, .
\end{equation}
Throughout this paper we will use the defining equation \eqref{eq:EFTstringnorm} extensively to find EFT string candidates, and test whether eq. \eqref{e.wi} is fulfilled for light towers, for any integer $w_i$. This is at the core of the bottom-up analysis of the integral scaling relations in section \ref{s:IntegralScalingProof}.

\subsection{Assumptions}\label{ss.assumptions}
In this paper, we will make three main assumptions:
\begin{enumerate}
    \item \textbf{Taxonomy rules}: We will assume that we are working in a principal plane, and that the brane-taxonomy dot product rules \eqref{e.radionrule} and \eqref{e.dilatonrule} apply to all towers and branes parametrically below the species scale.
    \item \textbf{Maximum spacetime dimensions}: We will assume that 11d is the maximum spacetime dimension allowed for decompactification limits, and that 10d is the maximum spacetime dimension allowed for perturbative string limits.
    \item \textbf{EFT string length formula}: We will assume that EFT strings have $\alpha$-vectors of length $\sqrt{2/n}$, where $n$ is a positive integer less than or equal to $7$.
\end{enumerate}

As explained in \cite{Etheredge:2024tok, Etheredge:2025ahf}, the first assumption actually can be derived from other conditions that frequently appear in the string landscape when the Emergent String Conjecture holds.\footnote{See also \cite{Aoufia:2026ztl}.} For example, this involves assumptions, such as that the decompactification limits of the Emergent String Conjecture decompactify to solutions satisfying Einstein's equations in vacuum, which is not always the case since there are counterexamples to this that result in a phenomenon called ``sliding" \cite{Etheredge:2023odp}. For simplicity, we assume only the taxonomy rules, rather than the detailed assumptions that can be used to derive the rules. In section \ref{sec:topdown}, we explore top-down UV complete examples of EFT strings, and we see that the physics of these examples align with these taxonomic assumptions, confirming that our ESC/taxonomy explanation of the ISC is capturing its essence.

The assumption on the maximum spacetime dimensions is supported by all known string theory examples. In certain contexts, it can be derived from the brane taxonomy rules \cite{Etheredge:2025ahf}, but this requires very strong assumptions about heavy towers and no sliding. Thus, these assumptions do not hold generally and do not apply to theories like bosonic string theory.

The assumption of EFT strings having $\alpha$-vectors of length $\sqrt{2/n}$, with $n\leq 7$ is consistent with previous literature on the subject, and is discussed in the previous subsection. It is also related to the maximum spacetime dimension being $11$.

With these assumptions made, we now show that the integral scaling conjecture can be proven for non-bound states.

The assumptions above are formulated in the lattice language of the taxonomy framework, and it is worth making their physical content as explicit as possible. Assumption 1 encodes several physical inputs. The ESC restriction on suitable towers within each duality frame, the requirement that decompactification limits correspond to vacua of the higher-dimensional theory (excluding the warped situations leading to sliding, see e.g., \cite{Etheredge:2023odp,Raucci:2026fzp}), the approximate flatness of the asymptotic slice of moduli space (which, in the language of section \ref{sec:topdown}, is guaranteed  inside each growth sector, since a single monomial dominates the K\"ahler potential), and the validity of the lattice rules for all states parametrically below the species scale. Assumption 2 is the lattice version of two familiar facts, namely the absence of supergravity theories above eleven dimensions and of weakly coupled critical strings above ten. Assumption 3 can be derived from the homogeneity and integrality of the asymptotic K\"ahler potential, with the degree bounded by the number of compact dimensions, i.e., $n\leq7$ (see also \cite{Grieco:2025bjy}).

Importantly, the taxonomy lattices themselves are universal, in the sense that they only depend on the frame simplex. The model-dependent information, such as the amount of supersymmetry, the compactification geometry or the available brane content, seems to be captured by which lattice sites are actually populated, by which kind of objects, and even whether such infinite distance directions in field space are obstructed. The examples of section \ref{sec:topdown} illustrate this. The same $(\infty,6)$ frame simplex is realized both in F-theory on $\mathbb{P}^3$ and in type IIA on the quintic, but the two theories populate different subsets of EFT string candidates, and even the same site can be populated by a BPS string in one realization and by a non-BPS string in another, as for the fundamental string at the site $(2,6)$. Moreover, in all our examples the EFT strings populating the string lattice sites arise from BPS branes wrapping suitable cycles, such as NS5-branes on Nef divisors or D3-branes on movable curves. This is the geometric data that defines the EFT string cone $\mathcal{C}_S^{\rm EFT}$. In this sense, the question of which lattice sites are populated encodes the information about the underlying compactification. Sharpening this dictionary, ideally to the point of predicting which sites are populated directly from the physical data, is a natural long-term goal of this approach.

\section{Integral Scaling from Brane Taxonomy}
\label{s:IntegralScalingProof}
This section contains the main results of this paper. It includes the systematic, bottom-up proof of different integral scaling relations along EFT string flows. We begin by outlining the logic that guides our analysis, which proceeds in a systematic and recursive manner by sequentially increasing the dimensionality of the asymptotic region of moduli space that we study. We then analyze 1d moduli and 2d moduli spaces in detail, and then summarize our findings for all possible higher-dimensional cases. We finish with a discussion of bound states and possible generalizations of the integral scaling relations considered here.

\subsection{Outline of Approach}    \label{ss.outline}
To systematically determine the validity of the integral scaling relations, \eqref{eq:IntegralScaling}, we follow these steps: 
\begin{itemize}
    \item[1.]{We begin by characterizing all the possible towers that are compatible with the \emph{taxonomy rules} introduced in \cite{Etheredge:2024tok} and summarized in section \ref{ss.TaxonomyRules}. To that end, we catalog all possible \emph{frame simplices}, which we divide in \emph{Geometric} and \emph{Stringy}.
    \begin{itemize}
        \item[(i)]{Geometric frame simplices are those where each principal tower is a KK-tower. They must not result in a duality frame that is for a theory with more than 11 dimensions.}
        \setlength{\arrayrulewidth}{0.2mm}
\renewcommand{\arraystretch}{1.3}
\begin{table}[tb]
\begin{center}
\begin{tabular}{|c|c|}
\hline
\rowcolor{lightgray} $N$ & Geometric frame simplices \\
\hline
1 & $1,2,3,4,5,6,7$ \\
2 & $(1, 1), (2, 1), (3, 1), (2, 2), (4, 1), (3, 2), (5, 1), (4, 2), (3, 3), (6, 1), (5, 2), (4, 3)$ \\
3 & $(1, 1, 1), (2, 1, 1), (3, 1, 1), (2, 2, 1), (4, 1, 1), (3, 2, 1), (2, 2, 2), (5, 1, 1), (4, 2, 1), (3, 3, 1), (3, 2, 2)$ \\
4 & $(1, 1, 1, 1), (2, 1, 1, 1), (3, 1, 1, 1), (2, 2, 1, 1), (4, 1, 1, 1), (3, 2, 1, 1), (2, 2, 2, 1)$ \\
5 & $(1, 1, 1, 1, 1), (2, 1, 1, 1, 1), (3, 1, 1, 1, 1), (2, 2, 1, 1, 1)$ \\
6 & $(1, 1, 1, 1, 1, 1), (2, 1, 1, 1, 1, 1)$ \\
7 & $(1, 1, 1, 1, 1, 1,1)$ \\
\hline
\end{tabular}
\caption{Geometric frame simplices for any $N$-dimensional slice of moduli space. A frame simplex is identified by the number of decompactifying dimensions $D_i-4$ along its generators.}
\label{tab:geom.simplices}
\end{center}
\end{table}

        \item[(ii)]{Stringy frame simplices are those where at least one principal tower is a string oscillator mode. They must not have a duality frame where the theory is more than 10 dimensional.} 
\setlength{\arrayrulewidth}{0.2mm}
\renewcommand{\arraystretch}{1.3}
\begin{table}[tb]
\begin{center}
\begin{tabular}{|c|c|}
\hline
\rowcolor{lightgray} $N$ & Stringy frame simplices \\
\hline
1 & $\infty$ \\
2 & $(\infty, 1), (\infty,2), (\infty,3), (\infty,4), (\infty,5), (\infty,6)$ \\
3 & $(\infty, 1, 1), (\infty, 2, 1), (\infty, 3, 1), (\infty, 2, 2), (\infty, 4, 1), (\infty, 3, 2), (\infty, 5, 1), (\infty, 4, 2), (\infty, 3, 3)$ \\
4 & $(\infty, 1, 1, 1), (\infty, 2, 1, 1), (\infty, 3, 1, 1), (\infty, 2, 2, 1), (\infty, 4, 1, 1), (\infty, 3, 2, 1), (\infty, 2, 2, 2)$ \\
5 & $(\infty, 1, 1, 1, 1), (\infty, 2, 1, 1, 1), (\infty, 3, 1, 1, 1), (\infty, 2, 2, 1, 1)$ \\
6 & $(\infty, 1, 1, 1, 1, 1), (\infty, 2, 1, 1, 1, 1)$ \\
7 & $(\infty, 1, 1, 1, 1, 1,1)$ \\
\hline
\end{tabular}
\caption{Stringy frame simplices for any $N$-dimensional slice of moduli space. A frame simplex is identified by the number of decompactifying dimensions $D_i-4$ along its generators, where an emergent string limit has $D_i=\infty$.\label{tab:str.simplices}}
\end{center}
\end{table}
    \end{itemize}
     This results in a finite number of possible duality frames, and thus lattices, to consider. The exact number is \emph{74 frames}, summarized in Tables \ref{tab:geom.simplices} and \ref{tab:str.simplices}, which contain the 44 geometric simplices and the 30 stringy simplices, respectively.
    }
    
    \item[2.]{Each of these 74 frames produces a lattice of string $\alpha$-vectors, which is constrained by the \emph{brane taxonomy} rules introduced in \cite{Etheredge:2025ahf}. After constructing this lattice, we record all lattice sites that satisfy eq. \eqref{eq:EFTstringnorm}, namely those that have length $\sqrt{2/n}$ for some integer, $n\leq7$, and we call these \emph{EFT string candidates}. Some geometric simplices do not include any EFT string candidates that become light, reducing the number of geometric simplices we need to scan over from 44 to 30. To the best of our knowledge, none of these 14 geometric simplices has a top-down realization in string theory. Furthermore, if we restrict to simplices that contain EFT strings \emph{within} the simplex, or at its boundaries, we further reduce the number of geometric simplices from 30 to 18 (recorded in Table \ref{tab:geomEFTstr}). The 30 original stringy simplices have EFT strings within or at the boundaries of the simplex, and these are recorded in Table \ref{tab:stringyEFTstr}
    }

    \item[3.]{Finally, for each EFT string candidate within or at the boundary of a given simplex, we check eq. \eqref{e.wi} for every tower generating the simplex. That is, we verify whether it can be fulfilled for each $\vec{\a}_i$ for an integer-valued $w_i$. This means that we investigate the integral scaling relation not only for the leading tower along the flow, but for all towers that are lighter than the corresponding species scale (i.e., with $\alpha$-vectors larger than the species scale vector \cite{Calderon-Infante:2023ler} along the flow). We do this for each EFT string candidate in each of the 48 simplices.
    } 
\end{itemize}

A few comments are in order. First, it is natural to question why we restrict ourselves to EFT strings with $\a$-vectors pointing inside or along the boundaries of a given frame simplex. The reason is that, by construction of the frame simplex (see \cite{Etheredge:2024tok}), following trajectories that leave it means exploring to a different duality frame. In other words, the set of towers that contribute to the species scale changes. Thus, if we want to study the flow associated to an EFT string outside a given simplex, two things can happen. On the one hand, it is possible that such field excursion cannot be followed because no perturbative duality frame exists, due to large quantum corrections or obstructions. If that is the case, we do not need to consider that EFT string flow. On the other hand, if another perturbative duality frame can describe such field excursion, then by construction it can be described by a trajectory contained in another frame simplex, which can be appropriately glued to the original one. If that is the case, the EFT string under consideration is also detected by analyzing the EFT strings inside or at the boundaries of the new frame simplex. Given that our analysis involves a full scan over all possible frame simplices, the flow associated to such EFT string is automatically included in the scan. The key point is that, by checking integral scaling along EFT string flows that are contained within the simplex (or at its boundaries), we keep track of integral scaling for all towers whose masses lie below (or parametrically at) the species scale along the flow. This is ensured by the construction of a frame simplex. The only light towers whose integral scaling we do not check explicitly along a given EFT string flow are those that become massless in 4d Planck units but are parametrically above the species scale, as their description above such quantum gravity cutoff is not reliable.

Second, and as explained in detail in sections \ref{ss.2dmodulispaces} and \ref{high_d_modsp}, frame simplices contain special loci inside which EFT strings are not directly detected by the procedure described in step 2 above. However, these special loci happen to be lower-dimensional frame simplices, and these extra EFT string candidates are obtained by applying step 2 directly to them. Since our procedure involves recursively analyzing all the frame simplices from 1d to 7d, we cover all possible EFT string candidates, including the ones that appear at said special loci.

Finally, it is important to highlight that our scan of string lattice sites is only guaranteed to capture \emph{elementary EFT strings}. As discussed in section \ref{ss.bound}, for EFT strings that are bound states, their $\alpha$-vectors can either slide to \emph{elementary} $\alpha$-vectors along their flow, or lie at pericenters of the convex hull of $\alpha$-vectors of strings that bind together to produce the bound state (this was already observed in certain top-down examples in \cite{Grieco:2025bjy}).\footnote{See also the discussion in \ref{ss.bound} for an extra possible complication if the K\"ahler potential does not fulfill certain assumptions.} In the former case, the above approach applies and one does not need to do any extra verification. In the latter case, we show in section \ref{ss.bound} that integral scaling along the elementary EFT string candidates guarantees integral scaling along the flows generated by the bound states.

In the following we systematically present the results of this approach applied to all the relevant frame simplices, starting from the 1d ones and going all the way up to the 7d ones.

\subsection{1-dimensional Moduli Spaces}        \label{1d_modsp}

We first classify 1-dimensional moduli-spaces. Here, there are eight duality frames to consider: decompactifications to 5-, 6-, $\dots$, 11-dimensions, which we call 1-, 2-, \dots, 7-frames, and a perturbative string limit, which we call an $\infty$-frame.

For a $(p-1)$-brane, the lattice rules are, for decompactifications and dilatons,
\begin{align}
\begin{aligned}
    -\nabla \log \frac{{\cal T}_p}{M_{\rm Pl}^p}&=\frac{p(D-2)-2P}{\sqrt{2(D-4)(D-2)}}\hat \rho \, ,& &\text{for radions,}\\
    -\nabla \log \frac{{\cal T}_p}{M_{\rm Pl}^p}& =\frac {2p-P}{\sqrt 2}\hat \phi ,&&\text{for dilatons,}
\end{aligned}
\end{align}
where $P$ is an integer. For decompactification limits, the KK-mode is given by $p=1$ and $P=0$, and for perturbative string limits, the string oscillator mode is given by $p=1$ and $P=1$.

For strings with $\alpha$-vectors of length $\sqrt{2/n}$ in 4d theories with a 1d radion lattice, their $n$-values must be
\be
n_\text{1d radion}= \frac{(D-4) (D-2)}{(D-2-P)^2} ,\label{e.n1d}
\ee
and requiring $n$ be an integer requires $P$ and $D$ to satisfy compatibility conditions: namely that $(D-2-P)^2$ divides $(D-4)(D-2)$. Meanwhile, for dilaton directions, strings of length $\sqrt{2/n}$ in 4d theories must have
\begin{align}
    n_\text{1d osc}=\frac 4{(4-P)^2},
\end{align}
which is equivalent to $P=2,3,5,6$.

Finally, for radions and dilatons, the scaling value $w$ is given by
\begin{align}
\begin{aligned}
    w_\text{KK}&=\frac{2\vec \alpha_2\cdot \vec \alpha_\text{KK}}{\alpha_2^2}=\frac{D-2}{D-P-2},\\
    w_\text{osc}&=\frac{2\vec \alpha_2\cdot \vec \alpha_\text{osc}}{\alpha_2^2}=\frac{2}{4-P}.
    \end{aligned}
\end{align}
Thus, integral scaling requires $\frac{D-2}{D-P-2}$ to be integer for KK-modes, and $\frac{2}{4-P}$ to be integer for string oscillator modes.

\subsubsection{1-frame}
We first consider a 1-frame, which is a decompactification limit to a 5d theory. The $\alpha$-vectors of branes are governed by the $D=5$ case of the radion lattice formula \eqref{e.radionrule}, which for $(p-1)$-branes gives
\begin{align}
    \vec \alpha_p &=\frac{3p-2P}{\sqrt 6}\hat \rho ,
\end{align}
where $P$ is an integer and $\hat \rho$ is the radion direction. For non-oscillator particles and strings, $p=1,2$ respectively, and the $\alpha$-vectors for particles and strings are given by
\begin{align}
    \vec \alpha_1=\frac{3-2P}{\sqrt 6}\hat \rho,\qquad \vec \alpha_2=\frac{6-2P}{\sqrt{6}}\hat \rho .
\end{align}
In particular, the $P=0$ case of the particle towers is the KK-mode, and it has $\vec \alpha_\text{KK}=\sqrt{\frac 32}\hat \rho$.

\begin{figure}
    \centering
    \includegraphics[width=0.7\linewidth]{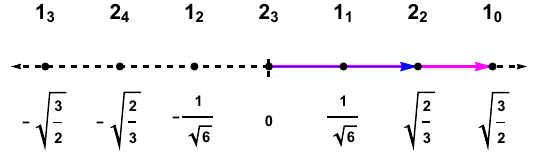}
    \caption{$\alpha$-vectors of the 1-frame. $\alpha$-vectors of strings and particles are labeled by $p_P$, where $p=1$ for particles and $p=2$ for strings. The magenta arrow points to the KK-mode at $1_0$, and the blue arrow points to the EFT-string candidate at $2_2$, which satisfies integral scaling: its oscillator mode, located at $1_1$, is a multiple $1/w$ of the $\alpha$-vector of the KK-mode, with $w=3$.}
    \label{f.4d5D}
\end{figure}

We are interested in EFT-string candidates, which have $\alpha$-vectors of length $\sqrt{2/n}$ with $n$ an integer. To obtain an integer $n$, we require the following to be an integer,
\begin{align}
    n=\frac{3}{(3-P)^2},
\end{align}
which requires $P=2,4$ and produces $n=3$. Only the $P=2$ vector becomes light in the direction of the decompactification. This can be seen in Figure \ref{f.4d5D}. 
The $\alpha$-vector of the KK-mode, which is labeled by $1_0$, is 3-times the length of the EFT-string candidate oscillator mode, which is at position $1_1$. This results in $w=3$.

\subsubsection{2-frame}
\begin{figure}
    \centering
    \includegraphics[width=0.7\linewidth]{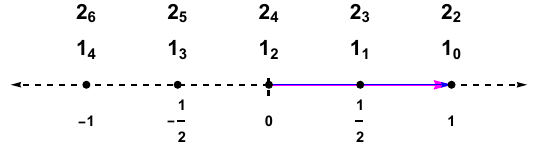}
    \caption{$\alpha$-vectors of the 2-frame. $\alpha$-vectors of strings and particles are labeled by $p_P$, where $p=1$ for particles and $p=2$ for strings. The magenta arrow points to the KK-mode at $1_0$, and the blue arrow points to the EFT-string candidate at $2_2$, which satisfies integral scaling: its oscillator mode, located at $1_1$, is a multiple $1/w$ of the $\alpha$-vector of the KK-mode, with $w=2$.}
    \label{f.4d6D}
\end{figure}

We next investigate 2-frames, corresponding to decompactifications from 4d to 6d. Here, the $\alpha$-vectors of non-oscillator particles and strings are given by
\begin{align}
    \vec \alpha_1=\frac{1}{4} (4-2 P),\qquad \vec \alpha_2=\frac{1}{4} (8-2 P).
\end{align}
EFT string candidates have $n$-values of
\begin{align}
    n=\frac{8}{(P-4)^2},
\end{align}
which is integer only if $P=2,6$, producing $n=2$, or $P=3,5$, producing $n=8$. The only EFT string candidates that become light in the decompactification limit are the $P=2,3$ cases. Furthermore, the integral scaling $w$-value is
\begin{align}
    w=\frac{4}{4-P}.
\end{align}
Here, we find that for $P=2$, $w=2$, and for $P=3$, $w=4$, i.e.
\begin{align}
    w_{n=2}=2,\qquad w_{n=4}=4.
\end{align}
So, only the $P=2$ and $w=2$ case is relevant. Also, in this example, $n>7$ if and only if $w>3$. This frame, and the $w=2$ EFT-string candidate $\alpha$-vector, are depicted in Figure \ref{f.4d6D}.

\subsubsection{3-frame}
We next investigate 3-frames, corresponding to decompactifications from 4d to 7d. Here, the $\alpha$-vectors of non-oscillator particles and strings are given by
\begin{align}
    \vec \alpha_1=\frac{5-2 P}{\sqrt{30}},\qquad \vec \alpha_2=\frac{10-2 P}{\sqrt{30}}.
\end{align}
Here, EFT string candidates have $n$-values of
\begin{align}
    n=\frac{15}{(P-5)^2},
\end{align}
which is integer only with $P=4,6$ and with $n=15$, and only the $P=4$ string becomes light in this limit. Furthermore, the integral scaling $w$-value for this string is
\begin{align}
    w=5
\end{align}
So, in this case, there is no EFT string with $n\leq 7$.

\subsubsection{4-frame}
\begin{figure}
    \centering
    \includegraphics[width=0.7\linewidth]{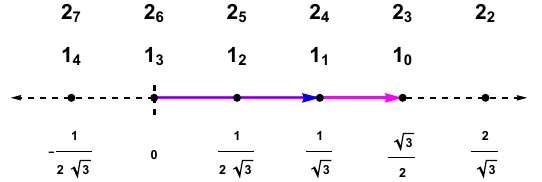}
    \caption{$\alpha$-vectors of the 4-frame. $\alpha$-vectors of strings and particles are labeled by $p_P$, where $p=1$ for particles and $p=2$ for strings. The magenta arrow points to the KK-mode at $1_0$, and the blue arrow points to the EFT-string candidate at $2_2$, which satisfies integral scaling: its oscillator mode, located at $1_1$, is a multiple $1/w$ of the $\alpha$-vector of the KK-mode, with $w=3$.}
    \label{f.4d8D}
\end{figure}

We next investigate 4-frames, corresponding to decompactifications from 4d to 8d. Here, the $\alpha$-vectors of non-oscillator particles and strings are given by
\begin{align}
    \vec \alpha_1=\frac{6-2 P}{4 \sqrt{3}},\qquad \vec \alpha_2=\frac{12-2 P}{4 \sqrt{3}}.
\end{align}
Here, EFT string candidates have $n$-values of
\begin{align}
    n=\frac{24}{(P-6)^2}.
\end{align}
The only strings that are light in the decompactification limit and have integer $n$ are the ones with $P=5$, where $n=24$, and $P=4$, where $n=6$. Furthermore, the integral scaling $w$-value for this string is
\begin{align}
    w_{n=24}=6, \qquad w_{n=6}=3.
\end{align}
So, in this case, integral scaling is preserved, and also $n>7$ iff $w>3$. The $w=3$ EFT string candidate is depicted in Figure \ref{f.4d8D}.

\subsubsection{5-frame}
We next investigate 5-frames, corresponding to decompactifications from 4d to 9d. Here, the $\alpha$-vectors of non-oscillator particles and strings are given by
\begin{align}
    \vec \alpha_1=\frac{7-2 P}{\sqrt{70}},\qquad \vec \alpha_2=\frac{14-2 P}{\sqrt{70}}.
\end{align}
Here, EFT string candidates have $n$-values of
\begin{align}
    n=\frac{35}{(P-7)^2},
\end{align}
The only string that becomes light in the decompactification limit and has integer $n$ are the ones with $P=6$, where $n=35$. Furthermore, the integral scaling $w$-value for this string is
\begin{align}
    w_{n=35}=7.
\end{align}
So, in this case, there are no EFT string candidates with $n\leq 7$.

\subsubsection{6-frame}
\begin{figure}
    \centering
    \includegraphics[width=0.7\linewidth]{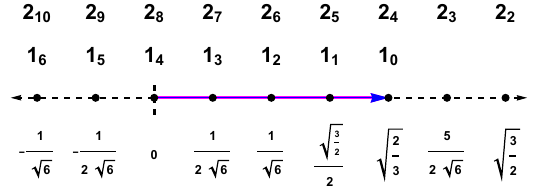}
    \caption{$\alpha$-vectors of the 6-frame. $\alpha$-vectors of strings and particles are labeled by $p_P$, where $p=1$ for particles and $p=2$ for strings. The magenta arrow points to the KK-mode at $1_0$, and the blue arrow points to the EFT-string candidate at $2_4$, which satisfies integral scaling: its oscillator mode, located at $1_2$, is a multiple $1/w$ of the $\alpha$-vector of the KK-mode, with $w=2$.}
    \label{f.4d10D}
\end{figure}

We next investigate 6-frames, corresponding to decompactifications from 4d to 10d. Here, the $\alpha$-vectors of non-oscillator particles and strings are given by
\begin{align}
    \vec \alpha_1=\frac{8-2 P}{4 \sqrt{6}},\qquad \vec \alpha_2=\frac{16-2 P}{4 \sqrt{6}}.
\end{align}
Here, EFT string candidates have $n$-values of
\begin{align}
    n=\frac{48}{(P-8)^2},
\end{align}
The only strings that are light in the decompactification limit with integer $n$ are the ones with  $P=4$, where $n=3$, $P=6$, where $n=12$, and $P=7$, where $n=48$. The integer scalings $w$-value for these strings are
\begin{align}
    w_{n=3}=2,\qquad w_{n=12}=6, \qquad w_{n=48}=8
\end{align}
So, in this case, the only candidate is $P=4$ with $n=2$ and $w=2$. The $w=2$ EFT string candidate is depicted in Figure \ref{f.4d10D}.

\subsubsection{7-frame}
\begin{figure}
    \centering
    \includegraphics[width=0.7\linewidth]{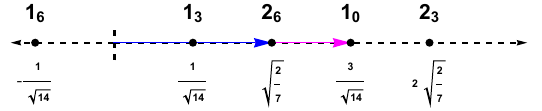}
    \caption{$\alpha$-vectors of the 7-frame. $\alpha$-vectors of strings and particles are labeled by $p_P$, where $p=1$ for particles and $p=2$ for strings. The magenta arrow points to the KK-mode at $1_0$, and the blue arrow points to the EFT-string candidate at $2_6$, which satisfies integral scaling: its oscillator mode, located at $1_3$, is a multiple $1/w$ of the $\alpha$-vector of the KK-mode, with $w=3$. Here, only the branes that can come from M-theory are depicted.}
    \label{f.4d11D}
\end{figure}

We next investigate 7-frames, corresponding to decompactifications from 4d to 11d. Here, the $\alpha$-vectors of non-oscillator particles and strings are given by
\begin{align}
    \vec \alpha_1=\frac{9-2 P}{3 \sqrt{14}},\qquad \vec \alpha_2=\frac{18-2 P}{3 \sqrt{14}}.
\end{align}
Here, EFT string candidates have $n$-values of
\begin{align}
    n=\frac{63}{(P-9)^2},
\end{align}
The only strings that are light in the decompactification limit with integer $n$ are the ones with $P=6$, where $n=7$, and $P=8$, where $n=63$. The integer scalings $w$-value for these strings are
\begin{align}
    w_{n=7}=3,\qquad w_{n=63}=9.
\end{align}
The $n\leq 7$ EFT strings are depicted in Figure \ref{f.4d11D}.

\subsubsection{$\infty$-frame}
\begin{figure}
    \centering
    \includegraphics[width=0.7\linewidth]{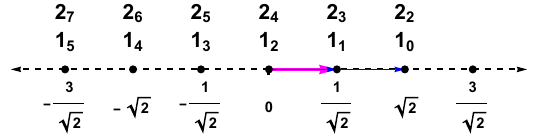}
    \caption{$\alpha$-vectors of the $\infty$-frame. $\alpha$-vectors of strings and particles are labeled by $p_P$, where $p=1$ for particles and $p=2$ for strings. The magenta arrow points to the string oscillator mode at $1_1$, which defines the frame, and the blue arrows point to the EFT-string candidates at $2_2$ and $2_3$, which satisfy integral scaling: their oscillator modes, located at $1_1$ and half of $1_1$, are multiples $1/w$ of the $\alpha$-vector of the fundamental string oscillator-mode, with $w=1$ and $w=2$.}
    \label{f.4di}
\end{figure}

We next investigate $\infty$-frames, corresponding to perturbative string limits. Here, the $\alpha$-vectors of non-oscillator particles and strings are given by
\begin{align}
    \vec \alpha_1=\frac{2-P}{\sqrt 2}\hat \phi,\qquad \vec \alpha_2=\frac{4-P}{\sqrt 2}\hat \phi.
\end{align}
Here, EFT string candidates have $n$-values of
\begin{align}
    n=\frac 4{(4-P)^2}.
\end{align}
The only strings that become light in the perturbative string limit with integer $n$ are the ones with $P=2$, where $n=1$, and $P=3$, where $n=4$. The integer scalings $w$-value for these strings are
\begin{align}
    w_{n=1}=1, \qquad w_{n=4}=2.
\end{align}
This is depicted in Figure \ref{f.4di}.

\subsection{2-dimensional Moduli Spaces}
\label{ss.2dmodulispaces}

We next investigate 2d moduli spaces. The approach is the same as in the previous section. Here, we construct frames, and find strings with $\alpha$-vectors that point in the direction of the frame simplex and have length $\sqrt 2/n$, and we test integer scaling with the principal towers of the frame.

The geometric frame simplices are constructed by two KK-modes and the stringy frame simplices are constructed by one KK-mode and one string oscillator mode. For the geometric frames, the first KK-mode decompactifies to $D_1$-dimensions, and the other to $D_2$-dimensions, and we will label these frames as $(D_1-4,D_2-4)$. The restriction that the maximum spacetime dimension of decompactification is 11 requires that $D_1+D_2-4\leq 11$, and this provides a finite number of geometric frames we can consider, and these are
\begin{align}
\text{set of 2d geometric frames}=\left\{\begin{matrix}(1,1),(1,2),(1,3),(1,4),(1,5),(1,6),\\(2,2),(2,3),(2,4),(2,5),(3,3),(3,4)\end{matrix}\right\}.
\end{align}
For the stringy frame, the KK-mode decompactifies to $D_1$-dimensions, and the string oscillator we can formally think of as decompactifying to $\infty$-dimensions. Following \cite{Etheredge:2025ahf}, requiring no perturbative string limits in $d>10$-dimensions, we require that $D_1\leq 10$. This gives a finite set of frames, which we label by $(D_1-4,\infty)$, and they are
\begin{align}
    \text{set of 2d stringy frames}=\{(1,\infty),(2,\infty),(3,\infty),(4,\infty),(5,\infty),(6,\infty)\}.
\end{align}

Each particle/string is labeled by lattice coordinates $(P_1,P_2)$, where $P_1$ and $P_2$ are the numbers appearing in the lattice dot product rule with the first and second principal towers. We then examine each frame, find all numbers $(P_1,P_2)$ such that the string $\alpha$-vectors have length $\sqrt{2/n}$, and we call these sites EFT string candidates. We then verify that each EFT-string candidate that points within the frame simplex satisfies the integer scaling relation $2\vec \alpha_\text{candidate}\cdot \vec \alpha_\text{principal}=w|\vec \alpha_\text{candidate}|^2$ for integer $w$. In doing so, we find that the $\alpha$-vectors of the oscillators of these strings are generators of the string and particle lattices---i.e. that in fact $2\vec \alpha_\text{candidate}\cdot \vec \alpha_\text{other}=w|\vec \alpha_\text{candidate}|^2$ holds for an integer $w$ for all other $\alpha$-vectors $\vec \alpha_\text{other}$ of strings and particles (we discuss this in detail in section \ref{ss.bound}). For $n\leq 7$, we find that the $w$ for principal towers satisfies $w\leq 3$.

\begin{figure}[tb]
    \centering
\includegraphics{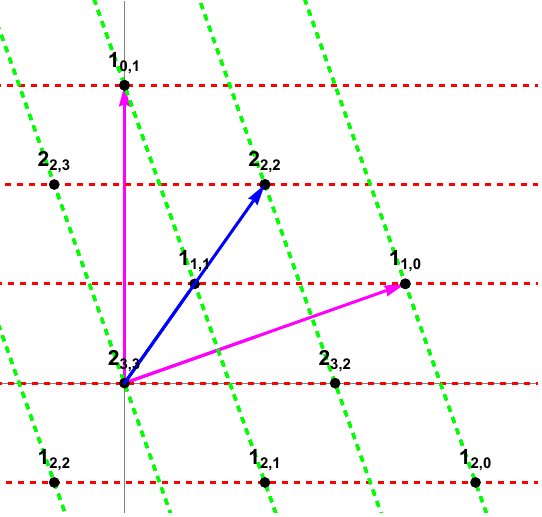}
    \caption{$\alpha$-vectors for the $(1,1)$-frame simplex, generated by the KK-modes in magenta. The lattice sites are for strings $p=2$ and for particles $p=1$, and labeled by $p_{P_1,P_2}$. Here, the EFT string candidate that points in the frame simplex is the $2_{2,2}$ string, labeled in blue. It satisfies integral scaling with the KK-modes.}
    \label{f.11}
\end{figure}

For example, consider first the $(1,1)$-frame. We have that the string with $\vec P\in\{(2,2)\}$ has length of the form $\sqrt{2/n}$ and points in the direction of the frame simplex. This string also satisfies integral scaling with $w=3$ for the $\vec P=(2,2)$ string. This string is depicted in \ref{f.11}.

\begin{figure}[tb]
\centering
\begin{subfigure}{.3\linewidth}
\centering
\includegraphics[scale=.9]{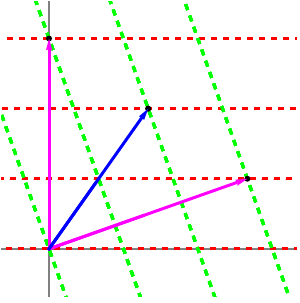}
\caption{$(1,1)$-frame.}\label{fig.11}
\end{subfigure}
\begin{subfigure}{.3\linewidth}
\centering
\includegraphics[scale=.9]{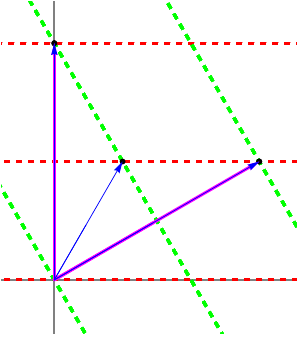}
\caption{$(2,2)$-frame.}
\end{subfigure}
\begin{subfigure}{.3\linewidth}
\centering
\includegraphics[scale=.9]{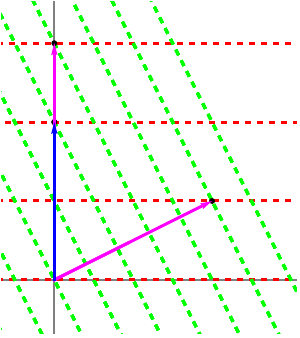}
\caption{$(3,1)$-frame.}
\end{subfigure}
\begin{subfigure}{.3\linewidth}
\centering
\includegraphics[scale=.9]{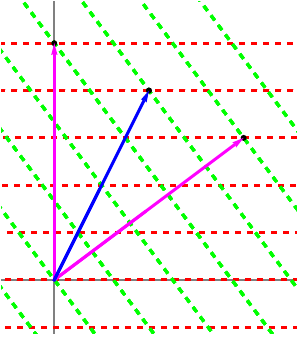}
\caption{$(3,3)$-frame.}
\end{subfigure}
\begin{subfigure}{.33\linewidth}
\centering
\includegraphics[scale=.9]{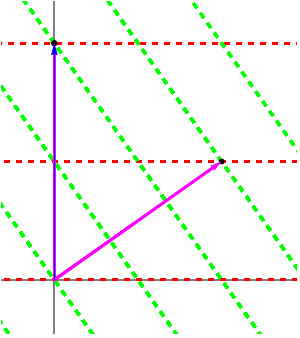}
\caption{$(4,2)$-frame.}    \label{fig.42}
\end{subfigure}
\begin{subfigure}{.3\linewidth}
\centering
\includegraphics[scale=.9]{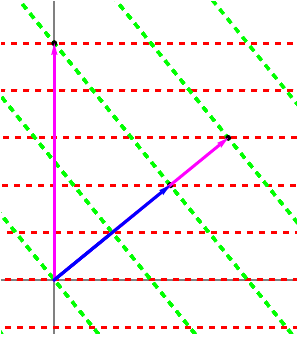}
\caption{$(4,3)$-frame.}
\end{subfigure}
\begin{subfigure}{.3\linewidth}
\centering
\includegraphics[scale=.9]{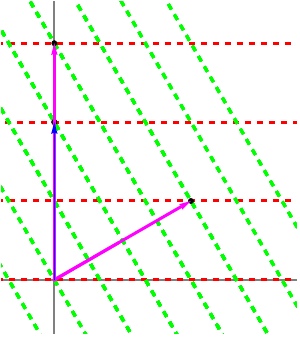}
\caption{$(6,1)$-frame.}  \label{fig.61}
\end{subfigure}
\caption{Geometric frame $\alpha$-vectors with EFT string candidates. Here, the frame simplices are generated by the magenta arrows, and the EFT string candidates are the blue arrows, all of which satisfy integer scaling. The dashed green and red lines denote the radion lattice conditions for the two principal towers.}
\label{f.2dgeo}
\end{figure}

\begin{figure}[tb]
\centering
\begin{subfigure}{.3\linewidth}
\centering
\includegraphics[scale=.9]{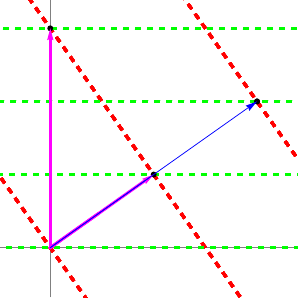}
\caption{$(\infty,1)$-frame.} \label{fig.i1}
\end{subfigure}
\begin{subfigure}{.3\linewidth}
\centering
\includegraphics[scale=.9]{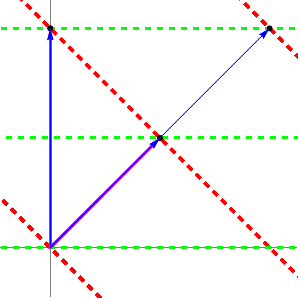}
\caption{$(\infty,2)$-frame.}\label{fig.i2}
\end{subfigure}
\begin{subfigure}{.3\linewidth}
\centering
\includegraphics[scale=.9]{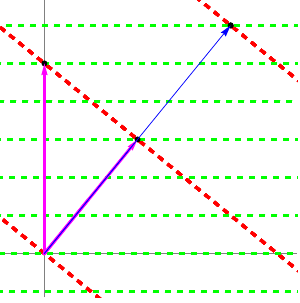}
\caption{$(\infty,3)$-frame.}
\end{subfigure}
\begin{subfigure}{.3\linewidth}
\centering
\includegraphics[scale=.9]{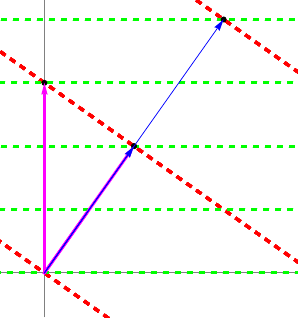}
\caption{$(\infty,4)$-frame.}\label{fig.i4}
\end{subfigure}
\begin{subfigure}{.33\linewidth}
\centering
\includegraphics[scale=.9]{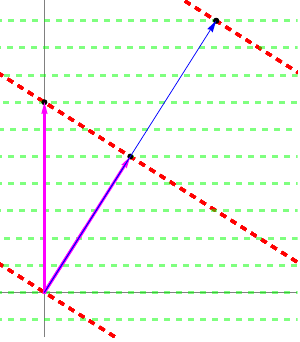}
\caption{$(\infty,5)$-frame.}
\end{subfigure}
\begin{subfigure}{.3\linewidth}
\centering
\includegraphics[scale=.9]{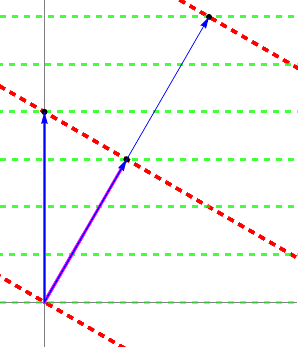}
\caption{$(\infty,6)$-frame.}\label{fig.i6}
\end{subfigure}
\caption{Stringy frame $\alpha$-vectors with EFT string candidates. Here, the frame simplices are generated by the magenta arrows, and the EFT string candidates are the blue arrows, all of which satisfy integer scaling. The dashed green and red lines denote the radion and dilaton lattice conditions for the two principal towers.}
\label{f.2dstr}
\end{figure}

One can compute all of the other frame simplices and their strings. Only a subset of the simplices have EFT-string candidates with $n\leq 7$ and that point in the direction of the frame simplex, and these are all of the stringy frames, together with the $(1,1)$, $(2,2)$, $(3,1)$, $(3,3)$, $(4,2)$, $(4,3)$ and $(6,1)$ frames. These are depicted in Figures \ref{f.2dgeo} and \ref{f.2dstr}.

\subsection{Higher-dimensional Moduli Spaces}       \label{high_d_modsp}

In this section we discuss how we can generalize the previous analysis to higher dimensional slices of moduli spaces. Suppose in our 4d EFT we have a $d$-dimensional slice of the moduli space, where we have $d$ independent principal towers. We can distinguish two cases: when all the principal towers are Kaluza-Klein modes we say that we are in a geometric frame, while if among the towers there is a stringy tower we call that a stringy frame. We will discuss these two cases separately.

\subsubsection{Geometric Frames}

 We start with the geometric frames, containing only Kaluza-Klein towers. We can take as basis of the tangent space to the moduli space at any point the $\a$-vectors of the KK towers that define the frame simplex, normalized to unit norm, and denote them by $\{\vec{e}_i\}, i=1,...,d$. Note that this basis need not be orthonormal and in general it will not be. One can then introduce the Gram matrix
\be
G_{ij} = \vec{e}_i \cdot \vec{e}_j =
\begin{cases}
1 \quad i=j\\
\cos \theta_{ij} \quad i \neq j
\end{cases} \, ,
\ee
where $\theta_{ij}$ is the angle between $\vec{e}_i$ and $\vec{e}_j$, given by the taxonomy rule \cite{Etheredge:2024tok}
\be		\label{cosij}
\cos \theta_{ij} = \sqrt{\frac{k_i k_j}{(k_i+2)(k_j+2)}} \, ,
\ee
where $k_i = D_i-4$ denotes the number of dimensions that are decompactifying along the direction $\vec{e}_i$. Given a tangent vector $\vec{v} = v^i \vec{e}_i$, the Gram matrix allows to switch easily between the components $v^i$ of the vector in terms of the basis $\{\vec{e}_i\}$ and the projections $p_i$ of such vector along the basis elements $\{\vec{e}_i\}$
\be
p_i \equiv \vec{v} \cdot \vec{e}_i = G_{ij} v^j \, , \qquad v^i = G^{ij} p_j \, .
\ee
This matrix is then particularly useful to compute scalar products between vectors, knowing the projections along the given basis, since this is precisely the information we get from the lattice rules from brane taxonomy \cite{Etheredge:2025ahf}. In particular, the norm of an $\a$-vector in terms of its projections $p_i$ is given by
\be
||\vec{\a}||^2 = G^{ij} p_i p_j \, ,
\ee
and to look for EFT string candidates in a given frame simplex we have to impose that this is equal to $2/n_{\bf e}$, for some integer $n_{\bf e} \in \{1,...,7\}$. Brane taxonomy applied to strings in a geometric frame simplex gives the following rules for the projections $p_i$ of their $\a$-vectors
\be
p_i = a_i \Delta_i \, , \qquad a_i \in \mathbb{Z} \, , \qquad \Delta_i = \sqrt{\frac{2}{k_i(k_i+2)}} \, ,
\ee
where $\Delta_i$ is the spacing of the string lattice along the $e_i$ direction, that can be read off from \eqref{e.radionrule}, and we have $a_i = D_i - 2 - P_i$ for radion directions. Taxonomy rules, in particular \eqref{tax_rules}, also tell us that the $\a$-vector of the KK tower associated to the $\vec{e}_i$ direction is
\be
\vec{\a}_{{\rm KK},i} = \sqrt{\frac{k_i+2}{2k_i}} \vec{e}_i \, ,
\ee
and we can easily compute the scalar product between such vector and the $\a$-vector of an EFT string candidate
\be
\vec{\a}_{\rm str} \cdot \vec{\a}_{{\rm KK},i} = p_i \sqrt{\frac{k_i+2}{2k_i}} = \frac{a_i}{k_i} \, ,
\ee
and recalling that the scaling weight $w$ of a particle tower along a given EFT string limit is given by $n_{\bf e} \, \vec{\a}_{\rm part} \cdot \vec{\a}_{\rm str}$, we get that the $i^{\rm th}$ KK tower has scaling weight
\be		\label{wi}
w_i = \frac{a_i}{k_i} n_{\bf e} \, .
\ee
The last equation allows to check very easily the scaling weight of all the KK towers in a given frame simplex along a given EFT string limit. We can then use Mathematica to build all the possible $N$-dimensional moduli space slices, with the constraints
\be
N \leq 7 \, , \qquad k_i \leq 7 \, , \qquad \sum_i k_i \leq 7 \, ,
\ee
and the corresponding frame simplices. For each frame simplex we look for EFT string candidates, satisfying the norm constraint $||\vec{\a}_{\rm str}||^2 = 2/n_{\bf e}$, and for each EFT string candidate we compute the scaling weights of the KK towers $w_i$. We find that all the principal KK towers have integer scaling weight, between 1 and 3, along all the candidate EFT string limits. In Tables \ref{tab:geom.simplices} and \ref{tab:geomEFTstr} we summarize all the possible geometric frame simplices and the candidate EFT strings that lie within the frame simplex (or at most on its boundaries). We restrict to these candidates because they are the only potential EFT strings whose flow can be well described within the perturbative region we are considering, as any string whose $\a$-vector points outside of the frame simplex takes us away from such region and possibly to a different duality frame. Let us notice that in 5d, 6d and 7d moduli spaces we find no such EFT string candidates.

\setlength{\arrayrulewidth}{0.2mm}
\renewcommand{\arraystretch}{1.2}
\begin{table}[tb]
\begin{center}
\begin{tabular}{|c|c|c|c|}
\hline
\rowcolor{lightgray} Frame simplex & Position on string lattice $(a_i)$ & $n_{\bf e}$ & $w_i$ \\
\hline
$1$ & $1$ & $3$ & $3$ \\
$2$ & $2$ & $2$ & $2$ \\
$4$ & $2$ & $6$ & $3$ \\
$6$ & $4$ & $3$ & $2$ \\
$7$ & $3$ & $7$ & $3$ \\
\hline
$(1,1)$ & $(1,1)$ & $2$ & $(2,2)$ \\
$(3,1)$ & $(1,1)$ & $3$ & $(1,3)$ \\
$(2,2)$ & $(1,1), (1,2)$ & $6, 2$ & $(3,3), (1,2)$ \\
$(4,2)$ & $(2,2)$ & $2$ & $(1,2)$ \\
$(3,3)$ & $(2,2)$ & $3$ & $(2,2)$ \\
$(6,1)$ & $(2,1)$ & $3$ & $(1,3)$ \\
$(4,3)$ & $(2,1)$ & $6$ & $(3,2)$ \\
\hline
$(2,1,1)$ & $(1,1,1)$ & $2$ & $(1,2,2)$\\
$(4,1,1)$ & $(2,1,1)$ & $2$ & $(1,2,2)$\\
$(2,2,2)$ & $(1,1,2)$ & $2$ & $(1,1,2)$\\
$(3,3,1)$ & $(1,1,1)$ & $3$ & $(1,1,3)$\\
$(3,2,2)$ & $(1,1,1)$ & $6$ & $(2,3,3)$\\
\hline
$(2,2,1,1)$ & $(1,1,1,1)$ & $2$ & $(1,1,2,2)$\\
\hline
\end{tabular}
\caption{Geometric moduli spaces, candidate EFT strings within the frame simplex and scaling weights of the principal KK towers (up to permutations of $k_i$ and $a_i$). The 5,6,7-dimensional moduli spaces do not appear here, because we find no EFT string candidates within the frame simplices in those cases. However, they appear in the tables of Appendix \ref{ap:tables}, where we collect all the EFT string candidates, including the ones outside of the frame simplex.}
\label{tab:geomEFTstr}
\end{center}
\end{table}
Notice that for all the limits in the table the lightest tower has $w>1$. This nicely matches the expectation that decompactification limits must have $w>1$ and $w=1$ limits can only be emergent string limits. Notice however that one could have an emergent string limit with $w>1$ if the emergent string is not the EFT string itself, but a lighter string. We will further comment on this in the following.

Actually, Table \ref{tab:geomEFTstr} is not the full story. Given a slice of a moduli space and the corresponding frame simplex, there are special submanifolds that must be treated as lower-dimensional moduli spaces, as they could host extra candidate EFT strings that are not captured by the previous analysis. This is because the strategy above is based on taxonomy rules, which assume that we are considering a so-called \textit{regular} infinite distance limit, see \cite{Etheredge:2024tok}. These are \textit{generic} trajectories in moduli space, which in practice means that they do not coincide with the following two types of special loci:
\begin{itemize}
    \item facets: they correspond to lower-dimensional boundaries of the original moduli space region or frame simplex. The corresponding lower-dimensional frame simplex is obtained by just dropping one or more of the entries of the original frame simplex;
    \item pericenters: they correspond to the loci where two or more principal towers decay with the same rate and combine in a tower of bound states. The corresponding frame simplex is obtained by summing the entries of the degenerate towers. The $\a$-vector of the pericenter can be written in terms of the vectors of the individual towers $\vec{\a}_i$ as \cite{Etheredge:2024tok}
    \be     \label{alpha_pc}
    \vec{\a}_{\rm pc} = \frac{1}{\sum_j k_j} \sum_i k_i \, \vec{\a}_i \, .
    \ee
\end{itemize}
As an example, consider the frame simplex $(6,1)$ with decompactification limits to $D_i=(10,5)$. The only candidate EFT string we find with the general analysis is the lattice point $(a_i)=(2,1)$, which takes us to the decompactification limit to 10d. Let us now look at the special loci we have described. First we consider the codimension-1 boundaries of this region of moduli space, which are the two decompactification limits that identify the frame simplex in the first place. These are described by lower-dimensional frame simplices of type $6$ and $1$ and they host the EFT string candidates corresponding to the fourth and first lines in Table \ref{tab:geomEFTstr}, respectively. Notice that the former is a new EFT string candidate, while the latter coincides with the lattice point $(2,1)$ found before. Let us now turn to the other special locus, namely the pericenter where the two Kaluza-Klein towers decay with the same rate. This is a decompactification limit to $D=11$ described by a lower-dimensional frame simplex of type $7$ that hosts the EFT string candidate appearing in the fifth line of Table \ref{tab:geomEFTstr}, which is also new. If we started with a higher dimensional moduli space, say 3d, then the special loci would be 2d submanifolds thereof. Then for each of them we would need to consider 1d subspaces and apply the same reasoning recursively to find all the EFT string candidates. The power of this method lies in the fact that the results in Table \ref{tab:geomEFTstr} already show that the leading tower has integral scaling along all limits, including the ones localized on special loci. The reason is the following: whenever we go to a facet, by the definition of frame simplex, all the principal towers that do not lie on the facet are at the species scale and then we are not interested in their scaling weight, while at pericenters all the degenerate towers become light at the same rate and then have the same scaling weight, which is the one we read off from the table.

\subsubsection{Stringy Frames}
\label{sss.stringyframes}

We now turn to stringy frames, where one of the principal towers corresponds to the tower of string oscillators of an emergent string, while the rest are KK towers, associated to decompactification limits to up to 10 dimensions. For the $\a$-vectors of KK towers the same rules as above hold, while for the stringy tower some of them must be modified. In particular, the angle between the $\a$-vector of the string oscillators and a KK tower decompactifying $k_i$ dimensions is
\be
\cos \theta_{{\rm str},i} = \sqrt{\frac{k_i}{k_i+2}} \, ,
\ee
which can be recovered from \eqref{cosij} by taking the $k_j \to \infty$ limit. In addition, we write the projection of $\a$-vectors along the string direction as, see \eqref{e.dilatonrule},
\be
p_{\rm str} = a_{\rm str} \D_{\rm str} = \frac{a_{\rm str}}{\sqrt{2}} \, ,
\ee
where $a_{\rm str} = P-4$, and now the $d$-dimensional moduli spaces (including the string direction) must satisfy the following constraints
\be
N \leq 7 \, , \qquad k_i \leq 6 \, , \qquad \sum_i k_i \leq 6 \, .
\ee
As for the scaling weights of the different towers, \eqref{wi} still holds for the KK towers, while for the tower of string oscillators we have
\be
\vec{\a}_{\rm osc} = \frac{1}{\sqrt{2}} \vec{e}_{\rm str} \, ,
\ee
which implies that along the flow of the string identified by the lattice site $(a_{\rm str},a_i)$, the oscillators of the string generating the frame simplex have a scaling weight
\be
w_{\rm osc} = \frac{a_{\rm str}}{2} n_{\bf e} \, .
\ee
In Tables \ref{tab:str.simplices} and \ref{tab:stringyEFTstr} we summarize all the stringy frame simplices and their candidate EFT strings whose $\a$-vectors lie inside the frame simplex, or at most on the boundaries. In this notation, the stringy generator of the frame simplex is given by $(a_{\rm str},a_i) = (2, k_i)$.

\setlength{\arrayrulewidth}{0.2mm}
\renewcommand{\arraystretch}{1.2}
\begin{table}[H]
\begin{center}
\begin{tabular}{|c|c|c|c|}
\hline
\rowcolor{lightgray} Frame simplex & Position on string lattice $(a_{\rm str}, a_i)$ & $n_{\bf e}$ & $(w_{\rm osc}, w_i)$ \\
\hline
$\infty$ & $1,2$ & $4,1$ & $2,1$ \\
\hline
$(\infty,1)$ & $(2,1)$ & $1$ & $(1,1)$ \\
$(\infty,2)$ & $(1,1), (1,2), (2,2)$ & $4, 2, 1$ & $(2,2), (1,2), (1,1)$ \\
$(\infty,3)$ & $(2,3)$ & $1$ & $(1,1)$ \\
$(\infty,4)$ & $(1,2), (2,4)$ & $4, 1$ & $(2,2), (1,1)$ \\
$(\infty,5)$ & $(2,5)$ & $1$ & $(1,1)$ \\
$(\infty,6)$ & $(1,3), (1,4), (2,6)$ & $4, 3, 1$ & $(2,2), (\frac{3}{2},2), (1,1)$ \\
\hline
$(\infty,1,1)$ & $(1,1,1), (2,1,1)$ & $2, 1$ & $(1,2,2), (1,1,1)$ \\
$(\infty,2,1)$ & $(2,2,1)$ & $1$ & $(1,1,1)$ \\
$(\infty,3,1)$ & $(2,3,1)$ & $1$ & $(1,1,1)$ \\
$(\infty,2,2)$ & $(1,1,1), (1,1,2), (2,2,2)$ & $4, 2, 1$ & $(2,2,2), (1,1,2), (1,1,1)$ \\
$(\infty,4,1)$ & $(2,4,1)$ & $1$ & $(1,1,1)$ \\
$(\infty,3,2)$ & $(2,3,2)$ & $1$ & $(1,1,1)$ \\
$(\infty,5,1)$ & $(2,5,1)$ & $1$ & $(1,1,1)$ \\
$(\infty,4,2)$ & $(1,2,1), (1,2,2), (2,4,2)$ & $4, 2, 1$ & $(2,2,2), (1,1,2), (1,1,1)$ \\
$(\infty,3,3)$ & $(1,2,2), (2,3,3)$ & $3, 1$ & $(\frac{3}{2},2,2), (1,1,1)$ \\
\hline
$(\infty,1,1,1)$ & $(2,1,1,1)$ & $1$ & $(1,1,1,1)$ \\
$(\infty,2,1,1)$ & $(1,1,1,1), (2,2,1,1)$ & $2, 1$ & $(1,1,2,2), (1,1,1,1)$ \\
$(\infty,3,1,1)$ & $(2,3,1,1)$ & $1$ & $(1,1,1,1)$ \\
$(\infty,2,2,1)$ & $(2,2,2,1)$ & $1$ & $(1,1,1,1)$ \\
$(\infty,4,1,1)$ & $(1,2,1,1), (2,4,1,1)$ & $2, 1$ & $(1,1,2,2), (1,1,1,1)$ \\
$(\infty,3,2,1)$ & $(2,3,2,1)$ & $1$ & $(1,1,1,1)$ \\
$(\infty,2,2,2)$ & $(1,1,1,1), (1,1,1,2), (2,2,2,2)$ & $4, 2, 1$ & $(2,2,2,2), (1,1,1,2), (1,1,1,1)$ \\
\hline
$(\infty,1,1,1,1)$ & $(2,1,1,1,1)$ & $1$ & $(1,1,1,1,1)$ \\
$(\infty,2,1,1,1)$ & $(2,2,1,1,1)$ & $1$ & $(1,1,1,1,1)$ \\
$(\infty,3,1,1,1)$ & $(2,3,1,1,1)$ & $1$ & $(1,1,1,1,1)$ \\
$(\infty,2,2,1,1)$ & $(1,1,1,1,1), (2,2,2,1,1)$ & $2, 1$ & $(1,1,1,2,2), (1,1,1,1,1)$ \\
\hline
$(\infty,1,1,1,1,1)$ & $(2,1,1,1,1,1)$ & $1$ & $(1,1,1,1,1,1)$ \\
$(\infty,2,1,1,1,1)$ & $(2,2,1,1,1,1)$ & $1$ & $(1,1,1,1,1,1)$ \\
\hline
$(\infty,1,1,1,1,1,1)$ & $(2,1,1,1,1,1,1)$ & $1$ & $(1,1,1,1,1,1,1)$ \\
\hline
\end{tabular}
\caption{1- to 7-dimensional stringy moduli spaces, candidate EFT strings within the frame simplex and scaling weights of the principal towers (up to permutations of $k_i$ and $a_i$).}
\label{tab:stringyEFTstr}
\end{center}
\end{table}

Let us make a couple of observations. First, in all the emergent string limits associated to the stringy generators of the frame simplices, all KK towers have $w=1$, so we always have that the string oscillators are the lightest tower and they are accompanied by the KK modes, that become light at the same rate. However for other strings, like the one in the lattice site $(1,1)$ in the $(\infty,2)$ frame simplex, we have that the string oscillators become light as quickly as the KK modes, but with $w=2$, which would go against the intuition that $w>1$ corresponds to a decompactification limit. The solution to this puzzle is that the EFT string is not always the lightest string along its own flow, as there can be a lighter string that plays the role of the emergent string. As we will see in the following, there are concrete string theory examples where this happens, see Section \ref{ss:IIAquintic}. In conclusion, we find that $w=1$ always corresponds to an emergent string limit where the emergent string is the EFT string itself, while $w>1$ might be a decompactification limit or an emergent string limit where the emergent string is lighter than the EFT string inducing the flow.

Another interesting observation is that for some towers we find half-integer scaling weight, in particular for the oscillators of the string that have $w_{\rm osc}=\frac{3}{2}$ along the flows generated by the EFT string candidates $(1,4)$ and $(1,2,2)$ in the frame simplices $(\infty,6)$ and $(\infty,3,3)$, respectively. This is not in contradiction with the ISC, since in both these cases the string oscillator modes are not the leading tower, which is instead given by the KK modes with $w=2 > \frac{3}{2}$. We postpone a more detailed discussion of these half-integral scaling weights, and of their interplay with the refined proposal of \cite{Grieco:2025bjy}, to section \ref{ss.bound}.

As for the geometric frames, one should treat facets and pericenters separately and also look for candidate EFT strings in these special loci. The procedure is completely analogous to the geometric case, with the only difference that whenever the towers that become light with the same rate include the string oscillators, the pericenter coincides with the emergent string direction itself, as one can recover by taking the formal limit $D \to \infty$ in \eqref{alpha_pc} for the stringy tower.

\subsection{Extensions of the Analysis: Bound States and Lattice Statements}
\label{ss.bound}

\subsubsection*{Bound States}
The above discussion holds for elementary EFT strings, namely those strings that are generators of the cone of EFT string charges \eqref{C_S^EFT}. For bound states of EFT strings with several charges turned on the discussion is more subtle, the reason being that their $\a$-vectors usually slide even if we restrict to a region of the moduli space where taxonomy rules apply. In order for them to be constant, usually we need to evaluate them along their own saxionic flow and treat these directions as lower-dimensional special loci, just like we do for boundaries and pericenters. In the following we discuss how they can be included in the analysis. From the formula of the tension of EFT strings, we can write the $\a$-vector of a bound state in terms of the elementary $\a$-vectors that contribute to it as follows
\be
\frac{{\cal T}_{\bf e}}{M_{\rm Pl}^2} = \sum_i e^i \ell_i \qquad \implies \qquad \vec{\a}_{\bf e} = \frac{1}{\sum_k e^k \ell_k} \sum_i e^i \ell_i \vec{\a}_i \, ,
\ee
where $\vec{\a}_i$ denotes the $i^{\rm th}$ elementary EFT string's $\a$-vector. So the bound state $\a$-vector is a weighted sum of the elementary ones, where the weights are the tensions of the elementary strings. The previous formula also implies that the bound state $\vec{\a}_{\bf e}$ lies on a hyperplane connecting the elementary $\vec{\a}_i$'s. Now we can distinguish two cases:
\begin{itemize}
        \item If the homogeneous function $P(s)$ appearing in the K\"ahler potential \eqref{Kahler_pot} only consists of one monomial, then we have
        \be
        \ell_i = \frac{n_i}{2s^i} \, ,
        \ee
        which implies
        \be
        e^i\ell_i \simeq \frac{n_i}{2\s}\, , \quad \s \to \infty \, ,
        \ee
        where $n_i$ is the exponent of $s^i$ in the K\"ahler potential. Then the weights in the weighted average are precisely the $n_i$'s and the $\a$-vector of the bound state reads
        \be     \label{alpha_bs}
        \vec{\a}_{\bf e} = \frac{1}{\sum_j n_j} \sum_i n_i \vec{\a}_i = \frac{1}{n_{\bf e}} \sum_i n_i \vec{\a}_i \, .
        \ee
        In this case, one can see that the integral scaling along the flow of the bound state follows from the integral scaling along the elementary flows
        \be
        \vec{\a}_{\rm tow} \cdot \vec{\a}_i = \frac{w_i}{2} \vec{\a}_i^2 \qquad \implies \qquad \vec{\a}_{\rm tow} \cdot \vec{\a}_{\bf e} = \frac{\sum_i w_i}{2} \frac{2}{n_{\bf e}} = \frac{w_{\bf e}}{2} \vec{\a}_{\bf e}^2\, ,
        \ee
        where the scaling weight along the bound state is $w_{\bf e} = \sum_i w_i$. Here one should be careful, as in general the lightest tower along the bound state flow might be subleading along the elementary flows or even lie outside of the frame simplex that contains the elementary EFT strings. If this is the case, our analysis does not guarantee that the $w_i$'s of such tower are integer. In particular they could be half-integer and result in a half-integer $w_{\bf e}$. However, in the examples we analyze they always add up to an integer, see \ref{ss:G2}. Finally let us mention that this argument still applies if the K\"ahler potential is more complicated, but there is a single leading monomial that dominates along all the infinite distance trajectories in the region of the moduli space we are considering;
    \item If $P(s)$ in the K\"ahler potential contains several monomials that compete with each other, the situation is more involved and we leave a complete analysis for future work.
\end{itemize}

\subsubsection*{Lattice Statements}
We now turn to a different kind of extension of our analysis, which promotes the integral scaling relations tested in the previous subsections to a statement about the full lattices of $\a$-vectors. The starting point is the observation made in section \ref{sss.stringyframes} that along the flows of the EFT string candidates $(1,4)$ and $(1,2,2)$ in the frame simplices $(\infty,6)$ and $(\infty,3,3)$, the oscillator modes of the corresponding emergent strings have the half-integral scaling weight $w_{\rm osc}=3/2$. In fact, as can be seen in the complete tables in appendix \ref{ap:tables}, for all the EFT string candidates within the frame simplices (or at their boundaries), the scaling weights of the KK towers are always integer, and half-integral values appear only for the oscillator modes of strings. Analogous half-integral weights were encountered in the analysis of \cite{Grieco:2025bjy}, where it was pointed out that they correspond to towers that do not generate the convex hull relevant for the Distance Conjecture, but instead lie on one of its facets. This observation led the authors of \cite{Grieco:2025bjy} to propose a refined version of the integral scaling relations, namely that they hold not only for the leading tower, but for every tower that generates the convex hull.

Our bottom-up scan suggests a natural lattice counterpart of this proposal. Given an EFT string candidate with $\a$-vector $\vec{\a}_{\rm str}$, we denote by $\vec{\a}_{\rm osc}=(1/2)\, \vec{\a}_{\rm str}$ the $\a$-vector associated to its oscillator modes.\footnote{We use \emph{oscillator modes} to refer to the mass scale given by the square root of the string tension in Planck units, without a priori assuming the existence of a perturbative oscillator spectrum, since the string may be strongly coupled.} We propose that \emph{the oscillator $\a$-vector of every EFT string candidate within a frame simplex (or at its boundary) is a generator of the lattices of particle and string $\a$-vectors}. In practice, this means that
\begin{equation} 
\label{eq:latticestatement}
\vec{\a}_{\rm osc}\cdot \vec{\a}_X = w_X\, |\vec{\a}_{\rm osc}|^2 \, , \qquad w_X\in \mathbb{Z}\, ,
\end{equation}
for every site $\vec{\a}_X$ of the particle and string lattices. Let us note that the restriction to candidates within the frame simplex is essential. For candidates outside it, such as the site $(0,2,1)$ of the $(\infty,4,1)$ frame simplex, even KK towers can display half-integral weights, as can be seen in the tables of appendix \ref{ap:tables}. In all the cases we have checked, $\vec{\a}_{\rm osc}$ itself sits at a site of these lattices, which justifies calling it a \emph{generator}. Notice that, by comparison with \eqref{e.wi}, the integer $w_X$ is nothing but the scaling weight of the tower associated to the site $\vec{\a}_X$ along the candidate flow. We have checked \eqref{eq:latticestatement} exhaustively for all the 1d and 2d frame simplex lattices (including sites $\vec{\a}_X$ lying outside the frame simplex), and verified it explicitly in several higher-dimensional frames, as well as in the top-down examples of section \ref{sec:topdown}. We thus find it natural to expect it to hold in general (see also the upcoming work \cite{GriecoRuizValenzuelaToAppear} for a complementary bottom-up perspective on how similar relations constrain towers of light states).

The statement \eqref{eq:latticestatement} is stronger than the convex hull refinement of \cite{Grieco:2025bjy}, and the precise relation between the two is the following. Convex hull generators that correspond to particle or string lattice sites automatically satisfy integral scaling if \eqref{eq:latticestatement} holds, so the lattice statement implies the convex hull one for all non-oscillator towers. The oscillator modes of strings sit instead at one half of the corresponding string lattice sites, so \eqref{eq:latticestatement} only fixes their scaling weights to be quantized in half-integral units. More concretely, along the flow of a candidate with lattice coordinates $(a_{\rm str},a_i)$ in a stringy frame simplex, the oscillator modes of the emergent string have $w_{\rm osc}=\frac{a_{\rm str}}{2}\, n_{\bf e}$, which is an integer precisely when the product $a_{\rm str}\, n_{\bf e}$ is even. The half-integral cases mentioned above correspond to odd products, and it is precisely in these cases that the top-down realizations feature an emergent string that is non-BPS and whose oscillator modes do not generate the convex hull, as for the fundamental string in the F-theory example of section \ref{ss:Fth_P3}. For even products, as for the D4-string on the quintic of section \ref{ss:IIAquintic}, the oscillator modes of the corresponding (BPS) emergent string do generate the convex hull and their scaling weight is integer. The convex hull criterion of \cite{Grieco:2025bjy} can thus be regarded as the selection rule determining which oscillator towers must display integral scaling, while \eqref{eq:latticestatement} fixes the allowed quantization for all of them.

It is worth remarking on the origin of this half-integral quantization. The masses of the oscillator modes of a string scale as the square root of its tension, which is just the statement that the worldvolume dimension of a string is $p=2$. As a consequence, oscillator $\a$-vectors sit at one half of the corresponding string lattice sites, and taking $\vec{\a}_{\rm osc}$ to generate the particle and string lattices automatically quantizes all the scaling weights in units of $1/2$. From this perspective, it is natural to expect suitable generalizations of EFT strings to $d$ spacetime dimensions to give rise to analogous lattice statements with scaling weights quantized in units of $1/p$, with $p$ the worldvolume dimension of the corresponding objects. Codimension-two objects, with $p=d-2$, are the natural candidates to play this role, since they are the ones implementing monodromies. We come back to this possibility in section \ref{sec:conclusions}.

Let us close by remarking that the half-integral quantization is also relevant for the bound state flows discussed above, as the sums $w_{\bf e}=\sum_i w_i$ can receive half-integral contributions from towers that are subleading along some of the elementary flows, and the lattice statement guarantees that such sums remain quantized in units of $1/2$, even though their integrality is not automatic.

\section{Comparison with Top-Down Examples}
\label{sec:topdown}

In this section we analyze some concrete top-down examples, in order to show how our strategy works in practice. In general, given a string theory compactification and the resulting 4d ${\cal N}=1$ theory, we focus on an asymptotic region of the moduli space which is approximately flat and where $\a$-vectors are constant. In some cases this happens for a whole slice of the moduli space, while in other cases we have to focus on a specific growth sector where the K\"ahler potential is dominated by one monomial. Once we have selected a particular moduli space region, we identify the frame simplices contained in it and apply the strategy outlined in the previous sections.

\subsection{Type IIA on $\mathbb{P}^{1,1,1,6,9}[18]$}       \label{ex1}

The first example we consider is type IIA string theory compactified on the Calabi--Yau threefold $X_3=\mathbb{P}^{1,1,1,6,9}[18]$, see \cite{Candelas:1994hw, Lanza:2021udy}, which is a smooth elliptic fibration over a $\mathbb{P}^2$ base and we focus on the vector multiplet sector of the moduli space. This geometry has two (string frame) K\"ahler moduli $\{v^1,v^2\}$, controlling the volumes of the elliptic fibre and the $\mathbb{P}^1$ curve in the base, respectively. The intersection polynomial in the dual basis of K\"ahler cone generators $\{J_1,J_2\}$ reads
\be
{\cal I}(X_3) = 9 J_1^3 + 3J_1^2 J_2 + J_1 J_2^2 \, .
\ee
In this setup, the saxionic cone coincides with the K\"ahler cone, and EFT strings are realized by NS5-branes wrapping Nef divisors, so that we have
\be     \label{cones_ex1}
\D = \{s^1,s^2>0\} \, , \qquad {\cal C}_S^{\rm EFT} = \{e^1,e^2 \in \mathbb{Z}_{\geq 0}\} \, ,
\ee
with $s^i=v^i$. In the large volume approximation the K\"ahler potential is given by
\be
K = - \log \cV_X \, , \qquad \cV_X = \frac{1}{6} \cK_{abc} s^a s^b s^c \, ,
\ee
where $\cV_X$ is the Calabi--Yau volume in string units, from which we can compute the saxionic metric \ref{Gij}, which in the two moduli case is flat and one can then find flat coordinates for the whole saxionic slice of the moduli space. In this example they read
\be
\begin{split}
x^1 = \frac{1}{\sqrt{6}} \log \cV_X \, ,\\
x^2 = \frac{1}{\sqrt{3}} \log \left[ 1+ \frac{2}{9} \frac{(s^2)^3+(s^2(3s^1+s^2))^\frac{3}{2}}{3(s^1)^3+3(s^1)^2 s^2+s^1 (s^2)^2} \right] \, .
\end{split}
\ee
One can then compute the $\a$-vectors of the relevant towers of light states, namely the D0-branes and the D2-branes wrapping the elliptic fibre, that in flat coordinates read
\be
\begin{split}
m_{\rm D0} = \frac{M_{\rm Pl}}{\sqrt{\cV_X}} \, , \qquad \vec{\a}_{\rm D0} = \left( \sqrt{\frac{3}{2}},0 \right)\, , \\
m_{{\rm D2},T^2} = M_{\rm Pl} \frac{\cV_{T^2}}{\sqrt{\cV_X}} \, , \qquad \vec{\a}_{{\rm D2}, T^2} = \left( \frac{1}{\sqrt{6}}, \frac{3s^1 +2s^2}{\sqrt{3s^2(3s^1+s^2)}} \right) \, .
\end{split}
\ee
It turns out that the $\a$-vectors of the light towers are not constant throughout this slice of the moduli space, so that we cannot directly apply the taxonomy rules. Rather we have to restrict to specific growth sectors:
\begin{itemize}
    \item If $s^1 \gg s^2$ the $\a$-vectors reduce to
    \be
    \vec{\a}_{\rm D0} = \left( \sqrt{\frac{3}{2}},0 \right)\, , \qquad \vec{\a}_{{\rm D2}, T^2} = \left( \frac{1}{\sqrt{6}}, \infty \right) \, ,
    \ee
    and the D2 tower slides to infinity and we effectively reduce to a one-dimensional moduli space. This can also be seen by the fact that the leading term of the K\"ahler potential only depends on $s^1$ in this limit. Hence, the frame simplex in this growth sector is generated by the D0 $\a$-vector alone and moving in its direction corresponds to a decompactification to 5d, so it is a limit with $D=5$. According to our analysis in Sections \ref{1d_modsp} and \ref{high_d_modsp}, there is one EFT string candidate in this frame simplex with $n_{\bf e}=3$, which is indeed realized by an NS5-brane wrapping the divisor $J_1$, whose $\a$-vector reduces to $\a_{{\rm NS5},J_1} = \left( \sqrt{\frac{2}{3}},0 \right)$.
    \item If $s^2 \gg s^1$ the $\a$-vectors reduce to
    \be
    \vec{\a}_{\rm D0} = \left( \sqrt{\frac{3}{2}},0 \right)\, , \qquad \vec{\a}_{{\rm D2}, T^2} = \left( \frac{1}{\sqrt{6}}, \frac{2}{\sqrt{3}} \right) \, ,
    \ee
    and they generate a frame simplex of type $(k_1,k_2)=(1,1)$, see Figure \ref{fig.11}. Our previous analysis (see Section \ref{high_d_modsp}) predicts a candidate EFT string with $n_{\bf e}=2$ and on the lattice site $a_i=(1,1)$, which is indeed realized by an NS5-brane wrapping $J_2$, whose $\a$-vector reads $\vec{\a}_{{\rm NS5},J_2} = \left( \sqrt{\frac{2}{3}}, \frac{1}{\sqrt{3}} \right)$. Notice that this is one of the cases where the general procedure already detects the EFT string candidate along the pericenter direction. Following our prescription, we would then analyze the 1d moduli space associated to such pericenter and find the same EFT string. If we were to analyze the 1d moduli spaces associated to the boundaries, we would find two more EFT string candidates corresponding to two copies of the first line of Table \ref{tab:geomEFTstr}. The one along the $\vec{\alpha}_{\rm D0}$ is not populated in this growth sector (it would correspond to the $\vec{\a}_{{\rm NS5},J_1}$ in the $s^1 \gg s^2$ growth sector above, which slides out of the frame simplex in the current growth sector). The one along $\vec{\alpha}_{{\rm D2}, T^2}$ lies outside this growth sector, constituting an example of a growth sector that does not cover the whole frame simplex. The reason is that this top-down description does not cover the whole frame simplex. However, the definition of frame simplex guarantees the existence of at least one duality frame that covers it entirely, which in this case would be F-theory on $X_3\times T^2$.
\end{itemize}

\subsection{Type IIA on $\mathbb{P}^{1,1,2,2,6}[12]$}

We now compactify type IIA string theory on the Calabi--Yau $X_3=\mathbb{P}^{1,1,2,2,6}[12]$, see \cite{Candelas:1994hw,Hosono:1993qy,Lee:2019wij}, namely a K3-fibration over a $\mathbb{P}^1$ base. As before, we have two K\"ahler moduli $\{v^1,v^2\}$, parametrizing the volumes of the fibre and the base, respectively, and the triple intersection polynomial reads
\be
{\cal I}(X_3) = 4 J_1^3 + 2 J_1^2 J_2 \, .
\ee
 The saxionic cone and the cone of EFT string charges are the same as in \eqref{cones_ex1} and the flat coordinates read
\be
\begin{split}
x^1 = \frac{1}{\sqrt{6}} \log \cV_X \, ,\\
x^2 = \frac{1}{\sqrt{3}} \log \left[ 1+ \frac{3s^2}{2 s^1} \right] \, .
\end{split}
\ee
 The relevant light towers are the D0-branes and the oscillator modes of an NS5-brane wrapping $J_2$, which corresponds to the K3 fibre, so that we have
 \be
\begin{split}
m_{\rm D0} = \frac{M_{\rm Pl}}{\sqrt{\cV_X}} \, , \qquad \vec{\a}_{\rm D0} = \left( \sqrt{\frac{3}{2}},0 \right)\, , \\
T_{\rm NS5,K3} = M_{\rm Pl}^2 \frac{\cV_{\rm K3}}{\cV_X} \, , \qquad \vec{\a}_{\rm NS5, osc} = \left( \frac{1}{\sqrt{6}}, \frac{1}{\sqrt{3}} \right) \, .
\end{split}
\ee
Since the $\a$-vectors are constant we can actually describe the whole slice of the moduli space with one frame simplex of type ($\infty,1$), see Figure \ref{fig.i1}. Our analysis finds an EFT string candidate with $n_{\bf e}=1$ corresponding to the emergent string, which is indeed realized as the NS5-brane wrapping the K3 fibre. In addition, if one treats the direction that decompactifies to 5d as a special locus, one also finds another candidate EFT string with $n_{\bf e}=3$, which is realized by the NS5-brane wrapping the divisor $J_1$. Notice that the $\alpha$-vector of this latter EFT string points in this direction only along its own flow, while it slides outside of the frame simplex as we explore other directions.

\subsection{F-theory on $B_3=\mathbb{P}^3$}     \label{ss:Fth_P3}

We now consider F-theory on an elliptic Calabi--Yau fourfold $X_4$, whose base $B_3$ is the complex projective space $\mathbb{P}^3$. The analysis is completely analogous to type IIB on a Calabi--Yau orientifold with one K\"ahler modulus. We consider the 2-dimensional slice of the moduli space spanned by the volume modulus of the base and the 10d dilaton $\phi$. In this setup, EFT strings are realized as D3-branes wrapping curves $C \subset B_3$ and D7-branes wrapping the whole base $B_3$, so that we have
\be
\D = \{s^0, s^1 >0\} \, , \qquad {\cal C}_S^{\rm EFT} = \{e^0,e^1 \in \mathbb{Z}_{\geq 0} \} \, ,
\ee
where the saxions are given by $s^0 = e^{-\phi}$ and $s^1 = \frac{1}{2} u^2$, with $u$ the 10d Einstein frame K\"ahler modulus of the base $B_3$, and $e^0$ and $e^1$ are the D7 and D3 charges, respectively. The K\"ahler potential in this case takes the simple form
\be
K = -\log s^0 - 3 \log s^1 \, ,
\ee
and the flat coordinates are
\be
x^0 = \frac{1}{\sqrt 2} \log s^0 \, , \qquad x^1 = \sqrt{\frac{3}{2}} \log s^1 \, .
\ee
Since we want to keep all the corrections under control, we focus on the large volume and weak coupling regime, where the light towers are given by the Kaluza-Klein modes along the base $B_3$ and the oscillators of the fundamental string and their $\a$-vectors are
\be
\begin{split}
m_{{\rm KK},6} = \frac{M_{\rm Pl}}{\cV_{B_3}^{2/3}} \, , \qquad \vec{\a}_{{\rm KK},6} = \left( 0, \sqrt{\frac{2}{3}} \right)\, , \\
T_{\rm F1} = M_{\rm Pl}^2 \frac{e^{\frac{1}{2}\phi}}{\cV_{B_3}} \, , \qquad \vec{\a}_{\rm F1, osc} = \frac{1}{2} \left( \frac{1}{\sqrt{2}}, \sqrt{\frac{3}{2}} \right) \, .
\end{split}
\ee
We are in a frame simplex of the type $(\infty,6)$ and, as one can see from Table \ref{tab:stringyEFTstr} and Figure \ref{fig.i6}, we have three EFT string candidates within the frame simplex, given by the lattice points $(1,3), (1,4), (2,6)$. However, only the one corresponding to the lattice site $(1,4)$ is realized as an elementary EFT string in this setup, namely the D3-brane wrapping a curve inside $B_3$. The string $(2,6)$ is the fundamental string, so it is realized but not as an EFT string, since it is non-BPS. Interestingly, due to the fact that this description extends outside the frame simplex, it can be seen that the bound states of D7-branes wrapping the whole $B_3$ (which provides an elementary EFT string outside the simplex) and the D3 EFT strings above, precisely populate the lattice site $(1,3)$ along their own flow. This corresponds to the pericenter of the D3 and D7 elementary EFT string $\a$-vectors, in agreement with \cite{Grieco:2025bjy} and the discussion in section \ref{ss.bound}. Finally, let us notice that by looking at the special loci in this frame simplex one would simply recover the same three EFT string candidates.

\subsection{Type IIA on the Quintic}  \label{ss:IIAquintic}
The same frame simplex appears if we consider type IIA string theory on the quintic, or more generally, on any Calabi--Yau (orientifold) with one K\"ahler modulus. In this new setup the lattice site $(1,3)$ is indeed populated by an EFT string, namely a D4-brane wrapping the "diagonal" three-cycle $\Sigma_3$ whose volume goes like $\cV_{\Sigma_3} \sim \sqrt{\cV_X}$. The fundamental string is also an EFT string populating the $(2,6)$ lattice site in the Calabi--Yau case, whereas the discussion above applies in the orientifold case, where it does not give rise to an EFT string. The string sitting at the lattice point $(1,4)$ is absent instead. The EFT string limit associated to the D4-string in this setup is an example of an emergent string limit with scaling weight $w=2$. This happens because the EFT string that induces the limit is the D4-string, but the emergent string is not the EFT string itself, it is instead the fundamental string which is lighter.

This setup is also an example where we can glue together two neighboring frame simplices. In particular, we have a frame simplex of type $(6,1)$, generated by the decompactification limits to 10d and to 5d M-theory, that shares the $6$ boundary with the one considered above. Here we find one EFT string candidate in the lattice point $(2,1)$, see Table \ref{tab:geomEFTstr} and Figure \ref{fig.61}, which corresponds to an NS5-brane wrapping a divisor in the Calabi-Yau $X_3$, which realizes the decompactification limit to 5d (unless obstructed by quantum corrections in the case of the orientifold \cite{Kaufmann:2026fli,Kaufmann:2026mha,Kaufmann:2026tsy}).

\subsection{F-theory on $\mathbb{P}^1_f \xhookrightarrow{} B_3 \rightarrow \mathbb{P}^2$}

We now consider F-theory on a slightly more complicated base $B_3$, namely a $\mathbb{P}^1$ fibration over a base $\mathbb{P}^2$, see also \cite{Lanza:2021udy}. We denote the fiber as $\mathbb{P}^1_f$ in order to distinguish it from the base curve $\mathbb{P}^1_b \subset \mathbb{P}^2$. Now we expand the 10d Einstein frame K\"ahler form as $J= u_i J^i$ and we choose a basis such that $u_1 = \cV_{\mathbb{P}^1_f}$ and $u_2 = \cV_{\mathbb{P}^1_b}$ in 10d Planck units. The triple intersection form then reads
\be
{\cal I}(B_3) = k^2 J_1^3 + k J_1^2 J_2 + J_1 J_2^2 \, ,
\ee
where the positive integer $k$ denotes the twisting of the fibration. For simplicity we fix $k=3$, because then the triple intersection form is the same as in \ref{ex1}, but the discussion can be generalized to different values of $k$ in a straightforward manner. In this case we focus only on the K\"ahler sector. The saxionic cone is spanned by the dual coordinates $s^i=\cK^{ijk} u_j u_k$, obtained by contracting the triple intersection numbers $\cK^{ijk}$ with the K\"ahler moduli $u_i$, that parametrize the volume of 4-cycles in 10d Planck units, and EFT strings are realized as D3-branes wrapping movable curves in the base $B_3$
\be
\D = \{s^1 \geq 3 s^2>0\} \, , \qquad {\cal C}_S^{\rm EFT} = \{(e^1,e^2) \in \mathbb{Z}^2 \big | e^1 \geq 3 e^2 \geq 0\} \, .
\ee
Again in the large volume approximation the K\"ahler potential reads
\be
K = - 2 \log \left( \cV_{B_3} (s^i) \right) = -2 \log \left( (s^1)^\frac{3}{2} - (s^1-3 s^2)^\frac{3}{2} \right) \, ,
\ee
and the flat coordinates are
\be
\begin{split}
x^1 = \sqrt{\frac{2}{3}} \log \cV_{B_3} \, ,\\
x^2 = \frac{1}{\sqrt{3}} \log \left[ 1+ \frac{2}{9} \frac{(u_2)^3+(u_2(3 u_1+u_2))^\frac{3}{2}}{ 3(u_1)^3+3(u_1)^2 u_2 + u_1 (u_2)^2} \right] \, .
\end{split}
\ee
The relevant light towers are the Kaluza-Klein modes along the base $\mathbb{P}^2$ and along the whole $B_3$ and the oscillator modes of a D3-brane wrapping $\mathbb{P}^1_f$
\be
\begin{split}
m_{{\rm KK},\mathbb{P}^2} = \frac{M_{\rm Pl}}{\cV_{\mathbb{P}^2}^{1/4} \sqrt{\cV_{B_3}}} \, , \qquad \vec{\a}_{{\rm KK},\mathbb{P}^2} = \left( \sqrt{\frac{2}{3}}, \frac{(u_2(3u_1 + u_2))^{\frac{3}{2}}}{2\sqrt{3} u_2^3} \right)\, , \\
m_{{\rm KK},B_3} = \frac{M_{\rm Pl}}{\cV_{B_3}^{2/3}} \, , \qquad \vec{\a}_{{\rm KK},B_3} = \left( \sqrt{\frac{2}{3}}, 0 \right)\, , \\
T_{{\rm D3},\mathbb{P}^1_f} = M_{\rm Pl}^2 \frac{\cV_{\mathbb{P}^1_f}}{\cV_{B_3}} \, , \qquad \vec{\a}_{\rm D3,osc} = \frac{1}{2} \left( \sqrt{\frac{2}{3}}, \frac{3u_1 + 2u_2}{\sqrt{3 u_2 (3u_1+u_2)}} \right) \, ,
\end{split}
\ee
where the $\a$-vectors are already rotated to the components in flat coordinates. Now we consider the growth sector $s^1 \gg s^2$, namely $u_2 \gg u_1$, where we have
\be
\vec{\a}_{{\rm KK},\mathbb{P}^2} = \left( \sqrt{\frac{2}{3}}, \frac{1}{2\sqrt{3}} \right) \, , \qquad \vec{\a}_{{\rm KK},B_3} = \left( \sqrt{\frac{2}{3}}, 0 \right) \, , \qquad \vec{\a}_{\rm D3,osc} = \frac{1}{2} \left( \sqrt{\frac{2}{3}}, \frac{2}{\sqrt{3}} \right) \, .
\ee
Within this growth sector we have a frame simplex of type $(\infty,4)$, see Figure \ref{fig.i4}, generated by $\vec{\a}_{{\rm KK},\mathbb{P}^2}$ and $\vec{\a}_{\rm D3,osc}$. Looking at Table \ref{tab:stringyEFTstr}, we see that we have two EFT string candidates in the lattice sites $(1,2)$ and $(2,4)$. The first one is not populated in this setup, while the second one is precisely the D3-brane realizing also the emergent string. The same growth sector also explores part of a neighbouring frame simplex, which is generated by $\vec{\a}_{{\rm KK},\mathbb{P}^2}$ and an additional KK tower associated to the decompactification of two dimensions, corresponding to Figure \ref{fig.42}. In this frame simplex $\vec{\a}_{{\rm KK},B_3}$ describes the tower of bound states of these two generators. Along this direction they become light at the same rate, signaling a total decompactification to 10d. However, to fully describe this second frame simplex, in general one needs to perform a change of duality frame. What we just discussed is an example of a growth sector that is not contained in a unique frame simplex, a phenomenon that was already observed in other recent works, see for instance \cite{Grieco:2025bjy}. Let us also stress that the other growth sector, namely $s^2 \gg s^1$, cannot be described within this duality frame. This is consistent with the fact that the growth sector $s^1 \gg s^2$ already leaves the original frame simplex. Furthermore, the limit of the growth sector $s^1 \gg s^2$ is expected to be associated with an EFT string bound state, for which $s^1\sim s^2\to \infty$. In this example, this direction coincides with the vector $\vec{\a}_{{\rm KK},B_3}$, which corresponds to an elementary EFT string, and also with all bound states between the latter and the EFT string associated to $\vec{\a}_{{\rm D3,osc}}$, which also lie along the direction $\vec{\a}_{{\rm KK},B_3}$. We expect this to be a concrete realization of a more general mechanism, namely that growth sectors are delimited by flows associated precisely to bound states of EFT strings.

\subsection{M-theory on $G_2$ manifolds}        \label{ss:G2}

Finally we consider M-theory compactified on a $G_2$ manifold $X_7$, obtained as the quotient of a 7-torus by a finite group $\Gamma$ \cite{Joyce:1,Joyce:2,Joyce:book}. We focus on a particular simple example of $G_2$ manifold that has recently been studied in \cite{Lanza:2021udy,Grieco:2025bjy} in the context of EFT strings. Neglecting the twisted sector arising from the resolution of orbifold singularities we have a 7-dimensional moduli space, parametrized by the saxions $s^i$, $i=1,...,7$, which control the volumes of the 3-cycles that survive the orbifold projection. The saxions in terms of the radii of the torus, measured in 11d Planck units, read
\be \begin{split}		\label{untw_sax}
s^1 &= R_1 R_2 R_7 \, , \quad s^2 = R_1 R_3 R_6 \, , \quad s^3 = R_1 R_4 R_5 \, , \quad s^4 = R_2 R_3 R_5 \, , \\
s^5 &= R_4 R_2 R_6 \, , \quad s^6 = R_3 R_4 R_7 \, , \quad s^7 = R_5 R_6 R_7 \, .
\end{split} \ee
We can also invert these relations to express the seven radii in terms of the saxions
\be \begin{split}		\label{radii(untw_sax)}
R_1 &= \frac{(s^1 s^2 s^3)^{\frac{1}{3}}}{(s^4 s^5 s^6 s^7)^{\frac{1}{6}}} \, , \quad
R_2 = \frac{(s^1 s^4 s^5)^{\frac{1}{3}}}{(s^2 s^3 s^6 s^7)^{\frac{1}{6}}} \, , \quad
R_3 = \frac{(s^2 s^4 s^6)^{\frac{1}{3}}}{(s^1 s^3 s^5 s^7)^{\frac{1}{6}}} \, , \quad
R_4 = \frac{(s^3 s^5 s^6)^{\frac{1}{3}}}{(s^1 s^2 s^4 s^7)^{\frac{1}{6}}} \, , \\
R_5 &= \frac{(s^3 s^4 s^7)^{\frac{1}{3}}}{(s^1 s^2 s^5 s^6)^{\frac{1}{6}}} \, , \quad
R_6 = \frac{(s^2 s^5 s^7)^{\frac{1}{3}}}{(s^1 s^3 s^4 s^6)^{\frac{1}{6}}} \, , \quad
R_7 = \frac{(s^1 s^6 s^7)^{\frac{1}{3}}}{(s^2 s^3 s^4 s^5)^{\frac{1}{6}}} \, . \\
\end{split} \ee
The K\"ahler potential reads
\be     \label{K_G2}
K = - \sum_{i=1}^7 \log(s^i) \, ,
\ee
from where we can derive the moduli space metric
\be
{\cal G}_{ij} = \frac{1}{2} \pa_{s^i} \pa_{s^j} K = \frac{1}{2 (s^i)^2} \d_{ij} \, ,
\ee
and the flat coordinates are given by
\be
y^i = \frac{1}{\sqrt{2}} \log s^i \, .
\ee
The saxionic cone and the cone of EFT string charges are simply
\be
\D = \{s^i > 0 \} \, , \qquad {\cal C}_S^{\rm EFT} = \{e^i \in \mathbb{Z}_{\geq 0} \} \, , \qquad i=1,...,7 \, ,
\ee
and EFT strings are realized as M5-branes wrapping 4-cycles. This 7d moduli space contains several different frame simplices/duality frames, which were analyzed in detail in \cite{Grieco:2025bjy}. Here, we just consider a couple of them to show how they fit into our general discussion. We first consider the frame simplex of type $(1,1,1,1,1,1,1)$ generated by the Kaluza-Klein towers associated to the seven radii, for which we have
\be
m_{{\rm KK}, R_i} = \frac{M_{\rm Pl}}{\sqrt{\cV_X}R_i} \, ,
\ee
\be     \label{KK_G2}
\begin{split}
\vec{\a}_{{\rm KK},R_1} &= \frac{1}{\sqrt{2}} (1,1,1,0,0,0,0) \, , \qquad \vec{\a}_{{\rm KK},R_2} = \frac{1}{\sqrt{2}} (1,0,0,1,1,0,0) \, , \\ \vec{\a}_{{\rm KK},R_3} &= \frac{1}{\sqrt{2}} (0,1,0,1,0,1,0)\, , \qquad \vec{\a}_{{\rm KK},R_4} = \frac{1}{\sqrt{2}} (0,0,1,0,1,1,0) \, , \\
\vec{\a}_{{\rm KK},R_5} &= \frac{1}{\sqrt{2}} (0,0,1,1,0,0,1) \, , \qquad \vec{\a}_{{\rm KK},R_6} = \frac{1}{\sqrt{2}} (0,1,0,0,1,0,1) \, , \\
\vec{\a}_{{\rm KK},R_7} &= \frac{1}{\sqrt{2}} (1,0,0,0,0,1,1) \, .
\end{split}
\ee
By applying the strategy of Section \ref{s:IntegralScalingProof}, in particular the part about lower-dimensional special loci, one finds many EFT string candidates for this frame simplex. The only one that is realized in this concrete setup is the one that appears in the total pericenter, namely a 1d moduli space with a decompactification limit to 11d. Such a string corresponds to an M5-brane with all the charges switched on $e^i > 0 \, , \forall i$. Indeed one can check that its $\a$-vector evaluated along its own flow takes the form
\be
(\a_{\rm M5})^i = \frac{\sqrt{2}}{\sum_{j=1}^7 \frac{e^j}{s^j}} \frac{e^i}{s^i} \qquad \implies \qquad \vec{\a}_{\rm M5} \simeq \frac{\sqrt{2}}{7} (1,1,1,1,1,1,1) \, .
\ee
Notice that this is a non-elementary string, as it is not one of the generators of the cone of EFT string charges, which are populated but lie outside of this frame simplex. This is an example of a bound state EFT string that is captured by our strategy. In general this is not the case, as we will show below. Furthermore,  all the KK towers in \eqref{KK_G2} satisfy integral scaling with $w=3$ for this string, as predicted by Table \ref{tab:geomEFTstr}. This can be understood as a consequence of the fact that all these towers also satisfy integral scaling with respect to the aforementioned elementary EFT strings, and thus the argument at the beginning of section \ref{ss.bound} applies directly.

Let us now consider the lower-dimensional frame simplex of type $(\infty,2)$, see Figure \ref{fig.i2}, generated by the pericenter of the first two KK towers above and the M2-string obtained by wrapping an M2-brane on the direction $R_7$, with tension and $\a$-vector
\be
T_{{\rm M2}, R_7} = M_{\rm Pl}^2 \frac{R_7}{\cV_X} \, , \qquad \vec{\a}_{{\rm M2},R_7} = \frac{1}{\sqrt{2}} (0,1,1,1,1,0,0) \, .
\ee
In terms of the saxions we are restricting to a submanifold obtained by fixing $s^2 = s^3 = s^4 = s^5$ and $s^6,s^7 = \text{const}$. In this frame simplex there are three EFT string candidates with lattice coordinates $(1,1),(1,2),(2,2)$ in Table \ref{tab:stringyEFTstr}. The first corresponds to an M5-string with charges $e^2,e^3,e^4,e^5>0$ and $\a$-vector
\be
\vec{\a}_{\rm M5} \simeq \frac{\sqrt{2}}{4} (0,1,1,1,1,0,0) \, ,
\ee
the second is not realized and the third is the M2-brane wrapping $R_7$ which is not an EFT string. However there is an additional EFT string which is also a bound state, namely the string with charges $e^1,...,e^5>0$ whose $\a$-vector along its own flow reads
\be
\vec{\a}_{\rm M5} \simeq \frac{\sqrt{2}}{5} (1,1,1,1,1,0,0) \, .
\ee
This string satisfies the $2/n_{\bf e}$ relation with $n_{\bf e}=5$ and also the integral scaling relation, since the KK tower associated to $R_1$ and $R_2$ has $w=3$ along its flow, however it does not appear in any of the tables above. In this particular example, since the K\"ahler potential \eqref{K_G2} only contains one monomial, one can actually characterize the $\a$-vectors of EFT string bound states as in \eqref{alpha_bs}. One could also predict the existence of this EFT string from the elementary ones, but this is not guaranteed to be the case for more general K\"ahler potentials. A remarkable observation, based on the examples we analyzed, is that the method described in section \ref{ss.outline} seems to capture bound states of EFT strings if they happen to be aligned with some principal tower. It would be nice to check or prove this more generally.

\section{Conclusions and Outlook}
\label{sec:conclusions}

In this work we have tested the Integral Scaling Conjecture from the bottom-up, using the framework of (brane) taxonomy \cite{Etheredge:2024tok,Etheredge:2025ahf}. Our analysis relies on two assumptions. First, that the taxonomy rules hold for all states parametrically below the species scale in each duality frame. Let us emphasize that these rules are rooted in the Emergent String Conjecture \cite{Lee:2019wij}, but they still incorporate extra assumptions. Second, that EFT string candidates correspond to string lattice sites whose $\alpha$-vectors have norm $\sqrt{2/n_{\bf e}}$ for a positive integer $n_{\bf e}\leq 7$, as follows from the homogeneity of the K\"ahler potential in the asymptotic regimes of 4d $\mathcal{N}=1$ theories. Under these assumptions, we have classified all the frame simplices that can arise in slices of the moduli space of a 4d EFT that correspond to decompactifications up to 11d, or weakly-coupled, critical string theories up to 10d, and we have tested the integral scaling relation \eqref{e.wi} for every EFT string candidate in the 48 (out of 74) frame simplices that contain them within the simplex or at its boundaries.

Our main findings can be summarized as follows:
\begin{itemize}
    \item Integral scaling holds, with integer $w\leq 3$, for all the leading towers along the flow of every EFT string candidate (see Tables \ref{tab:geomEFTstr} and \ref{tab:stringyEFTstr}). For subleading towers it also holds in all the geometric frame simplices, and in all the stringy ones except for the special cases with $w_{\rm osc}=3/2$ discussed in sections \ref{s:IntegralScalingProof} and \ref{sec:topdown}.
    \item Our bottom-up approach tests integral scaling for every EFT string candidate allowed by the lattice rules, independently of whether it is actually populated. Its conclusions are therefore robust under  subsets of candidates being populated or not in a given compactification. Which EFT string candidates are populated indeed varies among different top-down realizations of the same frame simplex.
    \item The $\alpha$-vectors of the oscillator modes of EFT string candidates act as generators of the lattices of particle and string $\alpha$-vectors. We checked this for all 1d and 2d frame simplices  explicitly, and verified it in many higher-dimensional ones, as well as in the top-down examples of section \ref{sec:topdown}. We thus expect it to hold in general, providing a lattice-level strengthening of the convex hull statements of \cite{Grieco:2025bjy} that is moreover compatible with the aforementioned half-integral weights.
    \item Whenever $w=1$, the flow corresponds to an emergent string limit in which the EFT string is itself the emergent string. The opposite is not true, since there exist emergent string limits with $w>1$ in which the emergent string is lighter than the BPS EFT string sourcing the flow (unless these limits turn out to be obstructed, as we discuss below).
    \item For bound states of EFT strings we find evidence that integral scaling is inherited from the elementary constituents, with additive scaling weights, whenever a single monomial of the K\"ahler potential dominates along the flow.
\end{itemize}

Our results also point to several directions that we find worth exploring in the future. Throughout our scan we find evidence that EFT string candidates with $n_{\bf e}>7$ always yield $w>3$ for the leading tower. Establishing this in full generality would imply that the ISC, combined with the taxonomy rules, bounds the degree of homogeneity of the asymptotic K\"ahler potential as $n\leq 7$, in accordance with eleven being the maximal decompactification dimension. \footnote{In \cite{Grieco:2025bjy}, they argue for the converse, namely that $n \leq 7$ implies $w \leq 3$, up to the possibility to have a limit with $w=4$ and $n \in \{6,7\}$ realized by a non-elementary EFT string, which is however not realized in any known top-down construction.} This would in turn justify a key assumption of the sharpened supersymmetric axion Weak Gravity Conjecture of \cite{Etheredge:2026rio}, and we plan to study this connection in more detail in the future.

A related question is whether every lattice site with $|\vec{\alpha}_{\rm str}|=\sqrt{2/n_{\bf e}}$ must be populated by an actual EFT string. The examples of section \ref{sec:topdown} show that this is not the case in general. It is interesting to understand whether some completeness-like principle (maybe along the lines of the Distant Axionic String Conjecture \cite{Lanza:2021udy} itself) requires a minimal subset of sites to be populated. The electromagnetic dual of this question, phrased in terms of instantons and axions, is closely related to sharpening the axion Weak Gravity Conjecture \cite{Etheredge:2026rio} (see also \cite{DiUbaldo:2026rly} for a recent derivation of the latter from positivity of the gravitational path integral).

Regarding bound states of EFT strings, our analysis covers the case in which a single monomial of the K\"ahler potential dominates along the flow, and the general case with several competing monomials remains open. We do not see clear evidence that the ISC would hold if such K\"ahler potentials were allowed in the landscape. Understanding it in detail requires keeping track of the sliding of $\alpha$-vectors along non-elementary flows (see e.g. \cite{Etheredge:2023odp} for related discussions), as well as clarifying under which conditions one can consistently restrict the analysis to lower-dimensional slices and special loci of a given frame simplex.

It would also be very interesting to promote our observations about generators to a precise and general lattice statement, and to explore its fate in dimensions other than four. Codimension-two objects remain the natural candidates to implement monodromies in higher dimensions, so it is natural to expect generalizations of EFT strings with worldvolume dimension $p>2$ to play their role there. As discussed in section \ref{ss.bound}, the factor of $1/2$ behind the half-integral weights can be traced back to the worldvolume dimension of the EFT string being two, and for higher-dimensional generalizations one may thus expect scaling weights quantized in units of $1/p$. First checks of these ideas could be performed in ten-dimensional string theories, where the $\alpha$-vectors of the fundamental string oscillators are expected to generate the corresponding brane lattices, and it would also be interesting to explore integral scaling relations for particles in 3d minimal supergravity. Let us also mention the recent observation of a universal co-scaling between towers of magnetic strings and electric particles in 5d \cite{Reece:2025zva}, which resonates with the relations studied here, and the upcoming bottom-up analysis of \cite{GriecoRuizValenzuelaToAppear}, where integral scaling relations are used to constrain the possible towers of light states and K\"ahler potentials.

As mentioned above, along some of the EFT string flows with $w>1$ the emergent string is lighter than the EFT string itself. It would be interesting to understand whether some of these limits are obstructed once quantum corrections are taken into account, possibly in connection with reductions of supersymmetry along the flow, in the spirit of the $\mathcal{N}=1$ obstructions recently studied in \cite{Kaufmann:2026fli,Kaufmann:2026mha,Kaufmann:2026tsy}.

Our top-down analysis in section \ref{sec:topdown} constitutes also a step towards a precise dictionary between the assumptions of the (brane) taxonomy framework, formulated in lattice language, and physical inputs such as the amount of supersymmetry, the compactification geometry and the available brane content. Extending this dictionary through broader top-down scans, in a way complementary to \cite{Grieco:2025bjy}, would help delimit the regime of validity of the lattice rules. If these turn out to be more generally applicable, the bottom-up rationale for integral scaling presented here would get closer to a derivation from first principles.

Another aspect that we leave for future work is the interplay between integral scaling and the species scale. Given the universal pattern \cite{Castellano:2023stg,Castellano:2023jjt} satisfied by the species scale vector \cite{Calderon-Infante:2023ler} and the tower vectors, it is natural to ask whether the scaling weights of the species scale along EFT string flows are constrained in a similar fashion, and whether they can be determined from the frame simplex data.

Let us finally comment on strings that are not EFT strings. As illustrated by the F-theory example of section \ref{sec:topdown}, the emergent string along a given limit can be non-BPS, and thus not an EFT string. It is thus natural to explore whether integral scaling relations extend to more general axionic strings, along the lines of \cite{Heidenreich:2021yda}, and ultimately to setups with reduced or no supersymmetry.

All in all, our results indicate that integral scaling is not an accident of particular compactifications, but a structural property of the lattices of towers and extended objects predicted by the Emergent String Conjecture at infinite distance. We regard this interplay between the physics of extended objects and the structure of infinite-distance limits as a promising window into the principles underlying quantum gravity constraints on EFTs.

\vspace{15pt}
\begin{center} 
{\textbf{Acknowledgements}}
\end{center}

We would like to thank Manuel Artime, Christian Aoufia, Matilda Delgado, Bernardo Fraiman, Alessandra Grieco, Ben Heidenreich, Sanjay Raman, Matthew Reece, Tom Rudelius, Ignacio Ruiz, Alex Stewart, Giulia Tazzoli, Irene Valenzuela and Matteo Zatti for useful conversations. We also thank Alessandra Grieco, Ignacio Ruiz and Irene Valenzuela for coordinating the submission date of their work \cite{GriecoRuizValenzuelaToAppear} with that of the present paper. The work of D.L. is supported by the German-Israel-Project (DIP) on Holography and the Swampland.

\newpage
\appendix
\section{$\alpha$-vectors for Candidate EFT Strings}
\label{ap:Homogeneousfunctions}
The goal of this appendix is to show that \eqref{eq:EFTstringnorm} for the norm of the EFT-string $\a$-vector holds at leading order in $\sigma$ along the flow \eqref{eq:EFTstringscalarflows}, using homogeneity of $P(s^i)$ in the K\"ahler potential \eqref{Kahler_pot}. We begin by rewriting \eqref{eq:EFTstringalphavector} using the explicit form of the K\"ahler potential \eqref{Kahler_pot} to get
\be
\label{eq:EFTstrnorm.app}
|\vec{\a}_{\rm str}| = \sqrt{2\left( 1-P(s) \frac{P_{ij} e^i e^j}{(P_k e^k)^2} \right)} \, .
\ee
where we have defined $P_j=\pa_{j}P(s^i)$ and $P_{jk}=\pa_{j}\pa_k P(s^i)$.
We can now use the fact that a homogeneous polynomial of integer degree $n$ in the saxions can be written as 
\be
P(s) = P_{i_1...i_n} s^{i_1} ... \, s^{i_n} \, ,
\ee
with $P_{i_1...\, i_n}$ a constant completely symmetric tensor. We then have
\begin{align}
P_j = \pa_j P(s) = n P_{j i_2...i_n} s^{i_2} ... \, s^{i_n} \, ,\\
P_{jk} = \pa_j \pa_k P(s) = n(n-1) P_{jk i_3...i_n} s^{i_3} ... \, s^{i_n} \, .
\end{align}
Evaluating these expressions along the EFT string flow \eqref{eq:EFTstringscalarflows} we find
\begin{align}
P(s) &= {\bf P}_{\bf e} \sigma^n + n \left( {\bf P}_{\bf e} \right)_{i} s^i_0 \sigma^{n-1} + ... \, ,\\
P_j e^j \sigma &= n {\bf P}_{\bf e} \sigma^n + n(n-1) \left( {\bf P}_{\bf e} \right)_{i} s^i_0 \sigma^{n-1} + ... \, ,\\
P_{jk} e^j e^k \sigma^2 &= n(n-1) {\bf P}_{\bf e} \sigma^n + n(n-1)(n-2) \left( {\bf P}_{\bf e} \right)_{i} s^i_0 \sigma^{n-1} + ... \, ,
\end{align}
where we have defined ${\bf P}_{\bf e} \equiv P_{i_1... \, i_n} e^{i_1} ... \, e^{i_n}$ and $\left( {\bf P}_{\bf e} \right)_{j} \equiv P_{j i_2 ... \, i_n} e^{i_2} ... \, e^{i_n}$. We finally arrive at
\be		\label{P_homog}
P_i e^i \sigma \simeq n_{\bf e} P \, , \qquad P_{ij} e^i e^j \sigma^2 \simeq n_{\bf e} \left( n_{\bf e}-1\right) P \, .
\ee
It is also possible to generalize \eqref{P_homog} to $P(s) = N(s)/D(s)$, where $N$ and $D$ are two homogeneous polynomials of the saxions and the degree of $N$ is higher than the one of $D$.
Inserting this into eq. \eqref{eq:EFTstrnorm.app} one arrives at 
\begin{equation}
    |\vec{\a}_{\rm str}|=\sqrt{\dfrac{2}{n_{\bf e}}}\, ,
\end{equation}
recovering eq. \eqref{eq:EFTstringnorm}

\section{Extra Tables}
\label{ap:tables}

Here we collect the complete tables with all the EFT string candidates in all the frame simplices we have analyzed, divided into geometric frames and stringy frames. We highlight in blue the candidates that are inside or at the boundaries of the frame simplices, which are the ones that appear in Tables \ref{tab:geomEFTstr} and \ref{tab:stringyEFTstr}.

\setlength{\arrayrulewidth}{0.2mm}
\renewcommand{\arraystretch}{1.5}
\begin{sidewaystable}
\caption{Geometric frame simplices, EFT string candidates and scaling weights of the KK towers (up to permutations of $k_i$ and $a_i$). In blue we highlight the ones that lie inside or on the boundary of the frame simplex and that appear in Table \ref{tab:geomEFTstr}.}
\tiny
\begin{center}
\resizebox{\textheight}{!}{%
\begin{tabular}{|c|c|c|c|}
\hline
\rowcolor{lightgray} $k_i$ & $a_i$ & $n_{\bf e}$ & $w_i$ \\
\hline
\textcolor{blue}{$(1,1)$} & \textcolor{blue}{$(1,1)$} & \textcolor{blue}{$2$} & \textcolor{blue}{$(2,2)$} \\
\textcolor{blue}{$(3,1)$} & \textcolor{blue}{$(1,1)$}, $(2,0)$ & \textcolor{blue}{$3$}, $3$ & \textcolor{blue}{$(1,3)$}, $(2,0)$ \\
\textcolor{blue}{$(2,2)$} & $(0,1)$, \textcolor{blue}{$(1,1)$}, \textcolor{blue}{$(1,2)$} & $6$, \textcolor{blue}{$6$}, \textcolor{blue}{$2$} & $(0,3)$, \textcolor{blue}{$(3,3)$}, \textcolor{blue}{$(1,2)$} \\
\textcolor{blue}{$(4,2)$} & $(2,0)$, \textcolor{blue}{$(2,2)$}, $(4,0)$ & $4$, \textcolor{blue}{$2$}, $1$ & $(2,0)$, \textcolor{blue}{$(1,2)$}, $(1,0)$ \\
\textcolor{blue}{$(3,3)$} & \textcolor{blue}{$(2,2)$} & \textcolor{blue}{$3$} & \textcolor{blue}{$(2,2)$} \\
\textcolor{blue}{$(6,1)$} & \textcolor{blue}{$(2,1)$}, $(3,0)$, $(6,0)$ & \textcolor{blue}{$3$}, $4$, $1$ & \textcolor{blue}{$(1,3)$}, $(2,0)$, $(1,0)$ \\
\textcolor{blue}{$(4,3)$} & $(0,3)$, \textcolor{blue}{$(2,1)$} & $1$, \textcolor{blue}{$6$} & $(0,1)$, \textcolor{blue}{$(3,2)$} \\
\hline
\textcolor{blue}{$(2,1,1)$} & $(1,0,0)$, \textcolor{blue}{$(1,1,1)$} & $6$, \textcolor{blue}{$2$} & $(3,0,0)$, \textcolor{blue}{$(1,2,2)$} \\
\textcolor{blue}{$(4,1,1)$} & $(2,0,0)$, \textcolor{blue}{$(2,1,1)$}, $(4,0,0)$ & $4$, \textcolor{blue}{$2$}, $1$ & $(2,0,0)$, \textcolor{blue}{$(1,2,2)$}, $(1,0,0)$ \\
$(3,2,1)$ & $(0,1,1)$, $(1,1,0)$, $(3,0,1)$ & $2$, $6$, $1$ & $(0,1,2)$, $(2,3,0)$, $(1,0,1)$ \\
\textcolor{blue}{$(2,2,2)$} & $(0,1,1)$, $(0,2,2)$, \textcolor{blue}{$(1,1,2)$} & $4$, $1$, \textcolor{blue}{$2$} & $(0,2,2)$, $(0,1,1)$, \textcolor{blue}{$(1,1,2)$} \\
$(5,1,1)$ & $(5,0,1)$ & $1$ & $(1,0,1)$ \\
$(4,2,1)$ & $(0,2,1)$, $(2,0,1)$, $(2,1,0)$, $(4,2,0)$ & $1$, $2$, $4$, $1$ & $(0,1,1)$, $(1,0,2)$, $(2,2,0)$, $(1,1,0)$ \\
\textcolor{blue}{$(3,3,1)$} & $(0,3,0)$, \textcolor{blue}{$(1,1,1)$}, $(1,2,0)$, $(3,3,0)$ & $1$, \textcolor{blue}{$3$}, $3$, $1$ & $(0,1,0)$, \textcolor{blue}{$(1,1,3)$}, $(1,2,0)$, $(1,1,0)$ \\
\textcolor{blue}{$(3,2,2)$} & \textcolor{blue}{$(1,1,1)$}, $(3,0,0)$ & \textcolor{blue}{$6$}, $1$ & \textcolor{blue}{$(2,3,3)$}, $(1,0,0)$ \\
\hline
$(3,1,1,1)$ & $(3,0,0,1)$ & $1$ & $(1,0,0,1)$ \\
\textcolor{blue}{$(2,2,1,1)$} & $(0,1,0,1)$, $(0,2,1,1)$, $(1,1,0,0)$, \textcolor{blue}{$(1,1,1,1)$}, $(2,2,0,0)$ & $2$, $1$, $4$, \textcolor{blue}{$2$}, $1$ & $(0,1,0,2)$, $(0,1,1,1)$, $(2,2,0,0)$, \textcolor{blue}{$(1,1,2,2)$}, $(1,1,0,0)$ \\
$(4,1,1,1)$ & $(0,1,1,1)$, $(2,0,0,1)$, $(4,0,1,1)$ & $1$, $2$, $1$ & $(0,1,1,1)$, $(1,0,0,2)$, $(1,0,1,1)$ \\
$(3,2,1,1)$ & $(0,2,0,1)$, $(3,0,0,0)$, $(3,2,0,1)$ & $1$, $1$, $1$ & $(0,1,0,1)$, $(1,0,0,0)$, $(1,1,0,1)$ \\
$(2,2,2,1)$ & $(0,0,2,1)$, $(0,1,1,1)$, $(1,1,1,0)$, $(2,2,2,0)$ & $1$, $2$, $4$, $1$ & $(0,0,1,1)$, $(0,1,1,2)$, $(2,2,2,0)$, $(1,1,1,0)$ \\
\hline
$(2,1,1,1,1)$ & $(0,1,1,1,1)$, $(1,0,0,0,1)$, $(2,0,0,1,1)$ & $1$, $2$, $1$ & $(0,1,1,1,1)$, $(1,0,0,0,2)$, $(1,0,0,1,1)$ \\
$(3,1,1,1,1)$ & $(0,0,1,1,1)$, $(3,0,0,0,0)$, $(3,0,1,1,1)$ & $1$, $1$, $1$ & $(0,0,1,1,1)$, $(1,0,0,0,0)$, $(1,0,1,1,1)$ \\
$(2,2,1,1,1)$ & $(0,0,1,1,1)$, $(0,2,0,0,1)$, $(1,1,0,0,1)$, $(2,2,0,1,1)$ & $1$, $1$, $2$, $1$ & $(0,0,1,1,1)$, $(0,1,0,0,1)$, $(1,1,0,0,2)$, $(1,1,0,1,1)$ \\
\hline
$(1,1,1,1,1,1)$ & $(0,0,1,1,1,1)$ & $1$ & $(0,0,1,1,1,1)$ \\
$(2,1,1,1,1,1)$ & $(0,0,0,1,1,1)$, $(2,0,0,0,0,1)$, $(2,0,1,1,1,1)$ & $1$, $1$, $1$ & $(0,0,0,1,1,1)$, $(1,0,0,0,0,1)$, $(1,0,1,1,1,1)$ \\
\hline
$(1,1,1,1,1,1,1)$ & $(0,0,0,0,1,1,1)$, $(0,1,1,1,1,1,1)$ & $1$, $1$ & $(0,0,0,0,1,1,1)$, $(0,1,1,1,1,1,1)$ \\
\hline
\end{tabular}
}
\end{center}
\end{sidewaystable}

\begin{landscape}
\scriptsize
\begin{longtable}{|c|P{7cm}|P{2cm}|P{7cm}|}
\caption{Stringy frame simplices, EFT string candidates and scaling weights of the towers (up to permutations of $k_i$ and $a_i$). In blue we highlight the ones that lie inside or on the boundary of the frame simplex and that appear in Table \ref{tab:stringyEFTstr}.}\\
\hline
\rowcolor{lightgray} $k_i$ & $(a_{\rm str},a_i)$ & $n_{\bf e}$ & $(w_{\rm osc},w_i)$ \\
\hline
\endfirsthead
\multicolumn{4}{c}{\tablename\ \thetable{} -- continued}\\
\hline
\rowcolor{lightgray} $k_i$ & $(a_{\rm str},a_i)$ & $n_{\bf e}$ & $(w_{\rm osc},w_i)$ \\
\hline
\endhead
\hline
\endfoot
%
%
\textcolor{blue}{$(\infty,1)$} & $(0,1)$, \textcolor{blue}{$(2,1)$} & $2$, \textcolor{blue}{$1$} & $(0,2)$, \textcolor{blue}{$(1,1)$} \\
\textcolor{blue}{$(\infty,2)$} & $(0,1)$, $(0,2)$, $(1,0)$, \textcolor{blue}{$(1,1)$}, \textcolor{blue}{$(1,2)$}, \textcolor{blue}{$(2,2)$} & $4$, $1$, $2$, \textcolor{blue}{$4$}, \textcolor{blue}{$2$}, \textcolor{blue}{$1$} & $(0,2)$, $(0,1)$, $(1,0)$, \textcolor{blue}{$(2,2)$}, \textcolor{blue}{$(1,2)$}, \textcolor{blue}{$(1,1)$} \\
\textcolor{blue}{$(\infty,3)$} & $(0,1)$, \textcolor{blue}{$(2,3)$} & $6$, \textcolor{blue}{$1$} & $(0,2)$, \textcolor{blue}{$(1,1)$} \\
\textcolor{blue}{$(\infty,4)$} & $(0,2)$, \textcolor{blue}{$(1,2)$}, \textcolor{blue}{$(2,4)$} & $2$, \textcolor{blue}{$4$}, \textcolor{blue}{$1$} & $(0,1)$, \textcolor{blue}{$(2,2)$}, \textcolor{blue}{$(1,1)$} \\
\textcolor{blue}{$(\infty,5)$} & \textcolor{blue}{$(2,5)$} & \textcolor{blue}{$1$} & \textcolor{blue}{$(1,1)$} \\
\textcolor{blue}{$(\infty,6)$} & $(0,2)$, $(1,0)$, $(1,2)$, \textcolor{blue}{$(1,3)$}, \textcolor{blue}{$(1,4)$}, $(1,6)$, \textcolor{blue}{$(2,6)$} & $3$, $1$, $3$, \textcolor{blue}{$4$}, \textcolor{blue}{$3$}, $1$, \textcolor{blue}{$1$} & $(0,1)$, $(\frac{1}{2},0)$, $(\frac{3}{2},1)$, \textcolor{blue}{$(2,2)$}, \textcolor{blue}{$(\frac{3}{2},2)$}, $(\frac{1}{2},1)$, \textcolor{blue}{$(1,1)$} \\
\hline
%
%
\textcolor{blue}{$(\infty,1,1)$} & $(0,0,1)$, $(0,1,1)$, $(1,0,0)$, $(1,0,1)$, \textcolor{blue}{$(1,1,1)$}, \textcolor{blue}{$(2,1,1)$} & $2$, $1$, $2$, $2$, \textcolor{blue}{$2$}, \textcolor{blue}{$1$} & $(0,0,2)$, $(0,1,1)$, $(1,0,0)$, $(1,0,2)$, \textcolor{blue}{$(1,2,2)$}, \textcolor{blue}{$(1,1,1)$} \\
\textcolor{blue}{$(\infty,2,1)$} & $(0,0,1)$, $(0,1,0)$, $(0,2,0)$, \textcolor{blue}{$(2,2,1)$} & $2$, $4$, $1$, \textcolor{blue}{$1$} & $(0,0,2)$, $(0,2,0)$, $(0,1,0)$, \textcolor{blue}{$(1,1,1)$} \\
\textcolor{blue}{$(\infty,3,1)$} & $(0,0,1)$, $(0,1,0)$, \textcolor{blue}{$(2,3,1)$} & $2$, $6$, \textcolor{blue}{$1$} & $(0,0,2)$, $(0,2,0)$, \textcolor{blue}{$(1,1,1)$} \\
\textcolor{blue}{$(\infty,2,2)$} & $(0,0,1)$, $(0,0,2)$, $(0,1,1)$, $(1,0,1)$, \textcolor{blue}{$(1,1,1)$}, \textcolor{blue}{$(1,1,2)$}, \textcolor{blue}{$(2,2,2)$} & $4$, $1$, $2$, $2$, \textcolor{blue}{$4$}, \textcolor{blue}{$2$}, \textcolor{blue}{$1$} & $(0,0,2)$, $(0,0,1)$, $(0,1,1)$, $(1,0,1)$, \textcolor{blue}{$(2,2,2)$}, \textcolor{blue}{$(1,1,2)$}, \textcolor{blue}{$(1,1,1)$} \\
\textcolor{blue}{$(\infty,4,1)$} & $(0,0,1)$, $(0,2,0)$, $(0,2,1)$, $(1,1,0)$, $(1,1,1)$, $(1,3,0)$, $(1,3,1)$, \textcolor{blue}{$(2,4,1)$} & $2$, $2$, $1$, $2$, $2$, $2$, $2$, \textcolor{blue}{$1$} & $(0,0,2)$, $(0,1,0)$, $(0,\frac{1}{2},1)$, $(1,\frac{1}{2},0)$, $(1,\frac{1}{2},2)$, $(1,\frac{3}{2},0)$, $(1,\frac{3}{2},2)$, \textcolor{blue}{$(1,1,1)$} \\
\textcolor{blue}{$(\infty,3,2)$} & $(0,0,1)$, $(0,0,2)$, $(0,1,0)$, \textcolor{blue}{$(2,3,2)$} & $4$, $1$, $6$, \textcolor{blue}{$1$} & $(0,0,2)$, $(0,0,1)$, $(0,2,0)$, \textcolor{blue}{$(1,1,1)$} \\
\textcolor{blue}{$(\infty,5,1)$} & $(0,0,1)$, $(1,0,0)$, $(1,0,1)$, $(1,5,0)$, $(1,5,1)$, \textcolor{blue}{$(2,5,1)$} & $2$, $1$, $1$, $1$, $1$, \textcolor{blue}{$1$} & $(0,0,2)$, $(\frac{1}{2},0,0)$, $(\frac{1}{2},0,1)$, $(\frac{1}{2},1,0)$, $(\frac{1}{2},1,1)$, \textcolor{blue}{$(1,1,1)$} \\
\textcolor{blue}{$(\infty,4,2)$} & $(0,0,1)$, $(0,0,2)$, $(0,2,0)$, $(1,0,0)$, $(1,0,2)$, $(1,2,0)$, \textcolor{blue}{$(1,2,1)$}, \textcolor{blue}{$(1,2,2)$}, $(1,4,0)$, $(1,4,2)$, \textcolor{blue}{$(2,4,2)$} & $4$, $1$, $2$, $1$, $1$, $2$, \textcolor{blue}{$4$}, \textcolor{blue}{$2$}, $1$, $1$, \textcolor{blue}{$1$} & $(0,0,2)$, $(0,0,1)$, $(0,1,0)$, $(\frac{1}{2},0,0)$, $(\frac{1}{2},0,1)$, $(1,1,0)$, \textcolor{blue}{$(2,2,2)$}, \textcolor{blue}{$(1,1,2)$}, $(\frac{1}{2},1,0)$, $(\frac{1}{2},1,1)$, \textcolor{blue}{$(1,1,1)$} \\
\textcolor{blue}{$(\infty,3,3)$} & $(0,0,1)$, $(0,1,1)$, $(1,0,0)$, $(1,0,3)$, $(1,1,1)$, $(1,1,2)$, \textcolor{blue}{$(1,2,2)$}, $(1,3,3)$, \textcolor{blue}{$(2,3,3)$} & $6$, $3$, $1$, $1$, $3$, $3$, \textcolor{blue}{$3$}, $1$, \textcolor{blue}{$1$} & $(0,0,2)$, $(0,1,1)$, $(\frac{1}{2},0,0)$, $(\frac{1}{2},0,1)$, $(\frac{3}{2},1,1)$, $(\frac{3}{2},1,2)$, \textcolor{blue}{$(\frac{3}{2},2,2)$}, $(\frac{1}{2},1,1)$, \textcolor{blue}{$(1,1,1)$} \\
\hline
%
%
\textcolor{blue}{$(\infty,1,1,1)$} & $(0,0,0,1)$, $(0,0,1,1)$, \textcolor{blue}{$(2,1,1,1)$} & $2$, $1$, \textcolor{blue}{$1$} & $(0,0,0,2)$, $(0,0,1,1)$, \textcolor{blue}{$(1,1,1,1)$} \\
\textcolor{blue}{$(\infty,2,1,1)$} & $(0,0,0,1)$, $(0,0,1,1)$, $(0,1,0,0)$, $(0,2,0,0)$, $(1,1,0,0)$, $(1,1,0,1)$, \textcolor{blue}{$(1,1,1,1)$}, \textcolor{blue}{$(2,2,1,1)$} & $2$, $1$, $4$, $1$, $2$, $2$, \textcolor{blue}{$2$}, \textcolor{blue}{$1$} & $(0,0,0,2)$, $(0,0,1,1)$, $(0,2,0,0)$, $(0,1,0,0)$, $(1,1,0,0)$, $(1,1,0,2)$, \textcolor{blue}{$(1,1,2,2)$}, \textcolor{blue}{$(1,1,1,1)$} \\
\textcolor{blue}{$(\infty,3,1,1)$} & $(0,0,0,1)$, $(0,0,1,1)$, $(0,1,0,0)$, \textcolor{blue}{$(2,3,1,1)$} & $2$, $1$, $6$, \textcolor{blue}{$1$} & $(0,0,0,2)$, $(0,0,1,1)$, $(0,2,0,0)$, \textcolor{blue}{$(1,1,1,1)$} \\
\textcolor{blue}{$(\infty,2,2,1)$} & $(0,0,0,1)$, $(0,0,1,0)$, $(0,0,2,0)$, $(0,1,1,0)$, $(0,1,1,1)$, \textcolor{blue}{$(2,2,2,1)$} & $2$, $4$, $1$, $2$, $1$, \textcolor{blue}{$1$} & $(0,0,0,2)$, $(0,0,2,0)$, $(0,0,1,0)$, $(0,1,1,0)$, $(0,\frac{1}{2},\frac{1}{2},1)$, \textcolor{blue}{$(1,1,1,1)$} \\
\textcolor{blue}{$(\infty,4,1,1)$} & $(0,0,0,1)$, $(0,0,1,1)$, $(0,2,0,0)$, $(0,2,0,1)$, $(1,0,0,0)$, $(1,0,0,1)$, $(1,0,1,1)$, $(1,2,0,0)$, $(1,2,0,1)$, \textcolor{blue}{$(1,2,1,1)$}, $(1,4,0,0)$, $(1,4,0,1)$, $(1,4,1,1)$, \textcolor{blue}{$(2,4,1,1)$} & $2$, $1$, $2$, $1$, $1$, $1$, $1$, $2$, $2$, \textcolor{blue}{$2$}, $1$, $1$, $1$, \textcolor{blue}{$1$} & $(0,0,0,2)$, $(0,0,1,1)$, $(0,1,0,0)$, $(0,\frac{1}{2},0,1)$, $(\frac{1}{2},0,0,0)$, $(\frac{1}{2},0,0,1)$, $(\frac{1}{2},0,1,1)$, $(1,1,0,0)$, $(1,1,0,2)$, \textcolor{blue}{$(1,1,2,2)$}, $(\frac{1}{2},1,0,0)$, $(\frac{1}{2},1,0,1)$, $(\frac{1}{2},1,1,1)$, \textcolor{blue}{$(1,1,1,1)$} \\
\textcolor{blue}{$(\infty,3,2,1)$} & $(0,0,0,1)$, $(0,0,1,0)$, $(0,0,2,0)$, $(0,1,0,0)$, $(1,0,0,0)$, $(1,0,0,1)$, $(1,0,2,0)$, $(1,0,2,1)$, $(1,3,0,0)$, $(1,3,0,1)$, $(1,3,2,0)$, $(1,3,2,1)$, \textcolor{blue}{$(2,3,2,1)$} & $2$, $4$, $1$, $6$, $1$, $1$, $1$, $1$, $1$, $1$, $1$, $1$, \textcolor{blue}{$1$} & $(0,0,0,2)$, $(0,0,2,0)$, $(0,0,1,0)$, $(0,2,0,0)$, $(\frac{1}{2},0,0,0)$, $(\frac{1}{2},0,0,1)$, $(\frac{1}{2},0,1,0)$, $(\frac{1}{2},0,1,1)$, $(\frac{1}{2},1,0,0)$, $(\frac{1}{2},1,0,1)$, $(\frac{1}{2},1,1,0)$, $(\frac{1}{2},1,1,1)$, \textcolor{blue}{$(1,1,1,1)$} \\
\textcolor{blue}{$(\infty,2,2,2)$} & $(0,0,0,1)$, $(0,0,0,2)$, $(0,0,1,1)$, $(1,0,0,0)$, $(1,0,0,2)$, $(1,0,1,1)$, $(1,0,2,2)$, \textcolor{blue}{$(1,1,1,1)$}, \textcolor{blue}{$(1,1,1,2)$}, $(1,2,2,2)$, \textcolor{blue}{$(2,2,2,2)$} & $4$, $1$, $2$, $1$, $1$, $2$, $1$, \textcolor{blue}{$4$}, \textcolor{blue}{$2$}, $1$, \textcolor{blue}{$1$} & $(0,0,0,2)$, $(0,0,0,1)$, $(0,0,1,1)$, $(\frac{1}{2},0,0,0)$, $(\frac{1}{2},0,0,1)$, $(1,0,1,1)$, $(\frac{1}{2},0,1,1)$, \textcolor{blue}{$(2,2,2,2)$}, \textcolor{blue}{$(1,1,1,2)$}, $(\frac{1}{2},1,1,1)$, \textcolor{blue}{$(1,1,1,1)$} \\
\hline
%
%
\textcolor{blue}{$(\infty,1,1,1,1)$} & $(0,0,0,0,1)$, $(0,0,0,1,1)$, \textcolor{blue}{$(2,1,1,1,1)$} & $2$, $1$, \textcolor{blue}{$1$} & $(0,0,0,0,2)$, $(0,0,0,1,1)$, \textcolor{blue}{$(1,1,1,1,1)$} \\
\textcolor{blue}{$(\infty,2,1,1,1)$} & $(0,0,0,0,1)$, $(0,0,0,1,1)$, $(0,1,0,0,0)$, $(0,2,0,0,0)$, \textcolor{blue}{$(2,2,1,1,1)$} & $2$, $1$, $4$, $1$, \textcolor{blue}{$1$} & $(0,0,0,0,2)$, $(0,0,0,1,1)$, $(0,2,0,0,0)$, $(0,1,0,0,0)$, \textcolor{blue}{$(1,1,1,1,1)$} \\
\textcolor{blue}{$(\infty,3,1,1,1)$} & $(0,0,0,0,1)$, $(0,0,0,1,1)$, $(0,1,0,0,0)$, $(1,0,0,0,0)$, $(1,0,0,0,1)$, $(1,0,0,1,1)$, $(1,0,1,1,1)$, $(1,3,0,0,0)$, $(1,3,0,0,1)$, $(1,3,0,1,1)$, $(1,3,1,1,1)$, \textcolor{blue}{$(2,3,1,1,1)$} & $2$, $1$, $6$, $1$, $1$, $1$, $1$, $1$, $1$, $1$, $1$, \textcolor{blue}{$1$} & $(0,0,0,0,2)$, $(0,0,0,1,1)$, $(0,2,0,0,0)$, $(\frac{1}{2},0,0,0,0)$, $(\frac{1}{2},0,0,0,1)$, $(\frac{1}{2},0,0,1,1)$, $(\frac{1}{2},0,1,1,1)$, $(\frac{1}{2},1,0,0,0)$, $(\frac{1}{2},1,0,0,1)$, $(\frac{1}{2},1,0,1,1)$, $(\frac{1}{2},1,1,1,1)$, \textcolor{blue}{$(1,1,1,1,1)$} \\
\textcolor{blue}{$(\infty,2,2,1,1)$} & $(0,0,0,0,1)$, $(0,0,0,1,1)$, $(0,0,1,0,0)$, $(0,0,2,0,0)$, $(0,1,1,0,0)$, $(0,1,1,0,1)$, $(1,0,0,0,0)$, $(1,0,0,0,1)$, $(1,0,0,1,1)$, $(1,0,2,0,0)$, $(1,0,2,0,1)$, $(1,0,2,1,1)$, $(1,1,1,0,0)$, $(1,1,1,0,1)$, \textcolor{blue}{$(1,1,1,1,1)$}, $(1,2,2,0,0)$, $(1,2,2,0,1)$, $(1,2,2,1,1)$, \textcolor{blue}{$(2,2,2,1,1)$} & $2$, $1$, $4$, $1$, $2$, $1$, $1$, $1$, $1$, $1$, $1$, $1$, $2$, $2$, \textcolor{blue}{$2$}, $1$, $1$, $1$, \textcolor{blue}{$1$} & $(0,0,0,0,2)$, $(0,0,0,1,1)$, $(0,0,2,0,0)$, $(0,0,1,0,0)$, $(0,1,1,0,0)$, $(0,\frac{1}{2},\frac{1}{2},0,1)$, $(\frac{1}{2},0,0,0,0)$, $(\frac{1}{2},0,0,0,1)$, $(\frac{1}{2},0,0,1,1)$, $(\frac{1}{2},0,1,0,0)$, $(\frac{1}{2},0,1,0,1)$, $(\frac{1}{2},0,1,1,1)$, $(1,1,1,0,0)$, $(1,1,1,0,2)$, \textcolor{blue}{$(1,1,1,2,2)$}, $(\frac{1}{2},1,1,0,0)$, $(\frac{1}{2},1,1,0,1)$, $(\frac{1}{2},1,1,1,1)$, \textcolor{blue}{$(1,1,1,1,1)$} \\
\hline
%
%
\textcolor{blue}{$(\infty,1,1,1,1,1)$} & $(0,0,0,0,0,1)$, $(0,0,0,0,1,1)$, \textcolor{blue}{$(2,1,1,1,1,1)$} & $2$, $1$, \textcolor{blue}{$1$} & $(0,0,0,0,0,2)$, $(0,0,0,0,1,1)$, \textcolor{blue}{$(1,1,1,1,1,1)$} \\
\textcolor{blue}{$(\infty,2,1,1,1,1)$} & $(0,0,0,0,0,1)$, $(0,0,0,0,1,1)$, $(0,1,0,0,0,0)$, $(0,2,0,0,0,0)$, $(1,0,0,0,0,0)$, $(1,0,0,0,0,1)$, $(1,0,0,0,1,1)$, $(1,0,0,1,1,1)$, $(1,0,1,1,1,1)$, $(1,2,0,0,0,0)$, $(1,2,0,0,0,1)$, $(1,2,0,0,1,1)$, $(1,2,0,1,1,1)$, $(1,2,1,1,1,1)$, \textcolor{blue}{$(2,2,1,1,1,1)$} & $2$, $1$, $4$, $1$, $1$, $1$, $1$, $1$, $1$, $1$, $1$, $1$, $1$, $1$, \textcolor{blue}{$1$} & $(0,0,0,0,0,2)$, $(0,0,0,0,1,1)$, $(0,2,0,0,0,0)$, $(0,1,0,0,0,0)$, $(\frac{1}{2},0,0,0,0,0)$, $(\frac{1}{2},0,0,0,0,1)$, $(\frac{1}{2},0,0,0,1,1)$, $(\frac{1}{2},0,0,1,1,1)$, $(\frac{1}{2},0,1,1,1,1)$, $(\frac{1}{2},1,0,0,0,0)$, $(\frac{1}{2},1,0,0,0,1)$, $(\frac{1}{2},1,0,0,1,1)$, $(\frac{1}{2},1,0,1,1,1)$, $(\frac{1}{2},1,1,1,1,1)$, \textcolor{blue}{$(1,1,1,1,1,1)$} \\
\hline
%
%
\textcolor{blue}{$(\infty,1,1,1,1,1,1)$} & $(0,0,0,0,0,0,1)$, $(0,0,0,0,0,1,1)$, $(1,0,0,0,0,0,0)$, $(1,0,0,0,0,0,1)$, $(1,0,0,0,0,1,1)$, $(1,0,0,0,1,1,1)$, $(1,0,0,1,1,1,1)$, $(1,0,1,1,1,1,1)$, $(1,1,1,1,1,1,1)$, \textcolor{blue}{$(2,1,1,1,1,1,1)$} & $2$, $1$, $1$, $1$, $1$, $1$, $1$, $1$, $1$, \textcolor{blue}{$1$} & $(0,0,0,0,0,0,2)$, $(0,0,0,0,0,1,1)$, $(\frac{1}{2},0,0,0,0,0,0)$, $(\frac{1}{2},0,0,0,0,0,1)$, $(\frac{1}{2},0,0,0,0,1,1)$, $(\frac{1}{2},0,0,0,1,1,1)$, $(\frac{1}{2},0,0,1,1,1,1)$, $(\frac{1}{2},0,1,1,1,1,1)$, $(\frac{1}{2},1,1,1,1,1,1)$, \textcolor{blue}{$(1,1,1,1,1,1,1)$} \\
\hline
\end{longtable}
\end{landscape}

\bibliography{ref.bib}

\providecommand{\href}[2]{#2}\begingroup\raggedright\begin{thebibliography}{10}

\bibitem{Vafa:2005ui}
C.~Vafa, \emph{{The String landscape and the swampland}},
  \href{https://arxiv.org/abs/hep-th/0509212}{{\ttfamily hep-th/0509212}}.

\bibitem{Brennan:2017rbf}
T.~D. Brennan, F.~Carta and C.~Vafa, \emph{{The String Landscape, the
  Swampland, and the Missing Corner}},
  \href{https://doi.org/10.22323/1.305.0015}{\emph{PoS} {\bfseries TASI2017}
  (2017) 015} [\href{https://arxiv.org/abs/1711.00864}{{\ttfamily
  1711.00864}}].

\bibitem{Palti:2019pca}
E.~Palti, \emph{{The Swampland: Introduction and Review}},
  \href{https://doi.org/10.1002/prop.201900037}{\emph{Fortsch. Phys.}
  {\bfseries 67} (2019) 1900037}
  [\href{https://arxiv.org/abs/1903.06239}{{\ttfamily 1903.06239}}].

\bibitem{vanBeest:2021lhn}
M.~van Beest, J.~Calder\'on-Infante, D.~Mirfendereski and I.~Valenzuela,
  \emph{{Lectures on the Swampland Program in String Compactifications}},
  \href{https://doi.org/10.1016/j.physrep.2022.09.002}{\emph{Phys. Rept.}
  {\bfseries 989} (2022) 1} [\href{https://arxiv.org/abs/2102.01111}{{\ttfamily
  2102.01111}}].

\bibitem{Grana:2021zvf}
M.~Gra\~na and A.~Herr\'aez, \emph{{The Swampland Conjectures: A Bridge from
  Quantum Gravity to Particle Physics}},
  \href{https://doi.org/10.3390/universe7080273}{\emph{Universe} {\bfseries 7}
  (2021) 273} [\href{https://arxiv.org/abs/2107.00087}{{\ttfamily
  2107.00087}}].

\bibitem{Agmon:2022thq}
N.~B. Agmon, A.~Bedroya, M.~J. Kang and C.~Vafa, \emph{{Lectures on the string
  landscape and the Swampland}},
  \href{https://arxiv.org/abs/2212.06187}{{\ttfamily 2212.06187}}.

\bibitem{Ooguri:2006in}
H.~Ooguri and C.~Vafa, \emph{{On the Geometry of the String Landscape and the
  Swampland}},
  \href{https://doi.org/10.1016/j.nuclphysb.2006.10.033}{\emph{Nucl. Phys.}
  {\bfseries B766} (2007) 21}
  [\href{https://arxiv.org/abs/hep-th/0605264}{{\ttfamily hep-th/0605264}}].

\bibitem{Klaewer:2016kiy}
D.~Klaewer and E.~Palti, \emph{{Super-Planckian Spatial Field Variations and
  Quantum Gravity}}, \href{https://doi.org/10.1007/JHEP01(2017)088}{\emph{JHEP}
  {\bfseries 01} (2017) 088}
  [\href{https://arxiv.org/abs/1610.00010}{{\ttfamily 1610.00010}}].

\bibitem{Etheredge:2022opl}
M.~Etheredge, B.~Heidenreich, S.~Kaya, Y.~Qiu and T.~Rudelius,
  \emph{{Sharpening the Distance Conjecture in diverse dimensions}},
  \href{https://doi.org/10.1007/JHEP12(2022)114}{\emph{JHEP} {\bfseries 12}
  (2022) 114} [\href{https://arxiv.org/abs/2206.04063}{{\ttfamily
  2206.04063}}].

\bibitem{Grimm:2018ohb}
T.~W. Grimm, E.~Palti and I.~Valenzuela, \emph{{Infinite Distances in Field
  Space and Massless Towers of States}},
  \href{https://doi.org/10.1007/JHEP08(2018)143}{\emph{JHEP} {\bfseries 08}
  (2018) 143} [\href{https://arxiv.org/abs/1802.08264}{{\ttfamily
  1802.08264}}].

\bibitem{Grimm:2018cpv}
T.~W. Grimm, C.~Li and E.~Palti, \emph{{Infinite Distance Networks in Field
  Space and Charge Orbits}},
  \href{https://doi.org/10.1007/JHEP03(2019)016}{\emph{JHEP} {\bfseries 03}
  (2019) 016} [\href{https://arxiv.org/abs/1811.02571}{{\ttfamily
  1811.02571}}].

\bibitem{Corvilain:2018lgw}
P.~Corvilain, T.~W. Grimm and I.~Valenzuela, \emph{{The Swampland Distance
  Conjecture for Kahler moduli}},
  \href{https://doi.org/10.1007/JHEP08(2019)075}{\emph{JHEP} {\bfseries 08}
  (2019) 075} [\href{https://arxiv.org/abs/1812.07548}{{\ttfamily
  1812.07548}}].

\bibitem{Font:2019cxq}
A.~Font, A.~Herr\'aez and L.~E. Ib\'a\~nez, \emph{{The Swampland Distance
  Conjecture and Towers of Tensionless Branes}},
  \href{https://doi.org/10.1007/JHEP08(2019)044}{\emph{JHEP} {\bfseries 08}
  (2019) 044} [\href{https://arxiv.org/abs/1904.05379}{{\ttfamily
  1904.05379}}].

\bibitem{Lee:2019wij}
S.-J. Lee, W.~Lerche and T.~Weigand, \emph{{Emergent strings from infinite
  distance limits}}, \href{https://doi.org/10.1007/JHEP02(2022)190}{\emph{JHEP}
  {\bfseries 02} (2022) 190}
  [\href{https://arxiv.org/abs/1910.01135}{{\ttfamily 1910.01135}}].

\bibitem{Lee:2018urn}
S.-J. Lee, W.~Lerche and T.~Weigand, \emph{{Tensionless Strings and the Weak
  Gravity Conjecture}},
  \href{https://doi.org/10.1007/JHEP10(2018)164}{\emph{JHEP} {\bfseries 10}
  (2018) 164} [\href{https://arxiv.org/abs/1808.05958}{{\ttfamily
  1808.05958}}].

\bibitem{Lee:2019apr}
S.-J. Lee, W.~Lerche and T.~Weigand, \emph{{Emergent Strings, Duality and Weak
  Coupling Limits for Two-Form Fields}},
  \href{https://arxiv.org/abs/1904.06344}{{\ttfamily 1904.06344}}.

\bibitem{Lee:2019tst}
S.-J. Lee, W.~Lerche and T.~Weigand, \emph{{Modular Fluxes, Elliptic Genera,
  and Weak Gravity Conjectures in Four Dimensions}},
  \href{https://doi.org/10.1007/JHEP08(2019)104}{\emph{JHEP} {\bfseries 08}
  (2019) 104} [\href{https://arxiv.org/abs/1901.08065}{{\ttfamily
  1901.08065}}].

\bibitem{Lee:2020gvu}
S.-J. Lee, W.~Lerche, G.~Lockhart and T.~Weigand, \emph{{Quasi-Jacobi Forms,
  Elliptic Genera and Strings in Four Dimensions}},
  \href{https://arxiv.org/abs/2005.10837}{{\ttfamily 2005.10837}}.

\bibitem{Lee:2021usk}
S.-J. Lee, W.~Lerche and T.~Weigand, \emph{{Physics of infinite complex
  structure limits in eight dimensions}},
  \href{https://doi.org/10.1007/JHEP06(2022)042}{\emph{JHEP} {\bfseries 06}
  (2022) 042} [\href{https://arxiv.org/abs/2112.08385}{{\ttfamily
  2112.08385}}].

\bibitem{Alvarez-Garcia:2021pxo}
R.~\'Alvarez-Garc\'\i{}a, D.~Kl\"awer and T.~Weigand, \emph{{Membrane Limits in
  Quantum Gravity}},  \href{https://arxiv.org/abs/2112.09136}{{\ttfamily
  2112.09136}}.

\bibitem{Lanza:2020qmt}
S.~Lanza, F.~Marchesano, L.~Martucci and I.~Valenzuela, \emph{{Swampland
  Conjectures for Strings and Membranes}},
  \href{https://doi.org/10.1007/JHEP02(2021)006}{\emph{JHEP} {\bfseries 02}
  (2021) 006} [\href{https://arxiv.org/abs/2006.15154}{{\ttfamily
  2006.15154}}].

\bibitem{Lanza:2021udy}
S.~Lanza, F.~Marchesano, L.~Martucci and I.~Valenzuela, \emph{{The EFT stringy
  viewpoint on large distances}},
  \href{https://doi.org/10.1007/JHEP09(2021)197}{\emph{JHEP} {\bfseries 09}
  (2021) 197} [\href{https://arxiv.org/abs/2104.05726}{{\ttfamily
  2104.05726}}].

\bibitem{Ferrara:1995ih}
S.~Ferrara, R.~Kallosh and A.~Strominger, \emph{{N=2 extremal black holes}},
  \href{https://doi.org/10.1103/PhysRevD.52.R5412}{\emph{Phys. Rev. D}
  {\bfseries 52} (1995) R5412}
  [\href{https://arxiv.org/abs/hep-th/9508072}{{\ttfamily hep-th/9508072}}].

\bibitem{Ferrara:1997tw}
S.~Ferrara, G.~W. Gibbons and R.~Kallosh, \emph{{Black holes and critical
  points in moduli space}},
  \href{https://doi.org/10.1016/S0550-3213(97)00324-6}{\emph{Nucl. Phys. B}
  {\bfseries 500} (1997) 75}
  [\href{https://arxiv.org/abs/hep-th/9702103}{{\ttfamily hep-th/9702103}}].

\bibitem{Sen:2005wa}
A.~Sen, \emph{{Black hole entropy function and the attractor mechanism in
  higher derivative gravity}},
  \href{https://doi.org/10.1088/1126-6708/2005/09/038}{\emph{JHEP} {\bfseries
  09} (2005) 038} [\href{https://arxiv.org/abs/hep-th/0506177}{{\ttfamily
  hep-th/0506177}}].

\bibitem{Bonnefoy:2019nzv}
Q.~Bonnefoy, L.~Ciambelli, D.~L\"ust and S.~L\"ust, \emph{{Infinite Black Hole
  Entropies at Infinite Distances and Tower of States}},
  \href{https://doi.org/10.1016/j.nuclphysb.2020.115112}{\emph{Nucl. Phys. B}
  {\bfseries 958} (2020) 115112}
  [\href{https://arxiv.org/abs/1912.07453}{{\ttfamily 1912.07453}}].

\bibitem{Cribiori:2022nke}
N.~Cribiori, D.~L{\"u}st and G.~Staudt, \emph{{Black hole entropy and
  moduli-dependent species scale}},
  \href{https://doi.org/10.1016/j.physletb.2023.138113}{\emph{Phys. Lett. B}
  {\bfseries 844} (2023) 138113}
  [\href{https://arxiv.org/abs/2212.10286}{{\ttfamily 2212.10286}}].

\bibitem{Calderon-Infante:2025pls}
J.~Calder{\'o}n-Infante, M.~Delgado, Y.~Li, D.~Lust and A.~M. Uranga,
  \emph{{Classical black hole probes of UV scales}},
  \href{https://doi.org/10.1007/JHEP06(2025)061}{\emph{JHEP} {\bfseries 06}
  (2025) 061} [\href{https://arxiv.org/abs/2502.03514}{{\ttfamily
  2502.03514}}].

\bibitem{Buratti:2021yia}
G.~Buratti, M.~Delgado and A.~M. Uranga, \emph{{Dynamical tadpoles, stringy
  cobordism, and the SM from spontaneous compactification}},
  \href{https://doi.org/10.1007/JHEP06(2021)170}{\emph{JHEP} {\bfseries 06}
  (2021) 170} [\href{https://arxiv.org/abs/2104.02091}{{\ttfamily
  2104.02091}}].

\bibitem{Buratti:2021fiv}
G.~Buratti, J.~Calder\'on-Infante, M.~Delgado and A.~M. Uranga,
  \emph{{Dynamical Cobordism and Swampland Distance Conjectures}},
  \href{https://doi.org/10.1007/JHEP10(2021)037}{\emph{JHEP} {\bfseries 10}
  (2021) 037} [\href{https://arxiv.org/abs/2107.09098}{{\ttfamily
  2107.09098}}].

\bibitem{Angius:2022aeq}
R.~Angius, J.~Calder\'on-Infante, M.~Delgado, J.~Huertas and A.~M. Uranga,
  \emph{{At the end of the world: Local Dynamical Cobordism}},
  \href{https://doi.org/10.1007/JHEP06(2022)142}{\emph{JHEP} {\bfseries 06}
  (2022) 142} [\href{https://arxiv.org/abs/2203.11240}{{\ttfamily
  2203.11240}}].

\bibitem{McNamara:2019rup}
J.~McNamara and C.~Vafa, \emph{{Cobordism Classes and the Swampland}},
  \href{https://arxiv.org/abs/1909.10355}{{\ttfamily 1909.10355}}.

\bibitem{Blumenhagen:2022mqw}
R.~Blumenhagen, N.~Cribiori, C.~Kneissl and A.~Makridou, \emph{{Dynamical
  cobordism of a domain wall and its companion defect 7-brane}},
  \href{https://doi.org/10.1007/JHEP08(2022)204}{\emph{JHEP} {\bfseries 08}
  (2022) 204} [\href{https://arxiv.org/abs/2205.09782}{{\ttfamily
  2205.09782}}].

\bibitem{Angius:2022mgh}
R.~Angius, M.~Delgado and A.~M. Uranga, \emph{{Dynamical Cobordism and the
  beginning of time: supercritical strings and tachyon condensation}},
  \href{https://doi.org/10.1007/JHEP08(2022)285}{\emph{JHEP} {\bfseries 08}
  (2022) 285} [\href{https://arxiv.org/abs/2207.13108}{{\ttfamily
  2207.13108}}].

\bibitem{Blumenhagen:2023abk}
R.~Blumenhagen, C.~Kneissl and C.~Wang, \emph{{Dynamical Cobordism Conjecture:
  solutions for end-of-the-world branes}},
  \href{https://doi.org/10.1007/JHEP05(2023)123}{\emph{JHEP} {\bfseries 05}
  (2023) 123} [\href{https://arxiv.org/abs/2303.03423}{{\ttfamily
  2303.03423}}].

\bibitem{Calderon-Infante:2023ler}
J.~Calder{\'o}n-Infante, A.~Castellano, A.~Herr{\'a}ez and L.~E.
  Ib{\'a}{\~n}ez, \emph{{Entropy bounds and the species scale distance
  conjecture}}, \href{https://doi.org/10.1007/JHEP01(2024)039}{\emph{JHEP}
  {\bfseries 01} (2024) 039}
  [\href{https://arxiv.org/abs/2306.16450}{{\ttfamily 2306.16450}}].

\bibitem{Angius:2023xtu}
R.~Angius, J.~Huertas and A.~M. Uranga, \emph{{Small black hole explosions}},
  \href{https://doi.org/10.1007/JHEP06(2023)070}{\emph{JHEP} {\bfseries 06}
  (2023) 070} [\href{https://arxiv.org/abs/2303.15903}{{\ttfamily
  2303.15903}}].

\bibitem{Huertas:2023syg}
J.~Huertas and A.~M. Uranga, \emph{{Aspects of dynamical cobordism in
  AdS/CFT}}, \href{https://doi.org/10.1007/JHEP08(2023)140}{\emph{JHEP}
  {\bfseries 08} (2023) 140}
  [\href{https://arxiv.org/abs/2306.07335}{{\ttfamily 2306.07335}}].

\bibitem{Angius:2023uqk}
R.~Angius, A.~Makridou and A.~M. Uranga, \emph{{Intersecting end of the world
  branes}}, \href{https://doi.org/10.1007/JHEP03(2024)110}{\emph{JHEP}
  {\bfseries 03} (2024) 110}
  [\href{https://arxiv.org/abs/2312.16286}{{\ttfamily 2312.16286}}].

\bibitem{Angius:2024zjv}
R.~Angius, \emph{{End of the world brane networks for infinite distance limits
  in CY moduli space}},
  \href{https://doi.org/10.1007/JHEP09(2024)178}{\emph{JHEP} {\bfseries 09}
  (2024) 178} [\href{https://arxiv.org/abs/2404.14486}{{\ttfamily
  2404.14486}}].

\bibitem{Huertas:2024mvy}
J.~Huertas and A.~M. Uranga, \emph{{End of the world brane dynamics in
  holographic 4d $ \mathcal{N} $ = 4 SU(N) with 3d $ \mathcal{N} $ = 2 boundary
  conditions}}, \href{https://doi.org/10.1007/JHEP01(2025)002}{\emph{JHEP}
  {\bfseries 01} (2025) 002}
  [\href{https://arxiv.org/abs/2410.05368}{{\ttfamily 2410.05368}}].

\bibitem{Angius:2024pqk}
R.~Angius, A.~M. Uranga and C.~Wang, \emph{{End of the world boundaries for
  chiral quantum gravity theories}},
  \href{https://doi.org/10.1007/JHEP03(2025)064}{\emph{JHEP} {\bfseries 03}
  (2025) 064} [\href{https://arxiv.org/abs/2410.07322}{{\ttfamily
  2410.07322}}].

\bibitem{Calderon-Infante:2026ymy}
J.~Calder{\'o}n-Infante, G.~Cheng, A.~Herr{\'a}ez and T.~Van~Riet,
  \emph{{End-of-the-World Singularities: The Good, the Bad, and the
  Heated-up}},  \href{https://arxiv.org/abs/2603.18133}{{\ttfamily
  2603.18133}}.

\bibitem{Makridou:2026jzy}
A.~Makridou and A.~J.~P. G{\'o}mez, \emph{{Sharpened Dynamical Cobordism}},
  \href{https://arxiv.org/abs/2605.06793}{{\ttfamily 2605.06793}}.

\bibitem{Lanza:2022zyg}
S.~Lanza, F.~Marchesano, L.~Martucci and I.~Valenzuela, \emph{{Large Field
  Distances from EFT strings}},  in \emph{{21st Hellenic School and Workshops
  on Elementary Particle Physics and Gravity}}, 5, 2022,
  \href{https://arxiv.org/abs/2205.04532}{{\ttfamily 2205.04532}}.

\bibitem{Heidenreich:2021yda}
B.~Heidenreich, M.~Reece and T.~Rudelius, \emph{{The Weak Gravity Conjecture
  and axion strings}},
  \href{https://doi.org/10.1007/JHEP11(2021)004}{\emph{JHEP} {\bfseries 11}
  (2021) 004} [\href{https://arxiv.org/abs/2108.11383}{{\ttfamily
  2108.11383}}].

\bibitem{Cota:2022yjw}
C.~F. Cota, A.~Mininno, T.~Weigand and M.~Wiesner, \emph{{The Asymptotic Weak
  Gravity Conjecture for Open Strings}},
  \href{https://arxiv.org/abs/2208.00009}{{\ttfamily 2208.00009}}.

\bibitem{Klaewer:2020lfg}
D.~Klaewer, S.-J. Lee, T.~Weigand and M.~Wiesner, \emph{{Quantum corrections in
  4d $N$ = 1 infinite distance limits and the weak gravity conjecture}},
  \href{https://doi.org/10.1007/JHEP03(2021)252}{\emph{JHEP} {\bfseries 03}
  (2021) 252} [\href{https://arxiv.org/abs/2011.00024}{{\ttfamily
  2011.00024}}].

\bibitem{Kaufmann:2026mha}
L.~Kaufmann, J.~Monnee, T.~Weigand and M.~Wiesner, \emph{{Quantum obstructions
  for $N=1$ infinite distance limits -- Part II: K{\"a}hler obstructions}},
  \href{https://doi.org/10.1103/ypyx-mg6r}{\emph{Phys. Rev. D} {\bfseries 113}
  (2026) 126028} [\href{https://arxiv.org/abs/2603.13470}{{\ttfamily
  2603.13470}}].

\bibitem{Kaufmann:2026tsy}
L.~Kaufmann, T.~Weigand and M.~Wiesner, \emph{{On Quantum Obstructions in Type
  IIA Orientifolds}},  \href{https://arxiv.org/abs/2604.25988}{{\ttfamily
  2604.25988}}.

\bibitem{Marchesano:2022avb}
F.~Marchesano and M.~Wiesner, \emph{{4d strings at strong coupling}},
  \href{https://doi.org/10.1007/JHEP08(2022)004}{\emph{JHEP} {\bfseries 08}
  (2022) 004} [\href{https://arxiv.org/abs/2202.10466}{{\ttfamily
  2202.10466}}].

\bibitem{Wiesner:2022qys}
M.~Wiesner, \emph{{Light Strings and Strong Coupling in F-theory}},
  \href{https://arxiv.org/abs/2210.14238}{{\ttfamily 2210.14238}}.

\bibitem{Martucci:2022krl}
L.~Martucci, N.~Risso and T.~Weigand, \emph{{Quantum Gravity Bounds on N=1
  Effective Theories in Four Dimensions}},
  \href{https://arxiv.org/abs/2210.10797}{{\ttfamily 2210.10797}}.

\bibitem{Martucci:2024trp}
L.~Martucci, N.~Risso, A.~Valenti and L.~Vecchi, \emph{{Wormholes in the
  axiverse, and the species scale}},
  \href{https://doi.org/10.1007/JHEP07(2024)240}{\emph{JHEP} {\bfseries 07}
  (2024) 240} [\href{https://arxiv.org/abs/2404.14489}{{\ttfamily
  2404.14489}}].

\bibitem{Marchesano:2022axe}
F.~Marchesano and L.~Melotti, \emph{{EFT strings and emergence}},
  \href{https://doi.org/10.1007/JHEP02(2023)112}{\emph{JHEP} {\bfseries 02}
  (2023) 112} [\href{https://arxiv.org/abs/2211.01409}{{\ttfamily
  2211.01409}}].

\bibitem{Marchesano:2023thx}
F.~Marchesano, L.~Melotti and L.~Paoloni, \emph{{On the moduli space curvature
  at infinity}}, \href{https://doi.org/10.1007/JHEP02(2024)103}{\emph{JHEP}
  {\bfseries 02} (2024) 103}
  [\href{https://arxiv.org/abs/2311.07979}{{\ttfamily 2311.07979}}].

\bibitem{Marchesano:2024tod}
F.~Marchesano, L.~Melotti and M.~Wiesner, \emph{{Asymptotic curvature
  divergences and non-gravitational theories}},
  \href{https://arxiv.org/abs/2409.02991}{{\ttfamily 2409.02991}}.

\bibitem{Casas:2024ttx}
G.~F. Casas, L.~E. Ib{\'a}{\~n}ez and F.~Marchesano, \emph{{Yukawa couplings at
  infinite distance and swampland towers in chiral theories}},
  \href{https://doi.org/10.1007/JHEP09(2024)170}{\emph{JHEP} {\bfseries 09}
  (2024) 170} [\href{https://arxiv.org/abs/2403.09775}{{\ttfamily
  2403.09775}}].

\bibitem{Grimm:2022sbl}
T.~W. Grimm, S.~Lanza and C.~Li, \emph{{Tameness, Strings, and the Distance
  Conjecture}}, \href{https://doi.org/10.1007/JHEP09(2022)149}{\emph{JHEP}
  {\bfseries 09} (2022) 149}
  [\href{https://arxiv.org/abs/2206.00697}{{\ttfamily 2206.00697}}].

\bibitem{Hassfeld:2025uoy}
B.~Hassfeld, J.~Monnee, T.~Weigand and M.~Wiesner, \emph{{Emergent strings in
  Type IIB Calabi-Yau compactifications}},
  \href{https://doi.org/10.1007/JHEP01(2026)140}{\emph{JHEP} {\bfseries 01}
  (2026) 140} [\href{https://arxiv.org/abs/2504.01066}{{\ttfamily
  2504.01066}}].

\bibitem{Monnee:2025ynn}
J.~Monnee, T.~Weigand and M.~Wiesner, \emph{{Physics and geometry of complex
  structure limits in type IIB Calabi-Yau compactifications}},
  \href{https://doi.org/10.1007/JHEP03(2026)063}{\emph{JHEP} {\bfseries 03}
  (2026) 063} [\href{https://arxiv.org/abs/2509.07056}{{\ttfamily
  2509.07056}}].

\bibitem{Grieco:2025bjy}
A.~Grieco, I.~Ruiz and I.~Valenzuela, \emph{{EFT strings and dualities in 4d
  {\ensuremath{\mathscr{N}}} = 1}},
  \href{https://doi.org/10.1007/JHEP06(2026)129}{\emph{JHEP} {\bfseries 06}
  (2026) 129} [\href{https://arxiv.org/abs/2504.16984}{{\ttfamily
  2504.16984}}].

\bibitem{GriecoRuizValenzuelaToAppear}
A.~Grieco, I.~Ruiz and I.~Valenzuela, \emph{{EFT (String) Tower Building, To
  appear}}, .

\bibitem{Stout:2022phm}
J.~Stout, \emph{{Infinite Distances and Factorization}},
  \href{https://arxiv.org/abs/2208.08444}{{\ttfamily 2208.08444}}.

\bibitem{Basile:2023blg}
I.~Basile, D.~L{\"u}st and C.~Montella, \emph{{Shedding black hole light on the
  emergent string conjecture}},
  \href{https://doi.org/10.1007/JHEP07(2024)208}{\emph{JHEP} {\bfseries 07}
  (2024) 208} [\href{https://arxiv.org/abs/2311.12113}{{\ttfamily
  2311.12113}}].

\bibitem{Bedroya:2024ubj}
A.~Bedroya, R.~K. Mishra and M.~Wiesner, \emph{{Density of states, black holes
  and the Emergent String Conjecture}},
  \href{https://doi.org/10.1007/JHEP01(2025)144}{\emph{JHEP} {\bfseries 01}
  (2025) 144} [\href{https://arxiv.org/abs/2405.00083}{{\ttfamily
  2405.00083}}].

\bibitem{Herraez:2024kux}
A.~Herr{\'a}ez, D.~L{\"u}st, J.~Masias and M.~Scalisi, \emph{{On the origin of
  species thermodynamics and the black hole - tower correspondence}},
  \href{https://doi.org/10.21468/SciPostPhys.18.3.083}{\emph{SciPost Phys.}
  {\bfseries 18} (2025) 083}
  [\href{https://arxiv.org/abs/2406.17851}{{\ttfamily 2406.17851}}].

\bibitem{Kaufmann:2024gqo}
L.~Kaufmann, S.~Lanza and T.~Weigand, \emph{{Asymptotics of 5d supergravity
  theories and the emergent string conjecture}},
  \href{https://doi.org/10.1007/JHEP06(2025)230}{\emph{JHEP} {\bfseries 06}
  (2025) 230} [\href{https://arxiv.org/abs/2412.12251}{{\ttfamily
  2412.12251}}].

\bibitem{Calderon-Infante:2020dhm}
J.~Calder{\'o}n-Infante, A.~M. Uranga and I.~Valenzuela, \emph{{The Convex Hull
  Swampland Distance Conjecture and Bounds on Non-geodesics}},
  \href{https://doi.org/10.1007/JHEP03(2021)299}{\emph{JHEP} {\bfseries 03}
  (2021) 299} [\href{https://arxiv.org/abs/2012.00034}{{\ttfamily
  2012.00034}}].

\bibitem{Cheung:2014vva}
C.~Cheung and G.~N. Remmen, \emph{{Naturalness and the Weak Gravity
  Conjecture}},
  \href{https://doi.org/10.1103/PhysRevLett.113.051601}{\emph{Phys. Rev. Lett.}
  {\bfseries 113} (2014) 051601}
  [\href{https://arxiv.org/abs/1402.2287}{{\ttfamily 1402.2287}}].

\bibitem{Dvali:2007hz}
G.~Dvali, \emph{{Black Holes and Large N Species Solution to the Hierarchy
  Problem}}, \href{https://doi.org/10.1002/prop.201000009}{\emph{Fortsch.
  Phys.} {\bfseries 58} (2010) 528}
  [\href{https://arxiv.org/abs/0706.2050}{{\ttfamily 0706.2050}}].

\bibitem{Dvali:2007wp}
G.~Dvali and M.~Redi, \emph{{Black Hole Bound on the Number of Species and
  Quantum Gravity at LHC}},
  \href{https://doi.org/10.1103/PhysRevD.77.045027}{\emph{Phys. Rev. D}
  {\bfseries 77} (2008) 045027}
  [\href{https://arxiv.org/abs/0710.4344}{{\ttfamily 0710.4344}}].

\bibitem{Dvali:2009ks}
G.~Dvali and D.~Lust, \emph{{Evaporation of Microscopic Black Holes in String
  Theory and the Bound on Species}},
  \href{https://doi.org/10.1002/prop.201000008}{\emph{Fortsch. Phys.}
  {\bfseries 58} (2010) 505} [\href{https://arxiv.org/abs/0912.3167}{{\ttfamily
  0912.3167}}].

\bibitem{Dvali:2010vm}
G.~Dvali and C.~Gomez, \emph{{Species and Strings}},
  \href{https://arxiv.org/abs/1004.3744}{{\ttfamily 1004.3744}}.

\bibitem{vandeHeisteeg:2022btw}
D.~van~de Heisteeg, C.~Vafa, M.~Wiesner and D.~H. Wu, \emph{{Moduli-dependent
  Species Scale}},  \href{https://arxiv.org/abs/2212.06841}{{\ttfamily
  2212.06841}}.

\bibitem{Castellano:2022bvr}
A.~Castellano, A.~Herr\'aez and L.~E. Ib\'a\~nez, \emph{{The Emergence Proposal
  in Quantum Gravity and the Species Scale}},
  \href{https://arxiv.org/abs/2212.03908}{{\ttfamily 2212.03908}}.

\bibitem{vandeHeisteeg:2023ubh}
D.~van~de Heisteeg, C.~Vafa and M.~Wiesner, \emph{{Bounds on Species Scale and
  the Distance Conjecture}},
  \href{https://arxiv.org/abs/2303.13580}{{\ttfamily 2303.13580}}.

\bibitem{vandeHeisteeg:2023dlw}
D.~van~de Heisteeg, C.~Vafa, M.~Wiesner and D.~H. Wu, \emph{{Species Scale in
  Diverse Dimensions}},  \href{https://arxiv.org/abs/2310.07213}{{\ttfamily
  2310.07213}}.

\bibitem{Castellano:2023aum}
A.~Castellano, A.~Herr\'aez and L.~E. Ib\'a\~nez, \emph{{On the Species Scale,
  Modular Invariance and the Gravitational EFT expansion}},
  \href{https://arxiv.org/abs/2310.07708}{{\ttfamily 2310.07708}}.

\bibitem{Calderon-Infante:2025ldq}
J.~Calder{\'o}n-Infante, A.~Castellano and A.~Herr{\'a}ez, \emph{{The double
  EFT expansion in quantum gravity}},
  \href{https://doi.org/10.21468/SciPostPhys.19.4.096}{\emph{SciPost Phys.}
  {\bfseries 19} (2025) 096}
  [\href{https://arxiv.org/abs/2501.14880}{{\ttfamily 2501.14880}}].

\bibitem{Castellano:2023stg}
A.~Castellano, I.~Ruiz and I.~Valenzuela, \emph{{Universal Pattern in Quantum
  Gravity at Infinite Distance}},
  \href{https://doi.org/10.1103/PhysRevLett.132.181601}{\emph{Phys. Rev. Lett.}
  {\bfseries 132} (2024) 181601}
  [\href{https://arxiv.org/abs/2311.01501}{{\ttfamily 2311.01501}}].

\bibitem{Castellano:2023jjt}
A.~Castellano, I.~Ruiz and I.~Valenzuela, \emph{{Stringy evidence for a
  universal pattern at infinite distance}},
  \href{https://doi.org/10.1007/JHEP06(2024)037}{\emph{JHEP} {\bfseries 06}
  (2024) 037} [\href{https://arxiv.org/abs/2311.01536}{{\ttfamily
  2311.01536}}].

\bibitem{Etheredge:2024tok}
M.~Etheredge, B.~Heidenreich, T.~Rudelius, I.~Ruiz and I.~Valenzuela,
  \emph{{Taxonomy of infinite distance limits}},
  \href{https://doi.org/10.1007/JHEP03(2025)213}{\emph{JHEP} {\bfseries 03}
  (2025) 213} [\href{https://arxiv.org/abs/2405.20332}{{\ttfamily
  2405.20332}}].

\bibitem{Etheredge:2025ahf}
M.~Etheredge, \emph{{Taxonomy of branes in infinite distance limits}},
  \href{https://doi.org/10.1007/JHEP10(2025)200}{\emph{JHEP} {\bfseries 10}
  (2025) 200} [\href{https://arxiv.org/abs/2505.10615}{{\ttfamily
  2505.10615}}].

\bibitem{Etheredge:2026rio}
M.~Etheredge, M.~Reece, T.~Rudelius and C.~Tudball, \emph{{Sharpening the
  Supersymmetric Axion Weak Gravity Conjecture}},
  \href{https://arxiv.org/abs/2605.22912}{{\ttfamily 2605.22912}}.

\bibitem{Etheredge:2023odp}
M.~Etheredge, B.~Heidenreich, J.~McNamara, T.~Rudelius, I.~Ruiz and
  I.~Valenzuela, \emph{{Running decompactification, sliding towers, and the
  distance conjecture}},
  \href{https://doi.org/10.1007/JHEP12(2023)182}{\emph{JHEP} {\bfseries 12}
  (2023) 182} [\href{https://arxiv.org/abs/2306.16440}{{\ttfamily
  2306.16440}}].

\bibitem{Aoufia:2026ztl}
C.~Aoufia, M.~Etheredge, B.~Fraiman, S.~Raman and A.~Stewart,
  \emph{{Moduli-Space Laplacians, Asymptotic Geometry, and the Emergent String
  Conjecture}},  \href{https://arxiv.org/abs/2607.20603}{{\ttfamily
  2607.20603}}.

\bibitem{Raucci:2026fzp}
S.~Raucci, I.~Ruiz and I.~Valenzuela, \emph{{Alice in Warpland: KK modes,
  Warped Compactifications and the Swampland}},
  \href{https://arxiv.org/abs/2603.11163}{{\ttfamily 2603.11163}}.

\bibitem{Candelas:1994hw}
P.~Candelas, A.~Font, S.~H. Katz and D.~R. Morrison, \emph{{Mirror symmetry for
  two parameter models. 2.}},
  \href{https://doi.org/10.1016/0550-3213(94)90155-4}{\emph{Nucl. Phys. B}
  {\bfseries 429} (1994) 626}
  [\href{https://arxiv.org/abs/hep-th/9403187}{{\ttfamily hep-th/9403187}}].

\bibitem{Hosono:1993qy}
S.~Hosono, A.~Klemm, S.~Theisen and S.-T. Yau, \emph{{Mirror symmetry, mirror
  map and applications to Calabi-Yau hypersurfaces}},
  \href{https://doi.org/10.1007/BF02100589}{\emph{Commun. Math. Phys.}
  {\bfseries 167} (1995) 301}
  [\href{https://arxiv.org/abs/hep-th/9308122}{{\ttfamily hep-th/9308122}}].

\bibitem{Kaufmann:2026fli}
L.~Kaufmann, J.~Monnee, T.~Weigand and M.~Wiesner, \emph{{Quantum obstructions
  for $N=1$ infinite distance limits -- Part I: $g_s$ obstructions}},
  \href{https://arxiv.org/abs/2603.12315}{{\ttfamily 2603.12315}}.

\bibitem{Joyce:1}
D.~Joyce, \emph{{Compact Riemannian 7-Manifolds with Holonomy
  {\ensuremath{G_2}} I}}, {\emph{J. Diff. Geom.} {\bfseries 43} (1996) 291}.

\bibitem{Joyce:2}
D.~Joyce, \emph{{Compact Riemannian 7-Manifolds with Holonomy
  {\ensuremath{G_2}} II}}, {\emph{J. Diff. Geom.} {\bfseries 43} (1996) 329}.

\bibitem{Joyce:book}
D.~Joyce, \emph{Compact Manifolds with Special Holonomy}. Cambridge University
  Press, 2000.

\bibitem{DiUbaldo:2026rly}
G.~Di~Ubaldo, L.~V. Iliesiu, H.~W. Lin and C.~Yan, \emph{{Positivity of the
  gravitational path integral implies the axionic weak gravity conjecture}},
  \href{https://arxiv.org/abs/2605.05305}{{\ttfamily 2605.05305}}.

\bibitem{Reece:2025zva}
M.~Reece, T.~Rudelius and C.~Tudball, \emph{{Co-scaling and alignment of
  electric and magnetic towers}},
  \href{https://doi.org/10.1007/JHEP09(2025)146}{\emph{JHEP} {\bfseries 09}
  (2025) 146} [\href{https://arxiv.org/abs/2505.22713}{{\ttfamily
  2505.22713}}].

\end{thebibliography}\endgroup
\bibliographystyle{JHEP}

\end{document}